\documentclass[a4paper,fleqn,usenatbib]{mnras}

\usepackage{newtxtext,newtxmath}
\usepackage[T1]{fontenc}
\usepackage{ae,aecompl}

\usepackage[version=3]{mhchem}
\usepackage{placeins}
\usepackage{enumitem}
\usepackage{rotating}
\usepackage{graphicx}	% Including figure files
\usepackage{amsmath}	% Advanced maths commands
\usepackage{amssymb}	% Extra maths symbols

\graphicspath{ {/Users/AislinnMcdonagh/Downloads/} }
\title[Modelling Observed Pollutant Abundances]{Polluted White Dwarfs: Constraints on the Origin and Geology of Exoplanetary Material}

\author[J. H. D. Harrison et al.]{
John H. D. Harrison,\thanks{E-mail: jhdh2@cam.ac.uk}
Amy Bonsor,
and Nikku Madhusudhan
\\
Institute of Astronomy, University of Cambridge, Madingley Road, Cambridge, CB3 0HA, UK\\
}

\date{Accepted 2018 June 21. Received 2018 June 21; in original form 2017 December 18}

\pubyear{2018}

\begin{document}
\label{firstpage}
\pagerange{\pageref{firstpage}--\pageref{lastpage}}
\maketitle

% Abstract of the paper
\begin{abstract}
White dwarfs that have accreted rocky planetary bodies provide unique insights regarding the bulk composition of exoplanetary material. The analysis presented here uses observed white dwarf atmospheric abundances to constrain both where in the planetary system the pollutant bodies originated, and the geological and collisional history of the pollutant bodies. At least one, but possibly up to nine, of the 17 systems analysed have accreted a body dominated by either core-like or mantle-like material. The approximately even spread in the core mass fraction of the pollutants and the lack of crust-rich pollutants in the 17 systems studied here suggest that the pollutants are often the fragments produced by the collision of larger differentiated bodies. The compositions of many pollutants exhibit trends related to elemental volatility, which we link to the temperatures and, thus, the locations at which these bodies formed. Our analysis found that the abundances observed in 11 of the 17 systems considered are consistent with the compositions of nearby stars in combination with a trend related to elemental volatility. The even spread and large range in the predicted formation location of the pollutants suggests that pollutants arrive in white dwarf atmospheres with a roughly equal efficiency from a wide range of radial locations. Ratios of elements with different condensation temperatures such as Ca/Mg, Na/Mg, and O/Mg distinguish between different formation temperatures, whilst pairs of ratios of siderophilic and lithophilic elements such as Fe/Mg, Ni/Mg and Al/Mg, Ca/Mg distinguish between temperature dependent trends and geological trends.

\end{abstract}

% Select between one and six entries from the list of approved keywords.
% Don't make up new ones.
\begin{keywords}
 white dwarfs -- stars: abundances -- planets and satellites: composition, formation -- minor planets, asteroids: general -- protoplanetary discs
\end{keywords}

%%%%%%%%%%%%%%%%%%%%%%%%%%%%%%%%%%%%%%%%%%%%%%%%%%

%%%%%%%%%%%%%%%%% BODY OF PAPER %%%%%%%%%%%%%%%%%%

\section{Introduction}

Our current knowledge of the bulk chemical composition of planets, including our own Earth, is limited. For an exoplanet, our only method of constraining the bulk composition comes from observations of the planet's mass and radius. Planetary structure models can produce mass-radius curves for a variety of chemical compositions, however many different compositions can produce the same total mass and total radius \citep{Seager2007}. In the solar system the composition of meteorites can be analysed to help resolve this issue.

Chondrites are primitive (unprocessed and unaltered post formation) planetesimals originating in the asteroid belt which fall to the Earth as meteorites, therefore, we can measure their chemical compositions in the laboratory \citep{Sears2005}. If it is assumed that the chondrites and the Earth formed out of the same protoplanetary disc, and that the formation of terrestrial planets mainly involves the aggregation of these planetesimals \citep{Chambers2004}, then the bulk composition of the chondrites should be similar to the bulk composition of Earth \citep{McDonough1995}. The chondritic reference model does not provide a perfect match to the Earth's observed composition, total mass, total radius, or its internal structure, and its validity is debated \citep{McDonough1995, Sears2005, Campbell2012}. However, to the accuracy to which the solar photospheric abundances can be measured, the Sun, the Earth, and the chondrites are a perfect match in the non-volatile elements \citep{McDonough1995}. The difference in the volatile elemental abundances found in the Sun, the Earth, and the chondrites are well explained by the condensation temperatures of the volatile elements \citep{Lodders2003, Lodders2010}. This suggests that the formation location of planetary bodies can be probed by analysing the depletion in the abundances of these species with respect to the host star's abundances, and that observations of the composition of rocky exoplanetary material could be critical in resolving the exoplanet mass-radius degeneracy problem.

From rock samples and seismic data we understand that the Earth is differentiated into a siderophile (iron-loving element) rich core and a lithophile (rock-loving element) rich crust and mantle \citep{DZIEWONSKI1981, McDonough2003}. However in this regard the Earth is not unique; all the terrestrial planets are differentiated, as well as many of the solar system's moons and asteroids \citep{SolarPlanets2011,SolarPlanets2015}. For smaller bodies whose mass is not substantial enough to produce the interior heat required for core formation, the isotope Aluminium 26 is key in driving the differentiation process, as its decay produces the heat required to allow the onset of core formation \citep{Ghosh2006}. The abundance of this isotope has been cited as an important factor in the differentiation of the asteroids Vesta and Ceres \citep{Ceres2005, Vesta2012}. Not all meteorites which fall to Earth are primitive, and thus their abundances cannot be solely explained by the solar photospheric abundances in combination with a volatility based depletion trend. Asteroidal achondrites are enhanced in lithophiles and depleted in siderophiles, whereas iron meteorites are enhanced in siderophiles and depleted in lithophiles \citep{WIIK1956, Scott1975}. These meteorite suites are understood to be the fragments of larger differentiated bodies, which were produced when differentiated bodies in the asteroid belt catastrophically collided \citep{McSween1987}. Collisional grinding of differentiated bodies has been shown to lead to fragments with a similar range of compositions as those observed in non-primitive meteorites \citep{Marcus2009, bonsorleinhardt, Carter2015}. Therefore, by analysing the abundances of the lithophilic and siderophilic species present in planetary bodies it should be possible to probe their geological and collisional history.

Currently the most direct way to observe the composition of exoplanetary material is by measuring the chemical abundances of the rocky bodies which accrete onto white dwarfs, and pollute their atmospheres. Strong metal absorption lines have been detected in many white dwarf spectra \citep{Weidemann1960, Zuckerman1998, Kepler2016}. The first such detection was Van Maanen's star, which was discovered in 1917 \citep{vanMaanen1917}, and was found to have strong calcium and iron absorption lines present in 1919 \citep{vanMaanen1919}. Since then many more polluted white dwarfs have been discovered, including many with several strong metal features \citep{Koester2014}. Metals in the atmospheres of these white dwarfs should sink and become unobservable on timescales of hundreds to millions of years for Helium dominated (DB) white dwarfs, and on timescales of days to thousands of years for Hydrogen dominated (DA) white dwarfs \citep{Koester09}; hence, due to the considerable cooling ages of these white dwarfs, these metals must have been accreted onto the white dwarfs relatively recently. A lack of correlation between the locations and velocities of polluted and non-polluted white dwarfs rules out the accretion of interstellar grains \citep{Farihi10ism}. The abundances observed in many polluted white dwarfs are inconsistent with the accretion of material from a companion star or with the radiative levitation of primordial metals from deeper inside of the white dwarf, although both of these processes have been observed in white dwarfs \citep{Jura2006, Gansicke2012, Koester2014}.

In many exoplanetary systems, when the host star leaves the main sequence it expands and becomes a red giant branch star, and eventually an asymptotic giant branch star, before ultimately becoming a white dwarf. The radius of the asymptotic giant branch star is of the order of astronomical units, and thus the star engulfs and vaporises any inner planets, whilst any outer planets present in the system survive \citep{VWood,Veras_review}. Adiabatic stellar mass loss and the shedding of the star's envelope on the giant branch leads to the expansion of planetary orbital radii, and the production of a planetary nebula and a white dwarf \citep{Veras_review}. Dynamical instabilities and planetesimal scattering during the white dwarf phase can lead to planetary bodies being perturbed onto star-grazing orbits \citep{DebesSigurdsson, bonsor11, debesasteroidbelt}, where they are likely to become tidally disrupted, form an accretion disc, and be accreted onto the white dwarf \citep{Jurasmallasteroid, Veras_tidaldisruption1, Veras_tidaldisrupt2}. The influence of wide binary companions \citep{bonsor_wdbinary, Hamers2016, Petrovich2017, Stephan2017} and the potential liberation of exo-moons \citep{Payne2017} have been suggested as alternate explanations for the pollution of white dwarfs. Although the exact mechanism is not certain, the prevailing explanation for the presence of metal lines in the spectra of many white dwarfs is the accretion of exoplanetary material \citep{JuraYoung2014}.

There are now multiple polluted white dwarfs which have been observed to contain many of the following features in their spectra: carbon (C), oxygen (O), nitrogen (N), sodium (Na), magnesium (Mg), aluminium (Al), silicon (Si), calcium (Ca), titanium (Ti), chromium (Cr), iron (Fe), and nickel (Ni) \citep{JuraYoung2014, Xu2017}. The spectral features, combined with various assumptions and white dwarf atmosphere models, can be manipulated to give the key rock forming and volatile elemental abundances of the pollutants which have accreted onto the white dwarfs \citep{Koester09, Koester2010}. All pollutants observed thus far have abundances dominated by Mg, Fe, Si, and O \citep{JuraYoung2014}, like the rocky bodies in the solar system. Evidence for the accretion of fragments of differentiated bodies onto white dwarfs has emerged due to the detection of multiple pollutant bodies with high, relative to the solar system, abundances in either the siderophilic elements or the lithophilic elements \citep{Zuckerman2011, Xu2013, Wilson2015}. ~\cite{Hollands2017} found evidence for a large spread in Ca, Fe, and Mg abundances in 230 white dwarf pollutants, which could possibly result from planetary differentiation and fragmentation. NLTT\,43806, analysed in \cite{Zuckerman2011},  has had its pollutant classified as requiring a large amount of crust-like material, which raises interesting questions about collisional processing in exoplanetary systems and the mechanism of white dwarf pollution. Evidence for the accretion of water ice in eight systems has been based on an excess abundance of observed oxygen compared with that which could be sequestered in the form of metal oxides using the observed metal abundances \citep{Klein2011, Farihi2011, Dufour2012, Gansicke2012, Raddi2015, Farihi2016, Xu2017}. The accretion of water ice is not surprising as ice species are expected to survive the post main sequence evolution of the host star \citep{JuraXu2010, Malamud2016}.~\cite{Xu2017} recently detected N for the first time, in the pollutant of WD\,1425+540. The abundances of C, N, and O in the pollutant hint that the system's white dwarf has potentially accreted an extrasolar volatile rich Kuiper belt analogue.

The aim of this work is to improve our understanding of how the abundances observed in polluted white dwarf atmospheres relate to the pollutant bodies' formation history, collisional history, and geological history. Thus, providing constraints regarding the nature and evolution of rocky bodies in exoplanetary systems. In Section \ref{methods} we outline the polluted white dwarf systems and data analysed in this work, along with the methods used to constrain the formation history of the pollutants. In Section \ref{results} we present the results of this analysis and compare our findings to the literature. Finally, in Section \ref{diss}, we discuss the caveats of our work and the implications of our results.

\begin{table*}
	\centering
	\caption{The stellar properties and characteristic sinking timescales for each of the white dwarf systems analysed in this work.}
	\label{tab:1a}
\begin{tabular}{ c c c c c c c}
\hline
System & Atmospheric Type & T/K & $\tau_{\rm{Fe}}\rm{/log(years)}$ & $\tau_{\rm{Mg}}\rm{/log(years)}$ & $\tau_{\rm{C}}\rm{/log(years)}$ & Reference \\
\hline
GD\,362 & Helium & 10500 & 5.04 & 5.34 & 5.32 & \protect\cite{Xu2013}\\
PG\,1225--079 & Helium & 10800 & 5.32 & 5.68 & 5.74 & \protect\cite{Xu2013} \\
SDSS\,J1242+5226 & Helium & 13000 & 6.04 & 6.36 & 6.36 & \protect\cite{Raddi2015}\\
SDSS\,J0738+1835 & Helium & 14000 & 5.05 & 5.26 & 5.24 & \protect\cite{Dufour2012}  \\
HS\,2253+8023 & Helium & 14400 & 4.71 & 4.93 & 5.05 & \protect\cite{Klein2011}\\
WD\,1425+540 & Helium & 14490 & 5.93 & 6.12 & 6.16 & \protect\cite{Xu2017}\\
G241-6 & Helium & 15300 & 5.75 & 6.08 & 6.04 & \protect\cite{Jura2012} \\
GD\,40 & Helium & 15300 & 5.75 & 6.08 & 6.04 & \protect\cite{Jura2012}\\
GD\,61 & Helium & 17300 & 4.56 & 4.90 & 4.97 & \protect\cite{Farihi2011}\\
SDSS\,J0845+2257 & Helium & 19780 & 3.94 & 4.08 & 4.18 &  \protect\cite{Wilson2015}\\
WD\,1536+520 & Helium & 20800 & 1.99 & 2.22 & 2.42 & \protect\cite{Farihi2016}\\
NLTT\,43806 & Hydrogen & 5900 & 4.00 & 4.28 & 4.31 & \protect\cite{Zuckerman2011}\\
G29-38 & Hydrogen & 11800 & -0.68 & -0.60 & -0.11 & \protect\cite{Xu2014a}\\
SDSS\,J1043+0855 & Hydrogen & 18330  &  -2.58 & -2.16 & -2.32 & \protect\cite{Dufour2016}\\
SDSS\,J1228+1040 & Hydrogen & 20900 & -1.66 & -1.53 & -1.18 & \protect\cite{Gansicke2012}\\
WD\,1929+012 & Hydrogen & 21200 & -1.66 & -1.53 & -1.18 & \protect\cite{Gansicke2012}\\
PG\,0843+516 & Hydrogen & 23100 & -1.61 & -1.50 & -1.12 &  \protect\cite{Gansicke2012}\\
\hline
\end{tabular}
\end{table*}

\begin{table*}
	\centering
	\caption{The atmospheric abundances derived for each of the systems. The abundances are in logarithmic number ratios relative to H or He depending on the systems atmospheric type. The abundances have been taken from the following papers: [1]: \protect\cite{Zuckerman07}, [2]: \protect\cite{Vennes2011b}, [3]: \protect\cite{Zuckerman2011}, [4]: \protect\cite{Klein2011}, [5]: \protect\cite{Farihi2011}, [6]: \protect\cite{Jura2012}, [7]: \protect\cite{Dufour2012}, [8]: \protect\cite{Gansicke2012}, [9]: \protect\cite{Xu2013}, [10]: \protect\cite{Xu2014a}, [11]: \protect\cite{Wilson2015}, [12]: \protect\cite{Raddi2015}, [13]: \protect\cite{Farihi2016}, [14]: \protect\cite{Dufour2016}, [15]: \protect\cite{Xu2017}.}
\label{tab:2a}
\begin{tabular}{ c c c c c c c c}
\hline
System & Al & Ti & Ca & Mg & Si & Ni & Fe\\
\hline
GD\,362\textsuperscript{[1,9]} & $ -6.40 \pm 0.20 $  & $ -7.95 \pm 0.10 $ & $ -6.24 \pm 0.10 $ & $ -5.98 \pm 0.25 $ & $ -5.84 \pm 0.30 $ & $ -7.07 \pm 0.15 $ & $ -5.65 \pm 0.10 $  \\
PG\,1225--079\textsuperscript{[4,9]} & $ < -7.84 $ & $ -9.45 \pm 0.02 $ & $ -8.06 \pm 0.03 $  & $ -7.5 \pm 0.20 $ & $ -7.45 \pm 0.10 $ & $ -8.76 \pm 0.14 $ & $ -7.42 \pm 0.07 $   \\
SDSS\,J1242+5226\textsuperscript{[12]} & $ < -6.50  $ & $ -8.20 \pm 0.20 $  & $ -6.53 \pm 0.10 $ & $ -5.26 \pm 0.15 $ & $ -5.30 \pm 0.06 $  & $ < - 7.30 $ & $ -5.90 \pm 0.15 $  \\
SDSS\,J0738+1835\textsuperscript{[7]} & $ -6.39 \pm 0.11 $  & $ -7.95 \pm 0.11 $ & $ -6.23 \pm 0.15 $  & $ -4.68 \pm 0.07 $ & $ -4.90 \pm 0.16 $ & $ -6.31 \pm 0.10 $ & $ -4.98 \pm 0.09 $   \\
HS\,2253+8023\textsuperscript{[4]} & $ < -6.70  $  & $ -8.74 \pm 0.02 $ & $ -6.99 \pm 0.03 $ & $ -6.10 \pm 0.04 $ & $ -6.27 \pm 0.03 $ & $ -7.31 \pm 0.10 $ & $ -6.17 \pm 0.03 $  \\
WD\,1425+540\textsuperscript{[15]} & --- & --- & $-9.26 \pm 0.10 $ & $-8.16 \pm 0.20 $ & $-8.03 \pm 0.31 $ & $ -9.67 \pm 0.20 $ & $ -8.15 \pm 0.14 $ \\
G241-6\textsuperscript{[6]} & $ < -7.70 $ & $ -8.97 \pm 0.10 $  & $ -7.30 \pm 0.20 $ & $ -6.26 \pm 0.10 $ & $ -6.62 \pm 0.20 $  & $ -8.15 \pm 0.40 $ & $ -6.82 \pm 0.14 $ \\
GD\,40\textsuperscript{[6]} & $ -7.35 \pm 0.12 $  & $ -8.61 \pm 0.20 $ & $ -6.90 \pm 0.20 $ & $ -6.20 \pm 0.16 $ & $ -6.44 \pm 0.30 $ & $ -7.84 \pm 0.26 $ & $ -6.47 \pm 0.12 $  \\
GD\,61\textsuperscript{[5]} & $ < -7.18  $  & $ < -9.08 $ & $ -7.88 \pm 0.19 $ & $ -6.63 \pm 0.18 $ & $ -6.83 \pm 0.08 $ & $ < -7.58 $ & $ -7.73 \pm 0.20 $ \\
SDSS\,J0845+2257\textsuperscript{[11]} & $ -5.70 \pm 0.15 $  & $ < -7.15 $ & $ -5.95 \pm 0.10 $ & $ -4.70 \pm 0.15 $ & $ -4.80 \pm 0.30 $  & $ -5.65 \pm 0.30 $ & $ -4.60 \pm 0.20 $\\
WD\,1536+520\textsuperscript{[13]} & $ -5.38 \pm 0.15 $  & $ -6.84 \pm 0.15 $ & $ -5.28 \pm 0.15 $ & $ -4.06 \pm 0.15 $ & $ -4.32 \pm 0.15 $ & --- & $ -4.50 \pm 0.15$ \\
NLTT\,43806\textsuperscript{[3]} & $ -7.60 \pm 0.17  $  & $  -9.55 \pm 0.14 $ & $ -7.90 \pm 0.19 $ & $ -7.10 \pm 0.13 $  & $ -7.20 \pm 0.14 $ & $ -9.10 \pm 0.17 $ & $ -7.80 \pm 0.17 $  \\
G29-38\textsuperscript{[10]} & $ < -6.10 $  & $ -7.90 \pm 0.16 $ & $ -6.58 \pm 0.12 $ & $ -5.77 \pm 0.13 $ & $ -5.60 \pm 0.17 $ & $ < -7.30 $ & $ -5.90 \pm 0.10 $  \\
SDSS\,J1043+0855\textsuperscript{[14]} & $ -7.06 \pm 0.30 $  & $ < -7.00 $ & $ -5.96 \pm 0.20 $ & $ -5.11 \pm 0.20 $  & $ -5.33 \pm 0.50 $ & $ -7.38 \pm 0.30 $ & $ -6.15 \pm 0.30 $  \\
SDSS\,J1228+1040\textsuperscript{[8]} & $ -5.75 \pm 0.20 $ & --- & $ -5.94 \pm 0.20 $  & $ -5.10 \pm 0.20 $ & $ -5.20 \pm 0.20 $ & $ < -6.50 $ & $ -5.20 \pm 0.30 $ \\
WD\,1929+012\textsuperscript{[2,8]} & $ -6.20 \pm 0.20$ & --- & $ -6.11 \pm 0.04 $ & $ -4.42 \pm 0.06 $ & $ -4.75 \pm 0.20 $ & $ -6.70 \pm 0.30 $ & $ -4.50 \pm 0.30 $\\
PG\,0843+516\textsuperscript{[8]} & $ -6.50 \pm 0.20 $ & --- & ---  & $ -5.00 \pm 0.20 $ & $ -5.20 \pm 0.20 $ & $ -6.30 \pm 0.30 $ & $ -4.60 \pm 0.20 $  \\
\hline
System & Cr & Na & P & S & O & C & N\\
\hline
GD\,362\textsuperscript{[1,9]} & $ -7.41 \pm 0.10 $ & $ -7.79 \pm 0.20$  & --- & $ < -6.70 $  & $ < -5.14 $ & $-6.70 \pm 0.30$ & $ < -4.14$\\
PG\,1225--079\textsuperscript{[4,9]} & $ -9.27 \pm 0.06 $ & $ < -8.26 $  & --- & $ < -9.50$ & $ < -5.54 $ &  $ -7.80 \pm 0.10 $ & --- \\
SDSS\,J1242+5226\textsuperscript{[12]} &  $ -7.50 \pm 0.20 $ & $ -7.20 \pm 0.20 $ &$ < -6.60$ & $ < -8.00$ & $ -4.30 \pm 0.10 $ & $ < -4.70 $ & $ < -5.00$\\
SDSS\,J0738+1835\textsuperscript{[7]} & $ -6.76 \pm 0.12 $ & $ -6.36 \pm 0.16$  & --- & --- & $ -3.81 \pm 0.19 $ & $< -3.80$ & --- \\
HS\,2253+8023\textsuperscript{[4]}  & $ -8.01 \pm 0.03 $ & $ < -6.80 $  & --- & --- & $ -5.37 \pm 0.07 $ & --- & ---\\
WD\,1425+540\textsuperscript{[15]} & --- & --- & --- & $ -8.36 \pm 0.11$  & $-6.62 \pm 0.23$ & $-7.29 \pm 0.17$ & $-8.09 \pm 0.10$\\
G241-6\textsuperscript{[6]} & $ -8.46 \pm 0.10 $ & --- & $ -9.04 \pm 0.13 $ & $ -7.07 \pm 0.30 $ & $ -5.64 \pm 0.11 $ & $ < -8.50 $ & $ < -8.90$\\
GD\,40\textsuperscript{[6]} & $ -8.31 \pm 0.16 $ & --- & $ -8.68 \pm 0.13 $ & $-7.80 \pm 0.20$  & $ -5.62 \pm 0.10 $ & $ -7.80 \pm 0.20 $ & $ < -8.80$\\
GD\,61\textsuperscript{[5]} & $ < -8.98 $ &  --- & --- & --- & $ -5.93 \pm 0.20 $ & $ < -8.93 $ & --- \\
SDSS\,J0845+2257\textsuperscript{[11]} & $ -6.40 \pm 0.30 $ & ---  & --- & $< -5.40$ & $ -4.25 \pm 0.20 $ & $ -4.90 \pm 0.20 $ & $ < -6.30$\\
WD\,1536+520\textsuperscript{[13]} & $ -5.93 \pm 0.15 $ & --- & $ <-7.10$ & $ < -5.40$ & $ -3.40 \pm 0.15 $  & $ < -4.20 $ & ---\\
NLTT\,43806\textsuperscript{[3]} & $ -9.55 \pm 0.22 $ & $ -8.10 \pm 0.14 $ & --- & --- & --- & --- & --- \\
G29-38\textsuperscript{[10]} & $ -7.51 \pm 0.12 $ & $ < -6.70 $ & --- & $ < -7.00$  & $ -5.00 \pm 0.12 $ & $ -6.90 \pm 0.12 $ & $ < -5.70$\\
SDSS\,J1043+0855\textsuperscript{[14]} & $ < -6.50 $ & --- &$ -7.40 \pm 0.30 $ & $ < -6.36$ & $ -4.90 \pm 0.20 $  & $  -6.15 \pm 0.30 $ & ---\\
SDSS\,J1228+1040\textsuperscript{[8]} & $ < -6.00 $ & --- & $ < -7.30$ & $ < -6.20$ & $ -4.55 \pm 0.20 $ & $ -7.50 \pm 0.20 $ & --- \\
WD\,1929+012\textsuperscript{[2,8]} &  $ -6.10 \pm 0.30 $ & --- & $ -7.00 \pm 0.30$  & $-6.60 \pm 0.20$ & $-4.10 \pm 0.30$ &  $-6.80 \pm 0.30$ & ---\\
PG\,0843+516\textsuperscript{[8]} & $ -5.80 \pm 0.30 $ & --- & $ -6.60 \pm 0.20 $ & $ -5.50 \pm 0.30 $ & $ -5.00 \pm 0.30 $  & $ -7.30 \pm 0.30 $ & --- \\
\hline
\end{tabular}
\end{table*}

\begin{table*}
	\centering
	\caption{The steady state pollutant chemical abundances in each system number ratioed to Mg. The abundances were derived in the following papers: [1]: \protect\cite{Zuckerman2011}, [2]: \protect\cite{Klein2011}, [3]: \protect\cite{Farihi2011}, [4]: \protect\cite{Dufour2012}, [5]: \protect\cite{Gansicke2012}, [6]: \protect\cite{Xu2013}, [7]: \protect\cite{Xu2014a}, [8]: \protect\cite{Wilson2015}, [9]: \protect\cite{Raddi2015}, [10]: \protect\cite{Farihi2016}, [11]: \protect\cite{Dufour2016}, [12]: \protect\cite{Xu2017}. All errors were given in the papers listed above and converted accordingly, with the exception of GD\,40 and G241-6 where the errors were originally given in \protect\cite{Jura2012}.}
	\label{tab:3a}
\begin{tabular}{ c c c c c c c }
\hline
 System & Al/Mg & Ti/Mg & Ca/Mg & Ni/Mg & Cr/Mg & Fe/Mg \\
\hline
GD\,362\textsuperscript{[6]} & $ 0.540 \pm 0.398 $ & $0.0244 \pm0.0151 $ & $1.237 \pm 0.767 $ & $0.182 \pm 0.122$ & $ 0.080\pm0.0498 $ & $ 4.35\pm2.70 $ \\
PG\,1225--079\textsuperscript{[6]} & $ < 0.611 $ & $0.0300 \pm0.01390 $ & $ 0.693\pm0.323 $ & $ 0.118\pm0.067 $  & $0.044 \pm0.021 $ & $ 2.80\pm1.37 $ \\
SDSS\,J1242+5226\textsuperscript{[9]} & $ < 0.062$  & $ 0.0029\pm0.0017 $ & $ 0.130\pm0.054 $ & $ < 0.018 $ & $0.014 \pm 0.008$ & $ 0.50\pm 0.25$ \\
SDSS\,J0738+1835\textsuperscript{[4]} & $ 0.020 \pm 0.0060$ & $0.0010 \pm 0.0003$ & $ 0.047 \pm 0.018 $ & $0.037 \pm 0.010 $ & $ 0.014 \pm 0.005 $ & $ 0.81 \pm 0.21 $ \\
HS\,2253+8023\textsuperscript{[2]} & $ <0.330 $ & $0.0040 \pm 0.0009 $ & $ 0.180\pm0.040 $ & $0.110 \pm0.040 $ & $0.020 \pm 0.006 $ & $ 1.48\pm 0.31$ \\
WD\,1425+540\textsuperscript{[12]} & ---  & --- & $ 0.125 \pm 0.064$ & $ 0.045 \pm 0.030$ & --- & $ 1.58 \pm 0.89$ \\
G241-6\textsuperscript{[6]} & $ < 0.037 $ & $0.0047 \pm 0.0015$ & $0.207 \pm 0.106$ & $ 0.024 \pm 0.023$ & $ 0.014 \pm 0.005$ & $0.58 \pm 0.23$\\
GD\,40\textsuperscript{[6]} & $ 0.074 \pm 0.034$ & $0.0096 \pm 0.0056$ & $0.464 \pm 0.274$ & $ 0.043 \pm 0.030$ & $ 0.018 \pm 0.009$ & $1.13 \pm 0.52$\\
GD\,61\textsuperscript{[3]} & --- & --- & $ 0.089\pm0.054$  & --- & --- & $ 0.17\pm0.11 $ \\							SDSS\,J0845+2257\textsuperscript{[8]} & $ 0.107\pm0.052 $  & $ < 0.0762$ & $ 0.072\pm0.030 $ & $ 0.311\pm0.240 $ &  $ 0.047\pm0.036 $ & $ 1.80\pm1.03 $ \\		
WD\,1536+520\textsuperscript{[10]} & $ 0.056\pm 0.027$ & $0.0019\pm 0.0009$ & $ 0.069\pm 0.034 $  & --- & $0.022 \pm0.011 $ & $ 0.62\pm0.30 $ \\		
NLTT\,43806\textsuperscript{[1]} & $ 0.355\pm 0.175$  & $ 0.0059\pm0.0026 $ & $ 0.216\pm0.115 $ & $ 0.020\pm0.010 $  & $ 0.006\pm0.004 $ & $ 0.38\pm0.19 $ \\
G29-38\textsuperscript{[7]} & $ <0.349 $ & $0.0073 \pm0.0034$ & $0.192 \pm0.078$ & $ <0.041 $ & $ 0.019\pm0.008 $ & $ 0.89\pm0.34 $  \\
SDSS\,J1043+0855\textsuperscript{[11]} & $ 0.013\pm0.011 $  & $ <0.0232 $ & $ 0.227\pm0.148 $ & $ 0.014\pm0.011 $ & $ > 0.084 $ & $ 0.24\pm0.20 $ \\
SDSS\,J1228+1040\textsuperscript{[5]} & $ 0.331\pm0.216 $  & --- & $ 0.297\pm 0.193$ & $ <0.129 $ & $ < 0.333$ & $ 2.33\pm1.94 $ \\
WD\,1929+012\textsuperscript{[5]} &  $0.019 \pm 0.009 $ & --- & $ 0.033 \pm 0.006 $ & $0.013 \pm 0.009$ & $0.044 \pm 0.031$ & $1.91 \pm 1.35$ \\
PG\,0843+516\textsuperscript{[5]} & $0.042 \pm0.027 $ & --- & ---  & $ 0.154\pm0.128$ & $ 0.398\pm 0.331$ & $ 6.92\pm4.51 $ \\			
\hline
 System & Si/Mg & O/Mg & C/Mg & Na/Mg & N/Mg & S/Mg\\
\hline
GD\,362\textsuperscript{[6]} & $ 2.49\pm2.24 $ & $ <6.68 $ & $0.202 \pm 0.182 $ & $ 0.016\pm0.012 $ & ---  & ---\\
PG\,1225--079\textsuperscript{[6]} & $ 1.85\pm0.95 $ & --- & $ 0.448\pm0.231 $ & $ < 0.196 $ & --- & ---\\
SDSS\,J1242+5226\textsuperscript{[9]} & $ 1.25\pm 0.47$ & $ 9.87\pm4.10$ & $ <3.807 $  & $ 0.013\pm0.007$ & --- & --- \\
SDSS\,J0738+1835\textsuperscript{[4]} & $ 0.62 \pm 0.25 $  & $7.68 \pm3.58 $ & $<7.810$ & $0.022 \pm 0.009 $ & --- & --- \\
HS\,2253+8023\textsuperscript{[2]} & $ 0.68\pm0.13 $ & $ 4.50 \pm 0.90 $ & $<0.010$  & --- & --- & ---\\
WD\,1425+540\textsuperscript{[12]} & $ 1.38 \pm 1.17$ & $34.75 \pm 24.39$ & $5.930 \pm 3.590$  & --- & $ 1.19 \pm 0.61$ & $0.758 \pm 0.398$ \\
G241-6\textsuperscript{[13]} & $0.50 \pm 0.26$ & $ 4.35 \pm 1.50$ & $ < 0.006 $ & --- & --- & $0.283 \pm 0.194$ \\
GD\,40\textsuperscript{[13]} & $0.69 \pm 0.52$ & $ 3.69 \pm 1.60 $ & $  0.026 \pm 0.015 $  & --- & --- & $0.046 \pm 0.026$ \\
GD\,61\textsuperscript{[3]} & $ 0.77\pm0.35 $ & $4.25 \pm 2.63$ & $< 0.003$ & --- & --- & ---\\							SDSS\,J0845+2257\textsuperscript{[8]} & $ 0.81\pm0.63 $ & $ 3.61\pm 2.08$ & $ 0.519\pm0.299 $ & --- & --- & ---\\	
WD\,1536+520\textsuperscript{[10]} & $ 0.74\pm0.36 $ & $2.51 \pm 1.23$ & $ < 0.461 $& --- & --- & --- \\					NLTT\,43806\textsuperscript{[1]} & $0.95 \pm0.42 $ & --- & --- & $0.096 \pm 0.042$ & --- & --- \\
G29-38\textsuperscript{[7]} & $0.83 \pm0.41 $ & $ 3.41\pm1.39 $ & $ 0.025\pm0.010 $ & $ <0.140 $ & --- & --- \\
SDSS\,J1043+0855\textsuperscript{[11]} & $ 0.83\pm1.03 $ & $ 3.35\pm2.18 $  & $ 0.126\pm0.105 $ & --- & --- & --- \\
SDSS\,J1228+1040\textsuperscript{[5]} & $ 1.29\pm0.84 $ & $ 12.78\pm 8.32$ & $ 0.008\pm0.005 $ & --- & --- & ---\\
WD\,1929+012\textsuperscript{[5]} &  $ 0.58 \pm 0.28 $ & $ 5.80 \pm 4.08 $ & $0.006 \pm 0.004 $ & --- & --- & $ 0.014 \pm 0.007 $ \\
PG\,0843+516\textsuperscript{[5]} & $ 0.92\pm 0.60$ & $ 3.15\pm2.62 $ & $0.008\pm0.006 $  & --- & --- & $0.664 \pm 0.398$\\		\hline
\end{tabular}
\end{table*}

\section{Methods}\label{methods}

\subsection{Polluted White Dwarf Data}\label{PWDD}

Our analysis focuses on 17 externally polluted white dwarfs with published measurements of at least five atmospheric elemental abundances. Our method is most successful at constraining the formation history of a white dwarf pollutant when the system has abundance measurements of at least one lithophile species, one siderophile species, one volatile species, in addition to Mg and Si. In order to accurately recreate the pollutant, and rule out alternative formation histories, it is beneficial to maximise the number of elements analysed. More elemental abundances allow the trends expected in certain formation scenarios to be investigated, and validated or dismissed. Our work could be readily extended to future observations.

The stellar properties and atmospheric chemical abundances of the white dwarfs considered have been collated from the literature and are listed in Table \ref{tab:1a} and Table \ref{tab:2a}. Table \ref{tab:3a} lists the abundances converted from atmospheric logarithmic number ratios to hydrogen/helium to number ratios with respect to Mg, which are adjusted to assume a steady state between accretion and diffusion. 

The errors given in Table \ref{tab:3a} were found by using Equation \ref{eq2}, which corresponds to using standard error propagation techniques on the errors in Table \ref{tab:2a}, assuming no error in the sinking timescales.
\begin{equation}\label{eq2}
\sigma_{C} = \frac{10^{f_{A}}}{10^{f_{B}}} \ln(10) \sqrt{ \sigma^{2}_{f_{A}} + \sigma^{2}_{f_{B}} }
\end{equation}
\noindent Where $f_{A} = \log_{10}\left(\frac{A}{H}\right)$, $f_{B} = \log_{10}\left(\frac{B}{H}\right)$, $C = \frac{A}{B}$, and $\sigma_{f_{A}}$, $\sigma_{f_{B}}$, and $\sigma_{C}$ are the errors in $f_{A}$, $f_{B}$, and $C$ respectively. The errors in Table \ref{tab:1a} were derived in the papers referenced in the table caption by analysing the noise in the original observations and the atmospheric model. We note here that this assumes independent errors, whereas systematics in the observations may mean this is not the case. We also note that degeneracies in the error calculations may exist due to the manner in which different groups derived their errors.

Heavy elements present in the atmospheres of white dwarfs are expected to sink quickly from the thin upper convective zone, due to the strong gravitational field of the white dwarf, and become unobservable. Different chemical elements will sink at different rates. Thus, the elemental abundance ratios of the pollutant body are not necessarily the same as the elemental abundance ratios present in the atmosphere today. A solution to this problem is to assume a steady state between accretion and diffusion through the atmosphere, as this allows the atmospheric abundances to be converted into pollutant abundances. This is the first approach that we consider in this work.

\begin{equation}\label{eq1}
\left(\frac{A}{B}\right)_{Pollutant,~SSP} = \left(\frac{A}{B}\right)_{Atmosphere} \left(\frac{t_{sink, B}}{t_{sink, A}}\right) 
\end{equation}

\noindent Equation \ref{eq1}, taken from \cite{Koester09}, shows how the pollutant elemental abundances can be related to the elemental abundances in the atmosphere for two elements, A and B, if it is assumed that the white dwarf is accreting in steady state. This analysis was performed in the papers referenced in Table \ref{tab:3a} for all the white dwarf systems.

We also consider the possibility that the pollutants were accreted relatively recently and, therefore, have not settled into a steady state of accretion yet. This assumption would suggest that the elemental abundances in the atmosphere, listed in Table \ref{tab:2a}, are identical to the elemental abundances of the pollutant body. \cite{Koester09} discusses how this phase of accretion is not likely for many polluted white dwarf systems given the length of time required for a system to settle into a steady state of accretion is of the order of a sinking timescale. We therefore only consider this to be a possibility for the nine systems which have sinking timescales larger than $10^{4.5}$\,years. We chose this value because white dwarfs with sinking timescales greater than this are expected to be in a pre-steady state phase for a time comparable to the duration of the steady state phase according to the disc lifetime estimates given in \cite{Girven2012}.

We also consider that the pollutants are not accreting in steady state phase, nor are in an early phase of accretion, but are instead in a declining phase, one in which accretion has stopped. Assuming the system is in a declining phase, the pollutant abundance ratio of two elements, A and B, would be related to the atmospheric abundance ratio of A and B by Equation \ref{eq1b}.

\begin{equation}\label{eq1b}
\resizebox{.9\hsize}{!}{$\left(\frac{A}{B}\right)_{Pollutant,~DP} = \left(\frac{A}{B}\right)_{Atmosphere} \left(\frac{t_{sink,B}}{t_{sink,A}}\right)e^{t\left(\frac{t_{sink, B}~ - ~ t_{sink, A}}{t_{sink, A} ~ t_{sink, B}}\right)}$}
\end{equation}

\noindent The declining phase is only expected to be observable for timescales of the order $10t_{sink}$ \citep{Koester09}. Therefore, if the sinking timescales are much shorter than the expected disc lifetimes then one would expect that it is unlikely to catch a system in the declining phase. Again by comparing the sinking timescales to the estimated disc lifetimes suggested in \cite{Girven2012} we find that a declining phase is only likely to be a possibility for the 11 systems which have sinking timescales larger than $10^{3.5}$\,years. We consider the possibility that these 11 systems are in a declining phase by converting the atmospheric abundances into pollutant abundances using Equation \ref{eq1b} for a range of t values, where t is the time passed after accretion has stopped. In general, if a system is in a declining phase a characteristic signature should be present in the abundance ratios. When ratioed to Mg one would expect the abundances to be extremely low in all elements except Al, Si, Na, O, C, and N, as these elements have sinking timescales which are comparable to or longer than Mg. This signature should allow systems which are in a declining phase to be identified, and allow the declining phase of accretion to be readily ruled out for many systems.

In summary, in this work we consider three potential accretion scenarios; steady state phase, pre-steady state phase, and declining phase. In Section \ref{individual} and Section \ref{diss} we discuss how the assumed phase of accretion influences the constraints we can place on the origin and geology of each of the systems' pollutants.

When presenting the data all the elemental abundances have been ratioed to Mg. We do this for five main reasons: 
\begin{itemize}[leftmargin=*]
\setlength\itemsep{1em}
\item We do not have an accurate total mass accreted onto the white dwarf in each system, as elements may be missing. Therefore, the most practical solution is to present the abundances as a ratio relative to another species.
\item Unlike Si, the uncertainty in the abundance of Mg is generally small in comparison to other observed elements. This is due to the nature of the spectral lines which need to be analysed in order to derive the respective abundances.
\item The sinking timescale of Mg is often shorter than those of H, O, N, and C, similar to those of Na, Al, and Si, and longer than those of Ca, Cr, Ti, Fe, and Ni. This spread allows one to easily pick up any signatures that indicate that the pollutant material is not currently accreting in a steady state or a pre-steady state phase and is in a declining phase.
\item Mg is a moderate volatile and thus Al, Ti, and Ca all have higher condensation temperatures, Si and Ni have similar condensation temperatures, and Fe, Cr, Na, O, C, and N all have lower condensation temperatures. This allows one to search for condensation temperature dependent formation signatures and condensation temperature dependent compositional evolution.
\item Geologically, the ratio of elements to Mg provide strong signatures of differentiation. On the Earth, and in the solar system, one can tell the nature (core-like, mantle-like, or crust-like) of a rock sample simply by observing the ratios of Al, Ti, Ca, Ni, Fe, and Si to Mg.
\end{itemize}

\subsection{What Determines the Composition of Pollutants?} \label{RPCIP}
The prevailing explanation for the presence of metals in the atmospheres of externally polluted white dwarfs is the accretion of rocky planetary debris \citep{JuraYoung2014}. In this work we assume that white dwarf pollutants are a single planetary body, this could be an exoplanet, an exo-moon, an exo-asteroid, an exo-comet, or a fragment of any of the above. In reality white dwarfs may be polluted by many bodies simultaneously \citep{JuraWD03,Wyattstochastic}, however, the abundances will be dominated by the most massive body.

Initially, our model only considers potential variations in the initial material that formed the planetary system. As a second step, we consider that the planetary bodies accreted by the white dwarfs could have formed at different temperatures and pressures, which we link to where in the planetary system they formed. Finally, we consider the potential that the planetesimals differentiated, forming a core and mantle (and possibly a crust), and that subsequent collisions distributed this material unequally between the collision fragments, which are the individual bodies accreted by white dwarfs. Such a range of models can, to first order, explain the elemental abundance patterns seen in the solar system's rocky bodies. Further details of the models and assumptions can be found in Sections \ref{HSCV}, \ref{PC}, and \ref{NPC}.

\begin{figure*}
	% To include a figure from a file named example.*
	% Allowable file formats are eps or ps if compiling using latex
	% or pdf, png, jpg if compiling using pdflatex   

	\includegraphics[width=\textwidth]{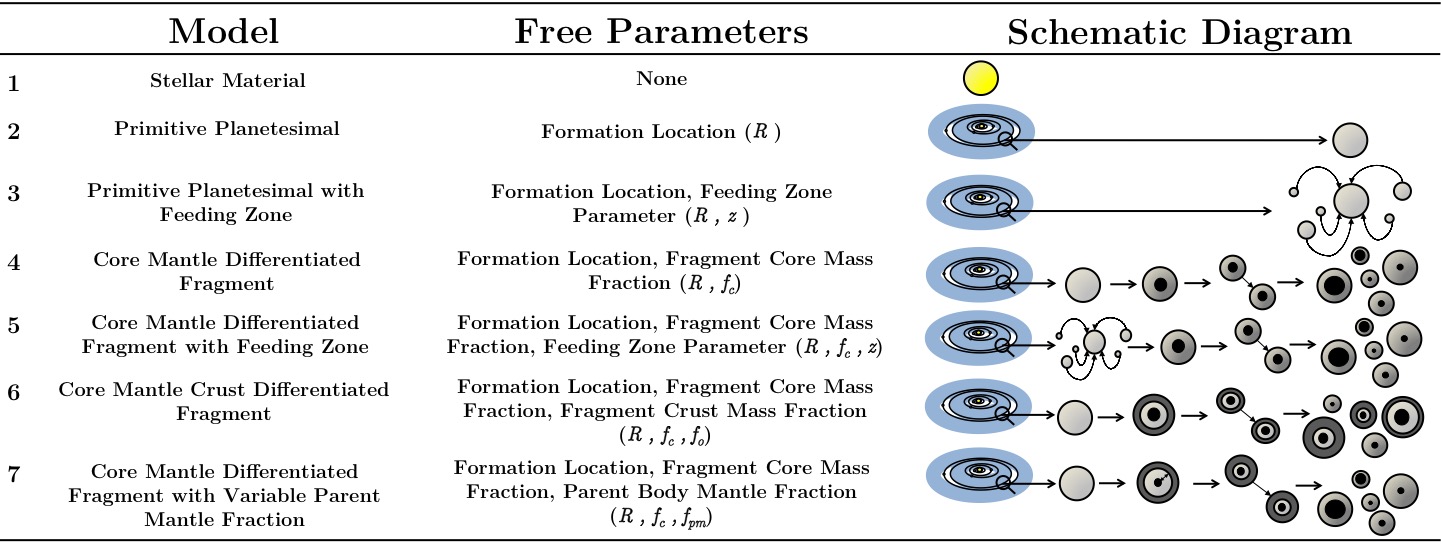}
 \caption{Details of the 7 models used to fit the abundances present in the atmospheres of polluted white dwarfs, alongside the free parameters of each model. For further details see Sections \protect\ref{HSCV} (model 1), \protect\ref{PC} (model 2 and 3), and \protect\ref{NPC} (model 4, 5, 6, and 7).}
    \label{fig:x2}
\end{figure*}

Figure \ref{fig:x2} displays a summary of the models investigated in this work. In the first model, we consider the possibility that the material in the atmospheres of the white dwarfs is compositionally similar to stellar material rather than planetary material. Models 2 through 7 all model the pollutant as a rocky planetary body whose composition is determined by the various free parameters shown in the central column. In the following sections we will present the methods used when determining the plausibility that each of the formation scenarios outlined in the seven models could reproduce the observed abundance patterns in the chosen white dwarf systems.

\subsubsection{The Initial Composition of the Planetesimal Forming Disc } \label{HSCV}
In order to assess whether the diversity in the chemical abundances observed in the atmospheres of polluted white dwarfs could arise due to chemical variation in the pollutants caused by variation in the chemical abundances of the systems' host stars, we considered a sample of nearby stars to be representative of the potential chemical diversity expected. \cite{FischerBrewer2016} observed the abundances of 1617 nearby FGK type stars. In this work nearby FGK stars were chosen in order to maintain the same approximate formation age as the polluted white dwarf progenitors (which were most likely earlier spectral types), thus, to first order the same diversity in chemical abundances. 

The 1617 stars in the catalogue were reduced to 960 by removing any star with $\log(g)$ less than 3.5 and any star whose signal to noise ratio was less than 100. This was done in order to avoid contamination by non-main sequence stars and by unreliable data. We chose to remove stars with $\log(g)$ less than 3.5 because stars with a $\log(g)$ value below this cut off are likely to be on the giant branches, therefore the dredging up of the products of nuclear fusion may be occurring and would give atmospheric abundances dissimilar to the original stellar nebula abundances.

\begin{figure}
	% To include a figure from a file named example.*  Brewer \& Fischer (2016)
	% Allowable file formats are eps or ps if compiling using latex
	% or pdf, png, jpg if compiling using pdflatex
	\includegraphics[width=\columnwidth]{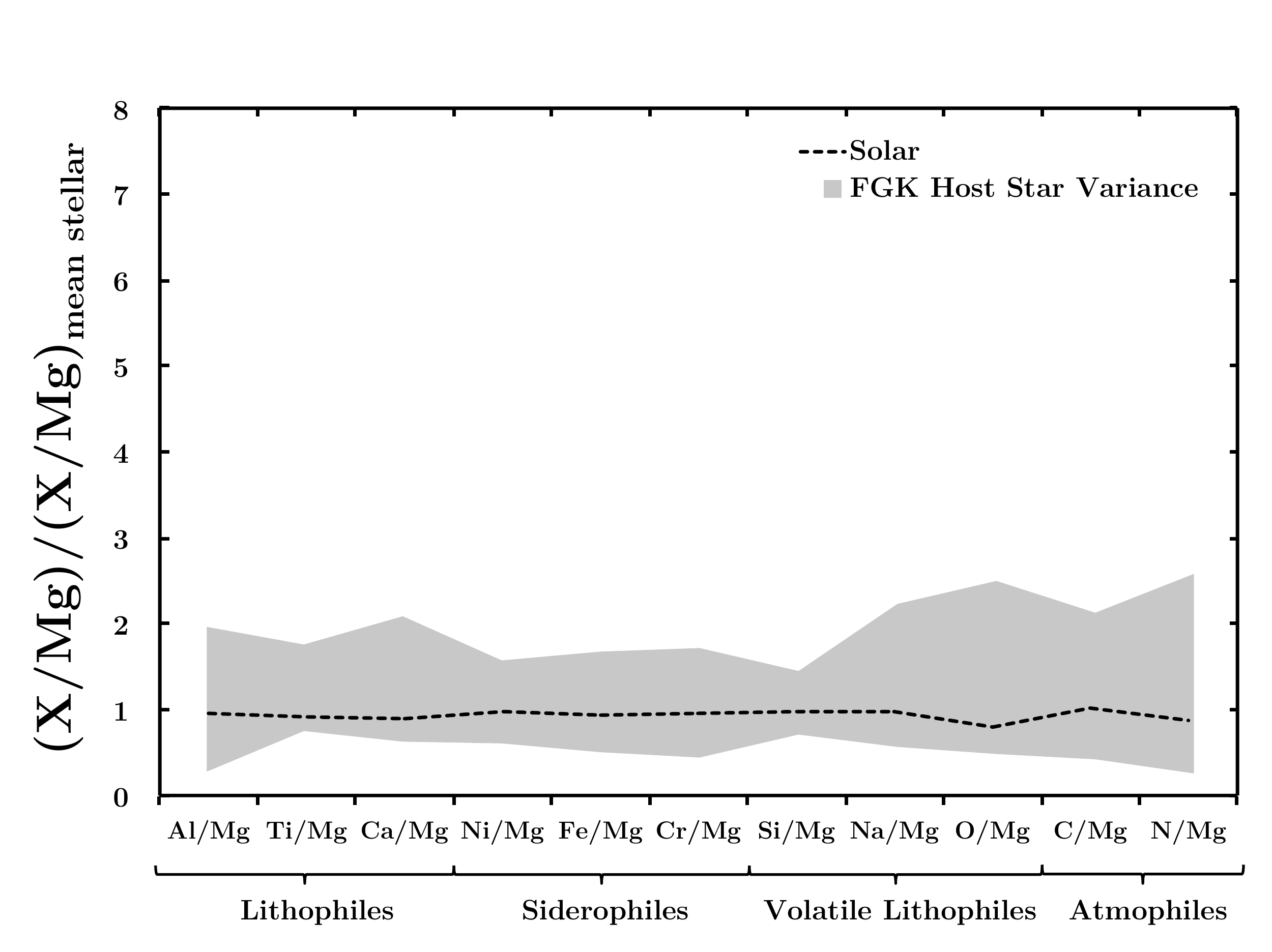}
    \caption{The range of host star chemical abundances in the \protect\cite{FischerBrewer2016} catalogue. The x axis displays the elements in volatility order, increasing in volatility from left to right (Si is shifted to allow the elements to be simultaneously grouped via their Goldschmidt classification). The y axis is normalised to the average stellar composition in the sample.}
    \label{fig:1}
\end{figure}

In model 1 we consider that the pollutants composition could be identical to that of the stars in the \cite{FischerBrewer2016} sample. This model is not expected to accurately reproduce the compositions seen in externally polluted white dwarfs, however, we perform the analysis to investigate to what level of statistical significance one can rule out pollutants with a stellar composition.

Figure \ref{fig:1} compares the range of abundances in the \cite{FischerBrewer2016} sample to the solar photosphere abundances. The x axis is displayed in volatility order, increasing from left to right, and grouped via the Goldschmidt classification. The y axis is normalised to the mean value of each elemental abundance in the \cite{FischerBrewer2016} sample. The figures in Appendix \ref{AppA} show histograms displaying how the ratios of Al, Ti, Ca, Si, Ni, Cr, Fe, Na, O, N, and C to Mg vary in the sample of stars observed by \cite{FischerBrewer2016}. Highlighting how the compositions of nearby stars are on average similar to the composition of the sun, however, individual elemental abundance ratios can often vary from as low as half solar to as high as twice solar.

\subsubsection{Heating During Formation or Subsequent Evolution } \label{PC}

Most pollutants of white dwarfs are depleted in the volatile elements, and some system's pollutants show evidence for enhancements in the refractory elements and depletion in the moderately volatile elements \citep{JuraYoung2014, Dufour2016}. Such trends are indicative of the temperature experienced by the material either during formation or afterwards. As a first approximation we assume this signature developed in chemical equilibrium. We can, therefore, employ a Gibbs free energy minimisation model to find the pressure-temperature space in which the pollutants compositions may be recreated. This space can then be used to constrain either where in the protoplanetary disc the planetesimal formed or possibly a location during the giant branch evolution of the host star. In this work we present a constraint on the highest possible temperature experienced by the pollutants and note the constraint on where in the protoplanetary disc the pollutant could have formed. We discuss the possibility the signatures developed post-formation in Section \ref{diss}.

We employed the same equilibrium chemistry model as \cite{Bond2010a} and \cite{Moriarty2014} as this model is effective in reproducing the bulk compositions of the rocky bodies in the solar system \citep{Moriarty2014}. We found the expected primitive planetesimal abundances by inputting the stellar elemental abundances from \cite{FischerBrewer2016}, as mentioned in Section \ref{HSCV}, into the program HSC chemistry version 8, a Gibbs free energy minimisation solver, and analysing the abundances of the solid species which condensed out of the gaseous nebula over the pressure-temperature space mapped out by the analytic disc model derived in \cite{Chambers2009}. The chosen disc model is an irradiated viscous disc model with an alpha parameterisation which models the evolving pressure-temperature space in an irradiated viscously heated protoplanetary disc around a solar mass star. The full specification of the viscous irradiated protoplanetary disc model and the equilibrium chemistry model are presented in Appendix \ref{disc} and Appendix \ref{gibbs} respectively.

\begin{figure}
\centering
	\includegraphics[width=\columnwidth]{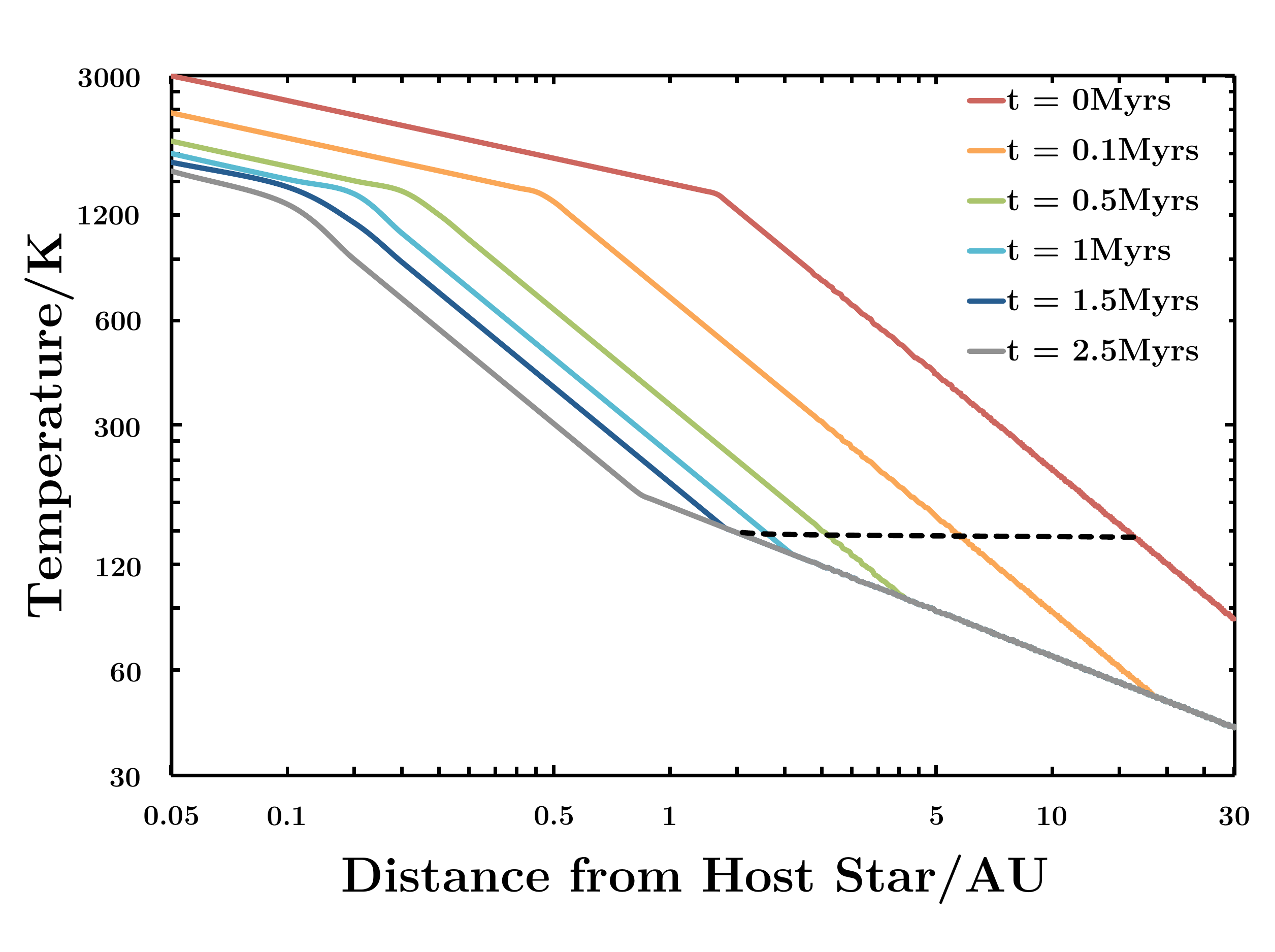} 
\caption{The temperature space mapped out by the \protect\cite{Chambers2009} protoplanetary disc model as a function of radius and time. The black dashed line indicates the position of the water ice line as a function of time.}
    \label{fig:A1}
\end{figure}

There is good evidence to support the equilibrium chemistry model's validity to first order, which mainly stems from its ability to recreate the abundances present in the solar system's rocky bodies \citep{Ciesla2004}. Evidence from Earth suggests a nebula origin to the volatile depletion trend observed \citep{Palme2003}, as volatilisation and recondensation would produce enrichments and isotope fractions different to those observed \citep{Cassen2000}. This, along with the requirement of oxidizing conditions in a recondensation model, which were not present in the nebula, suggest that a nebula based model for the composition of rocky bodies is a valid one \citep{Cassen2000}. However, the existence of pre-solar grains and the low pressures and temperatures present in the outer disc could be crucial in determining the chemical abundance patterns of planetesimals, and this has not been taken into consideration in this model. Migration of planetary bodies after formation and the migration of dust during formation \citep{Desch2017b,Desch2017a} has not been modelled in this work. This is a major limitation to our model, however the inclusion of a feeding zone parameter should allow some of the effects of migration to be accounted for.

Figure \ref{fig:A1} displays the temperature space mapped out by the chosen protoplanetary disc model as a function of radius and time. In this work we discuss constraints on both temperature and distance from the host star. However, because the formation distance is heavily model dependent and is degenerate in formation time, the constraints on the formation temperature will be a more robust prediction. Therefore, we will focus on formation temperature, however, for reference we often quote initial equivalent distances to the host star in the \cite{Chambers2009} protoplanetary disc model because this provides an estimate for the furthest possible distance the pollutant could have formed from the host star.

\begin{figure}
	% To include a figure from a file named example.*
	% Allowable file formats are eps or ps if compiling using latex
	% or pdf, png, jpg if compiling using pdflatex
	\includegraphics[width=\columnwidth]{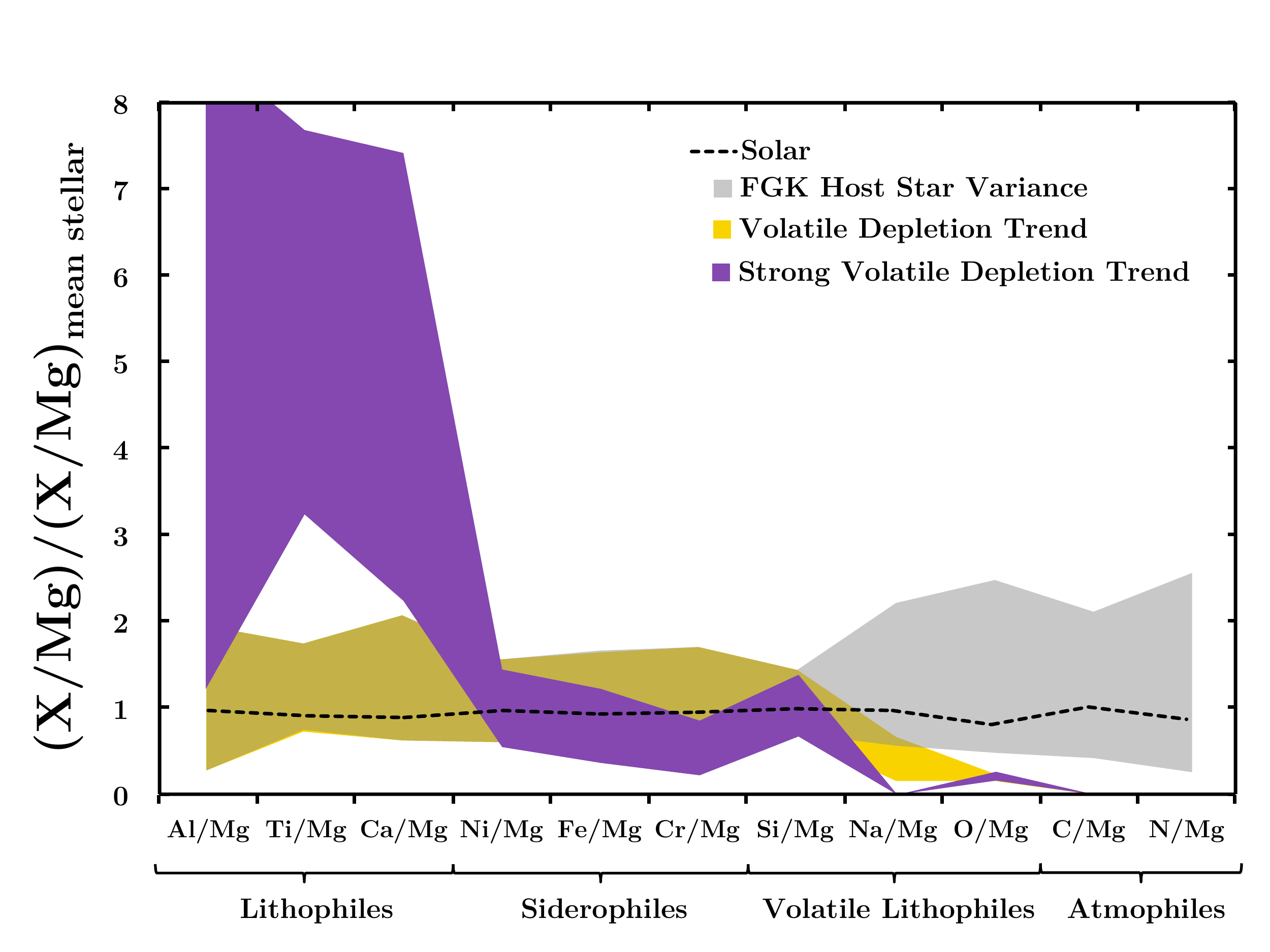}
    \caption{Two characteristic volatile depletion trends created when the models corresponding to the formation of a planetesimal at t=0\,Myrs at a distance of 1.35\,AU / 2\,AU from the host star with a feeding zone parameter of 0.15\,AU / 0\,AU (purple/orange) are applied to all the stars in the sample and the new abundance ranges are over plotted onto Figure \protect\ref{fig:1}.}
    \label{fig:2}
\end{figure}

In model 2 we consider that the abundances observed in the white dwarf atmospheres originated from a pollutant which formed at a single pressure and temperature, corresponding to a single radial location in the protoplanetary disc. In reality a large planetesimal or minor planet will incorporate material from a range of formation locations. In order to account for this, we consider a model (model 3 and model 5) in which the material that forms a given planetesimal originates from a range of formation locations described by a Gaussian distribution centred at distance R and with a width of z. Thus, in model 3 we have two free parameters, the formation location, R, which is equivalent to the mean of the normal distribution, and the feeding zone parameter, z, which is equivalent to the standard deviation of the normal distribution. Model 2 is identical to the scenario in model 3 where the standard deviation is zero.

Figure \ref{fig:2} displays the region in chemical abundance space possible for planetesimals which formed at t=0\,Myrs, R=1.35\,AU, with a feeding zone parameter of 0.15\,AU (purple region) and t=0\,Myrs, R=2\,AU, with a feeding zone parameter of 0\,AU (orange region) over plotted onto Figure \ref{fig:1}.

\subsubsection{Differentiation, Collisions, and Fragmentation} \label{NPC}

Many pollutants of white dwarfs show enhancement or depletion in siderophile and/or lithophile elements in comparison to bulk Earth \citep{JuraYoung2014}. This has been cited as evidence for the accretion of fragments of differentiated bodies \citep{Zuckerman2011, Xu2013, Wilson2015}. Collisions during planet formation and in the protoplanetary disc phase are common and often disruptive \citep{Vries2016}. This collisional processing is expected to continue into the debris disc phase \citep{wyattreview}. Disruptive collisions between planetary bodies that have differentiated can lead to the creation of fragments with non-primitive chemical abundance patterns \citep{Marcus2009, bonsorleinhardt, Carter2015}. As all bodies over 1000\,km in size are expected to differentiate due to formational impact heating (differentiation could occur in smaller bodies due to the presence of Aluminium 26) non-primitive planetesimals are expected to be commonplace in exoplanetary systems \citep{Lodders1998,Ghosh2006}.

To recreate the expected non-primitive planetesimal chemical abundances present in extrasolar systems we multiplied the abundances relative to Mg of each element derived in Section \ref{PC} by an element specific enhancement factor which was dependent on the geological and collisional history of the body. This is equivalent to allowing primitive planetesimals to condense out of a protoplanetary disc, differentiate in a similar manner to the Earth, then collide and produce fragments of various compositions. In models 4, 5, 6, and 7 these fragments are the bodies which pollute the white dwarf atmospheres. We note here that planetesimals may differentiate under different conditions, such that the composition of their mantles and cores differ from that of Earth, however to a first approximation using Earth's composition allows us to provide important first insights that should broadly hold in a more general case, without introducing more free parameters.

\begin{figure}
	% To include a figure from a file named example.*
	% Allowable file formats are eps or ps if compiling using latex
	% or pdf, png, jpg if compiling using pdflatex
	\includegraphics[width=\columnwidth]{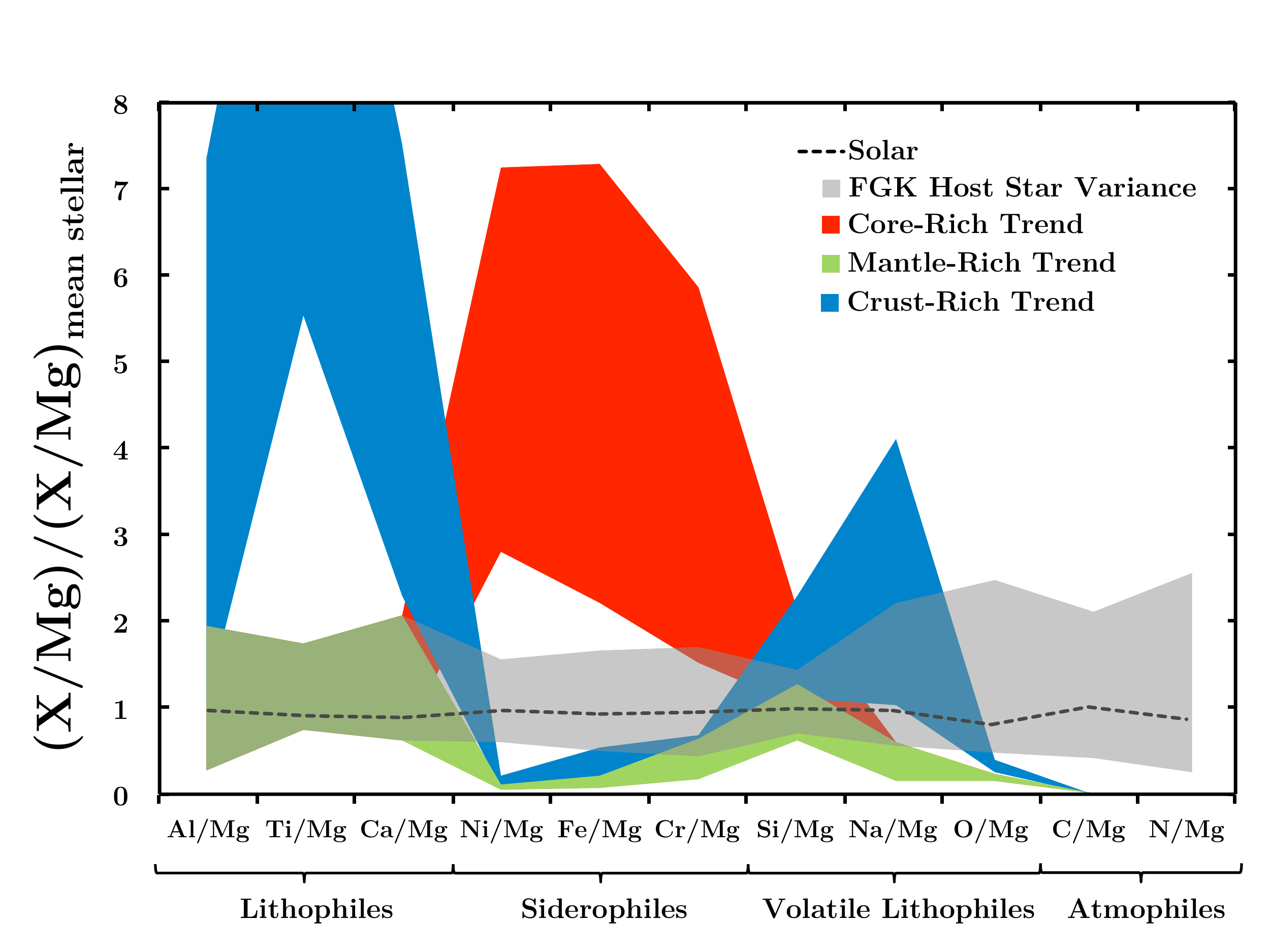}
    \caption{Example modelled abundance trends of a core-rich, mantle-rich, and crust-rich pollutant. The core-rich model is a fragment of a planetesimal which is 70 percent core material by mass and 30 percent mantle material by mass. The mantle-rich model is 100 percent mantle material by mass. The crust-rich model is 50 percent crust material and 50 percent mantle material by mass. All models plotted were created from planetesimals which formed 2\,AU from the host star at t=0\,Myrs with a feeding zone parameter of 0\,AU. The chosen three models have been applied to all the stars in the sample and the new abundance ranges are over plotted onto Figure \protect\ref{fig:1}.}
    \label{fig:3}
\end{figure}

In models 4 and 5 we considered a grid defined by $f_{\rm{core}} + f_{\rm{mantle}} =1$ and found the enhancement factor $E_{\rm{factor,X}}$ for each element $X$ at a single location in the grid using Equation \ref{E1}. 
\begin{equation} \label{E1}
\resizebox{.9\hsize}{!}{$E_{\rm{factor,X}}{\left(\frac{X}{\rm{Mg}}\right)_{\oplus}} = \left(\frac{f_{\rm{core}}X_{\rm{core},\oplus}\,+\,f_{\rm{mantle}}X_{\rm{mantle},\oplus}}{f_{\rm{core}}\rm{Mg}_{\rm{core},\oplus}\,+\,\bf{\it{f}}_{\rm{mantle}}\rm{Mg}_{\rm{mantle},\oplus}}\right)$}
\end{equation}
\noindent Where $f_{\rm{core,mantle}}$ is the fraction of core-like and mantle-like material in the fragment respectively, X\textsubscript{core,mantle$\oplus$} is the abundance of an element $X$ in the Earth's core or mantle respectively, Mg\textsubscript{core,mantle$\oplus$} is the abundance of magnesium  in the Earth's core or mantle respectively, and $\left(\frac{X}{\rm{Mg}}\right)_{\oplus}$ is the element $X$'s ratio to Mg in bulk Earth. The chemical abundance data for each element in each layer of the Earth and in bulk Earth is taken from \cite{McDonough2003}. The enhancement factors for each element at one location in the grid defined in model 4 could then be multiplied by the expected primitive abundances, derived in Section \ref{PC}, to find the expected abundances of fragments of planetesimals which had that specific core mass fraction at that formation location.

 In model 6 we considered a grid defined by $f_{\rm{core}} + f_{\rm{mantle}} + f_{\rm{crust}} =1$ and found the enhancement factor $E_{\rm{factor,X}}$ for each element $X$ at a single location in the grid using Equation \ref{E2}. This was motivated by the conclusions reached in \cite{Zuckerman2011}, which suggested the best explanation for the pollutant of NLTT\,43806 was an extrasolar lithosphere. 
\begin{equation} \label{E2}
\resizebox{.9\hsize}{!}{$E_{\rm{factor,X}}{\left(\frac{X}{\rm{Mg}}\right)_{\oplus}} = \left(\frac{f_{\rm{core}}X_{\rm{core},\oplus}\,+\,f_{\rm{mantle}}X_{\rm{mantle},\oplus}\,+\,f_{\rm{crust}}X_{\rm{crust},\oplus}}{f_{\rm{core}}\rm{Mg}_{\rm{core},\oplus}\,+\,\bf{\it{f}}_{\rm{mantle}}\rm{Mg}_{\rm{mantle},\oplus}\,+\,\bf{\it{f}}_{\rm{crust}}\rm{Mg}_{\rm{crust},\oplus}}\right)$}
\end{equation}
\noindent Where $f_{\rm{core,mantle,crust}}$ is the fraction of core-like, mantle-like, and oceanic crust-like material in the fragment respectively, X\textsubscript{core,mantle,crust$\oplus$} is the abundance of an element $X$ in the Earth's core, mantle, and oceanic crust respectively, Mg\textsubscript{core,mantle,crust$\oplus$} is the abundance of magnesium  in the Earth's core, mantle, and oceanic crust respectively, and $\left(\frac{X}{\rm{Mg}}\right)_{\oplus}$ is the element $X$'s ratio to Mg in bulk Earth. The chemical abundance data for each element in each layer of the Earth is taken from \cite{McDonough2003} (bulk, mantle, and core), and \cite{White2014} (oceanic crust). The enhancement factors for each element at one location in the grid could then be multiplied by the expected primitive abundances, derived in Section \ref{PC}, to find the expected abundances of fragments of planetesimals which had the specific core and crust fractions at that formation location. The bodies which fragment to produce non-primitive planetesimals are not expected to be large enough to have plate tectonics, therefore, we have only considered oceanic crust; however the model could be extended to include continental crust.

Model 7 incorporates a more asteroidal-like differentiation process.~\cite{Ceres2005} and \cite{Vesta2012} indicate that asteroids and dwarf planets often have non-Earth-like core:mantle:crust mass ratios, suggesting that the depletion to the mantle caused by differentiation in these bodies may be different to the depletions seen in the Earth. The final model was calculated by fixing the bulk composition of the body and the composition of its crust and core to that of the Earth as before, but in this case we varied the parent mantle mass fraction and, thus, forced the composition of the mantle to change to one which could be more or less heavily depleted in lithophiles and/or siderophiles in comparison to the Earth. The enhancement factor was then calculated in the same way as suggested in Equation \ref{E1}, but with the new values of X\textsubscript{mantle} and Mg\textsubscript{mantle}. 

Figure \ref{fig:3} displays the abundance patterns created when the enhancement factors for material primarily composed of crust, mantle, and core are applied to planetesimals which formed at t=0\,Myrs, with a feeding zone parameter of 0\,AU, 2\,AU from the host stars in the \cite{FischerBrewer2016} sample over plotted onto Figure \ref{fig:1}.

\subsection{Statistically Constraining the Origin of Pollutants} \label{stats}

The aim of the following section is to not only select the most probable origin of the pollutants using our models (outlined in Section \ref{RPCIP}) but to also statistically rule out models, and areas of parameter space within our models, which cannot accurately recreate the observed pollutant abundances in the atmospheres of the chosen white dwarfs.

In order to achieve this we use a chi-squared parameterisation ($\chi^{2}$) to assess how well a particular model and set of free parameters fits the observations, where:
\begin{equation} \label{M}
\resizebox{.9\hsize}{!}{$\chi_{j}^{2} = \sum\limits_{i=1}^{\rm{N_{star}}} \sum\limits_{X} ^{\rm{Al,Ti,...}}{\frac{\left( X_{\rm{star(i),model(j)}} \,-\,  X_{\rm{observations}}\right)} {\sigma^{2}}^{2}}$}
\end{equation}
\noindent for all j models considered. Where $X_{\rm{observations}}$ is the observed abundance of an element X converted to assume either a pre-steady state phase, a steady state phase, or a declining phase of accretion, $X_{\rm{star(i),model(j)}}$ is the modelled abundance of an element X for a specific star and set of free parameters and $\sigma$ is the propagated uncertainty on the observationally derived abundance measurement. As Equation \ref{M} shows, we calculate $\chi^{2}$ values for every element with a measured abundance, for a number of initial conditions (number of stars in the sample) and sum them. We do this for every value of the model free parameters considered. The total number of models considered, j, is determined by how fine a grid we choose for our free parameters. In this work we chose to vary R from 0.05 to 33\,AU in steps of 0.05\,AU, z from 0 to 0.5\,AU in steps of 0.02\,AU, $f_{c}$ from 0 to 1 in steps of 0.01, $f_{o}$ from 0 to 1 in steps of 0.01, and $f_{pm}$ from 0.68 to 0.51 in steps of 0.005. Therefore, model 4, for example, contains $j = 66660$ different possible model fits to the observations for each star in the chosen sample.

We then calculate:
\begin{equation} \label{M1}
\left<\chi_{j}^{2}\right> = \frac{\chi_{j}^{2}}{\rm{N_{star}}}
\end{equation}
\noindent where $\rm{N_{star}}$ is the number of different initial conditions (stars) the $\chi^{2}$ parameterisation is found for and summed over. The value computed in Equation \ref{M1} is the average $\chi^{2}$ value per star, and is calculated because it allows the quality of the fit to be weighted depending on how common the required stellar conditions are. In this work $\rm{N_{star}}$ is generally chosen to be all 960 of the stars in our sample, however, for three systems we refine the sample as it is clear that only stars with specific abundance patterns can reproduce the pollutant composition. Our reasoning in each case is discussed in Section \ref{individual}.

We use a standard p-value to assess how confidently we can rule out models and areas of parameter space based on their average $\chi^{2}$ value. We convert the average $\chi^{2}$ value for each model into a p-value using: 
\begin{equation} \label{P}
p_{j,\left<\chi_{j}^{2}\right>,d} = {\left(2^{\left(\frac{d}{2}\right)} \Gamma_{\left(\frac{d}{2}\right)}\right)}^{-1} \int_{\left<\chi_{j}^{2}\right>}^{\infty} t^{\frac{d}{2}-1} e^{-\frac{t}{2}} dt
\end{equation}
\noindent where $\Gamma_{x} = \int_{0}^{\infty} t^{x-1} e^{-t} dt$ and d is the number of degrees of freedom (the number of observed elemental abundances included in the calculation minus the number of free parameters in the model). We then use the following standard convention to calculate which models and parameter values can be ruled out to a given level of statistical significance. A model which produces a fit with a p-value less than 0.317 can be ruled out with a confidence of 1$\sigma$, a model which produces a fit with a p-value less than 0.046 can be ruled out with a confidence of 2$\sigma$, and a model which produces a fit with a p-value less than 0.003 can be ruled out with a confidence of 3$\sigma$. The optimum model and set of free parameters is chosen to be the model and set of free parameters which produces the highest p-value. 

\subsection{Testing our Model on the Solar System} \label{test}

\begin{figure}
	% To include a figure from a file named example.*
	% Allowable file formats are eps or ps if compiling using latex
	% or pdf, png, jpg if compiling using pdflatex
	\includegraphics[width=\columnwidth]{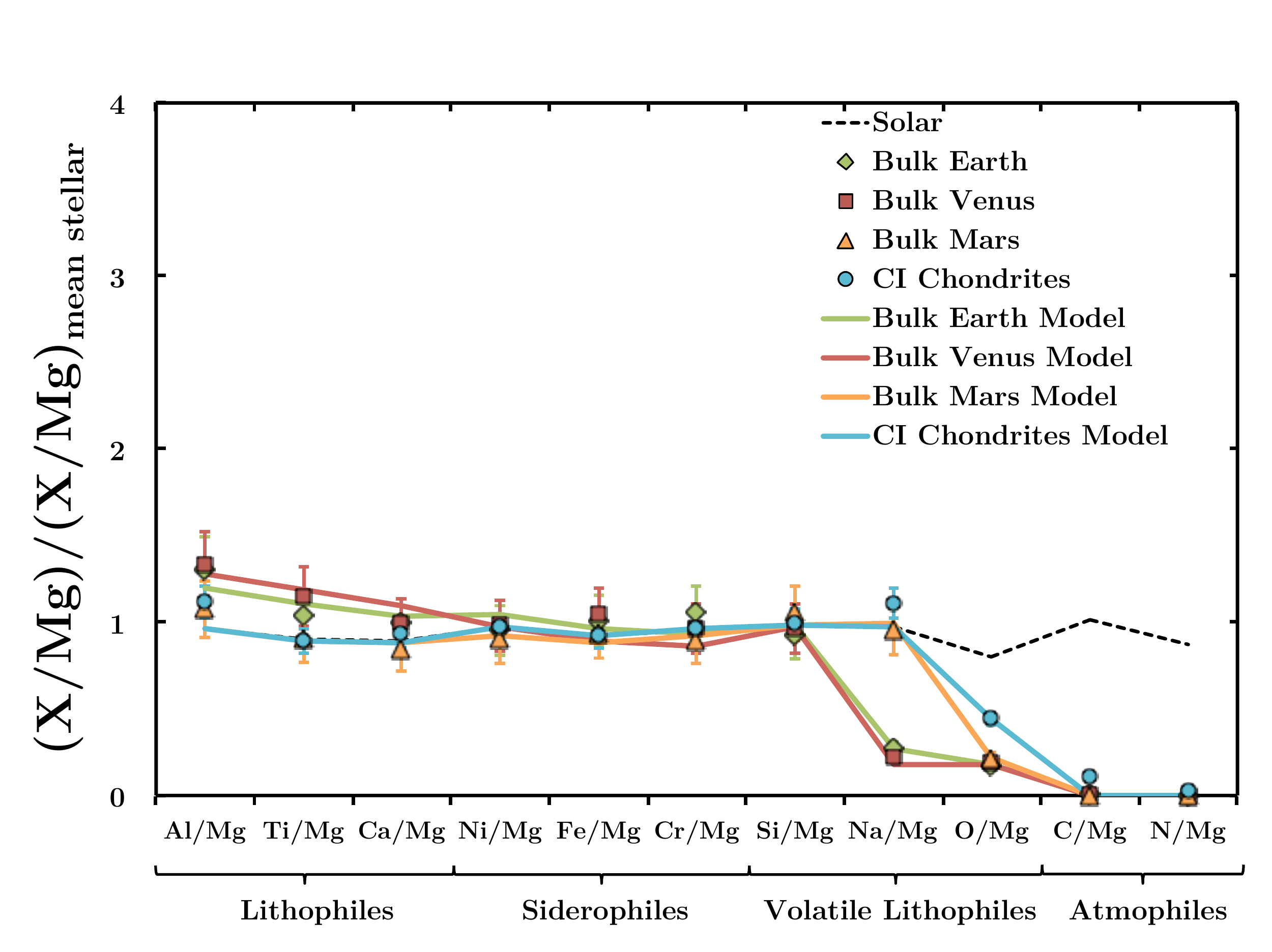}
    \caption{The abundances of the terrestrial planets and CI chondrites, taken from \protect\cite{McDonough2003} and \protect\cite{Lodders1998}, are shown by solid points. The solid lines show the best-fit models to describe the primitive solar system bodies abundances. The fits were determined using the model outlined in Section \protect\ref{PC} to modify the solar abundances. We find the predicted formation location order of the solar system bodies are as expected.}
    \label{fig:4}
\end{figure}

\begin{figure}
	% To include a figure from a file named example.*
	% Allowable file formats are eps or ps if compiling using latex
	% or pdf, png, jpg if compiling using pdflatex
	\includegraphics[width=\columnwidth]{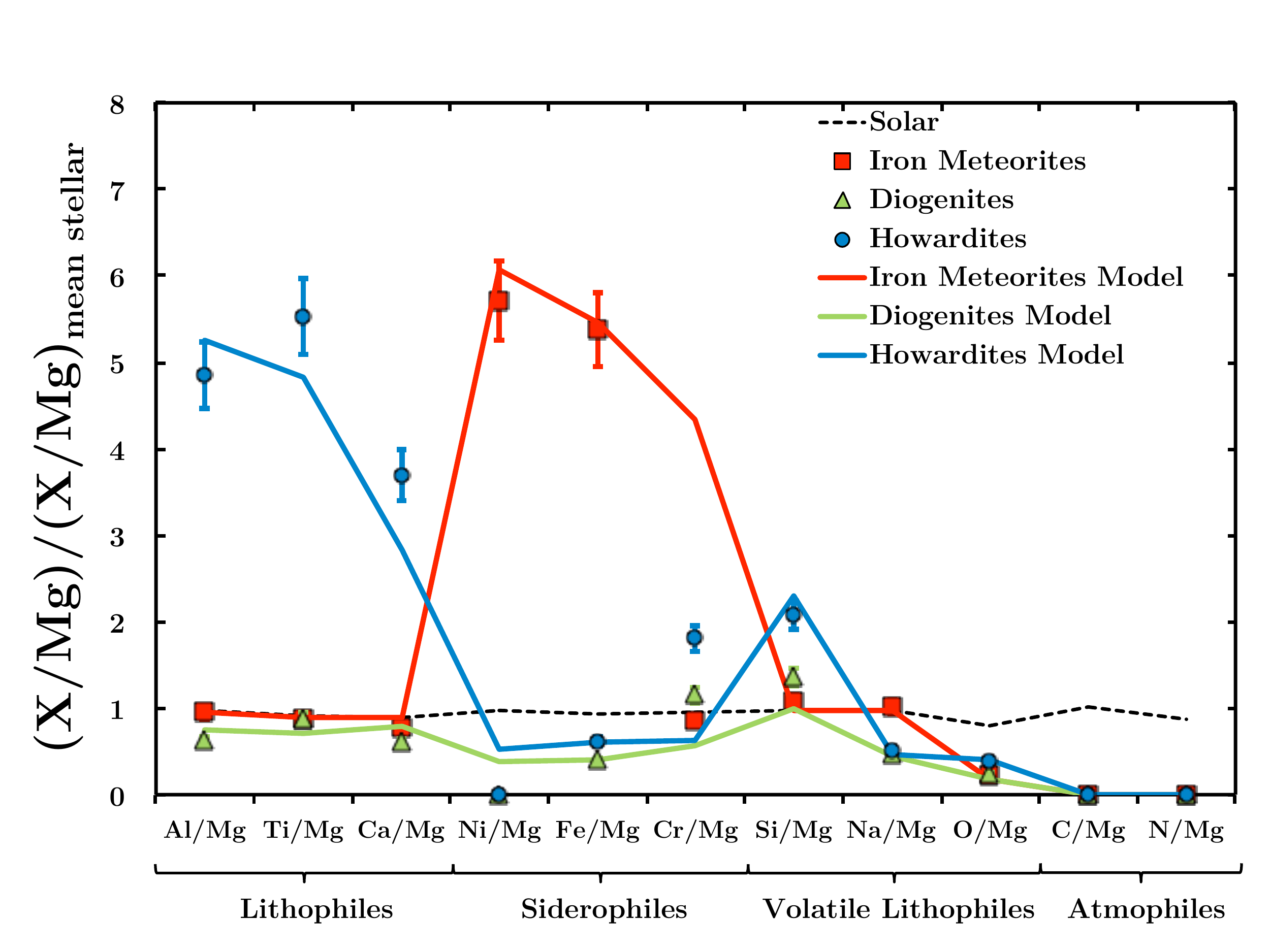}
    \caption{The abundances of the non-primitive meteorites in the solar system, taken from \protect\cite{Lodders1998}, are shown by solid points. The solid lines show the best-fit models to describe the non-primitive solar system bodies abundances. The fits were determined using the models outlined in Sections \protect\ref{PC} and \protect\ref{NPC} to modify the solar abundances. As the iron meteorites have no Mg present, the data displayed is a body composed of 50 percent iron meteorite material and 50 percent O chondrite material.}
    \label{fig:5}
\end{figure}

We tested our method by applying it to the observed abundances of the CI chondrite meteorites, the Diogenite meteorites, the Howardite meteorites, the iron meteorites, and the terrestrial planets (whose abundances have been derived using their bulk properties and the abundances measured in various meteoritic groups). We find that the best fit model selected by our analysis accurately recreates the abundances observed and finds the expected formation order in terms of temperature, or radial location, of the primitive solar system bodies (Figure \ref{fig:4}). The fits of our models to the non-primitive meteorites are less accurate, however to first order the expected trends are well reproduced and are done so via the expected scenarios (Figure \ref{fig:5}). This suggests our model is too simple, and to accurately recreate these bodies we would have to include differentiation involving non-Earth-like core and crust abundances. This is not unexpected as some elements differentiate differently in the lower pressure conditions present inside smaller bodies, thus, the abundance patterns in the fragments of smaller asteroids, like the non-primitive meteorites analysed here, are not necessarily mapped out well by our model. The main elements affected by this are Ni, and Cr. Ni is more siderophilic in low pressure conditions. Thus, abundances in mantle-rich asteroids are lower than expected. Cr is more lithophilic in lower pressure conditions, thus, abundances are higher than expected for the Diogenites, but lower than expected for the iron meteorites. Due to the uncertainties on the polluted white dwarf abundances this effect should not provide a major hindrance when analysing the data.

\subsection{Methods Summary}
We hypothesise that the diversity in the abundances observed in white dwarf atmospheres results from two key processes which alter the pollutant abundances from the initial composition of the planetesimal forming disc; namely, the highest temperature experienced by the body, and core formation followed by collisional erosion. Our model suggests that the chemical abundance patterns present in rocky planetary bodies (assumed to be the pollutants of white dwarfs) should offer insights into the formation history of the bodies. This is because the formation location, collisional evolution, and geological evolution produce clear distinguishable signatures in the abundance patterns (Figure \ref{fig:2} and Figure \ref{fig:3}). Our model accurately recreates the abundance patterns of the rocky bodies in the solar system finding reasonable formation histories for each of the them (Figure \ref{fig:4} and Figure \ref{fig:5}).
\FloatBarrier

\section{Results} \label{results}

\subsection{Polluted White Dwarf Results Overview}
We minimised $\chi^{2}$ (as defined in Section \ref{stats}) for all seven models (presented in Section \ref{RPCIP}) and for all possible values of their free parameters for the 17 white dwarf systems listed in Table \ref{tab:1a}. We analsyed these systems assuming they were in one of three possible phases of accretion; a pre-steady state phase, a steady state phase, or a declining phase (as outlined in Section \ref{PWDD}). In order to asses the quality of the fits across the different models we used a standard p-value convention as described in Section \ref{stats}.

\begin{figure*}
	% To include a figure from a file named example.*
	% Allowable file formats are eps or ps if compiling using latex
	% or pdf, png, jpg if compiling using pdflatex
	\includegraphics[width=\textwidth]{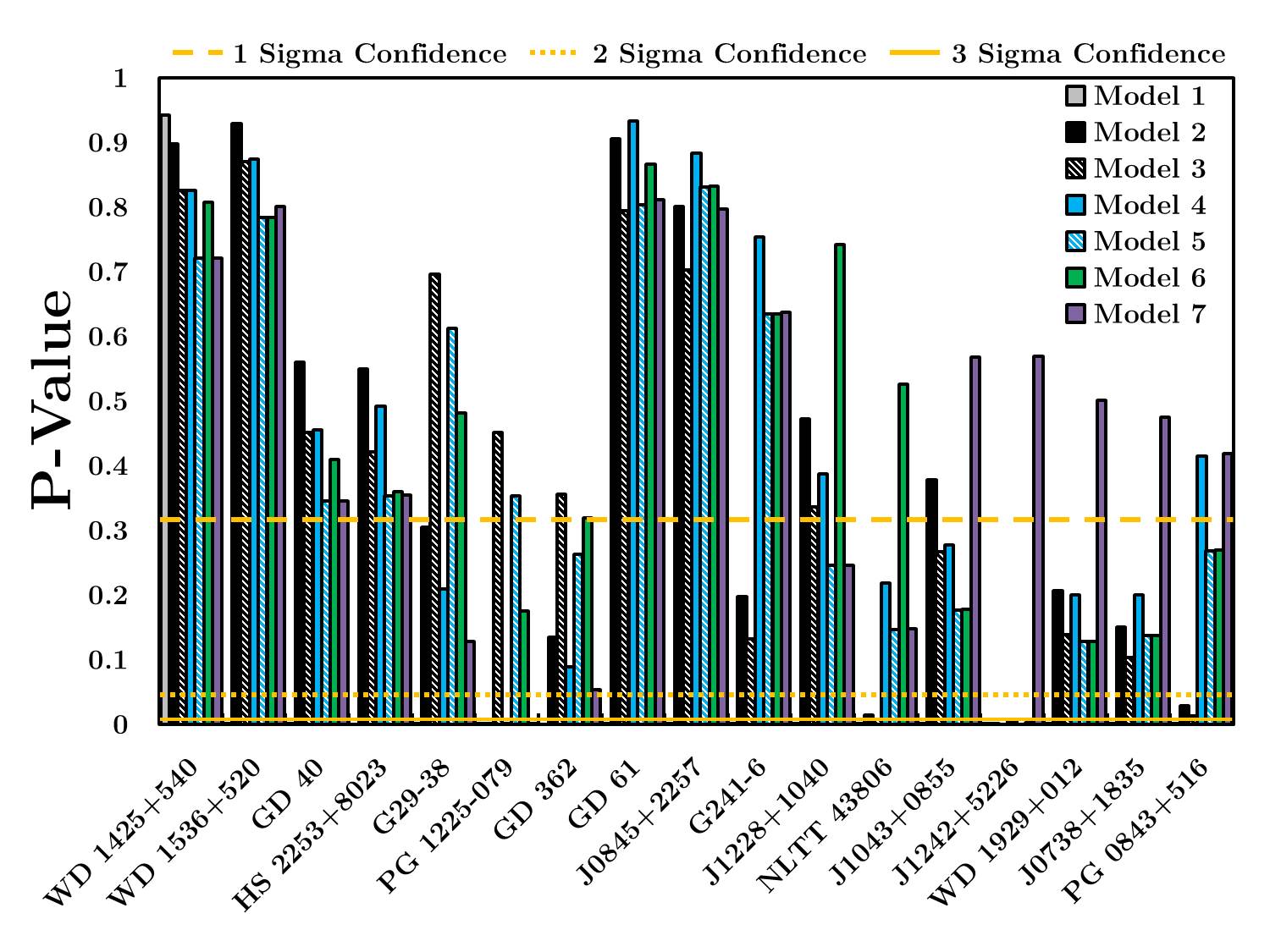}
    \caption{ The maximised p-values for the fits of each of the seven models analysed in this work to the abundance data assuming three possible accretion scenarios. Only the accretion scenario which produces the highest p-value is plotted for each model. The orange lines represent the p-values models must be greater than to not be ruled out to with a confidence of 1$\sigma$ (dashed), 2$\sigma$ (dotted), and 3$\sigma$ (solid).}
    \label{fig:6}
\end{figure*}

Figure \ref{fig:6} shows the maximised p-values for each of the seven models for each of the systems considered and highlights which models can be ruled out to what level of statistical significance for each system. For each model the p-value plotted is the maximum p-value the model can achieve when fitting the pollutant abundances for all possible phases of accretion. \footnote{For HS\,2253+8023 the values shown correspond to the p-values produced when the observed Ca abundance is ignored. For GD\,362 and PG\,1225--079 the values shown correspond to the p-values produced when only stars with above average Ca/Mg ratios are included as possible initial conditions. For WD\,1425+540 the values shown correspond to the p-values produced when only stars with O/Mg ratios greater than that of the pollutant are included as initial conditions. The reasoning behind these constraints are discussed in Section \ref{individual}.}

Figure \ref{fig:6} highlights that we can rule out pollutants with abundances similar to stellar material (model 1) with a confidence of at least 3$\sigma$ for the majority of systems analysed in this work. However, we cannot rule out pollutants with abundances similar to what one would expect from rocky exoplanetary material (models 2 to 7). This result is as expected and reinforces the hypothesis that these white dwarfs have been externally polluted by rocky exoplanetary material \citep{JuraYoung2014}.

\subsection{Polluted White Dwarf Individual System Results} \label{individual}

\begin{figure*}
	% To include a figure from a file named example.*
	% Allowable file formats are eps or ps if compiling using latex
	% or pdf, png, jpg if compiling using pdflatex
	\includegraphics[height=0.95\textheight]{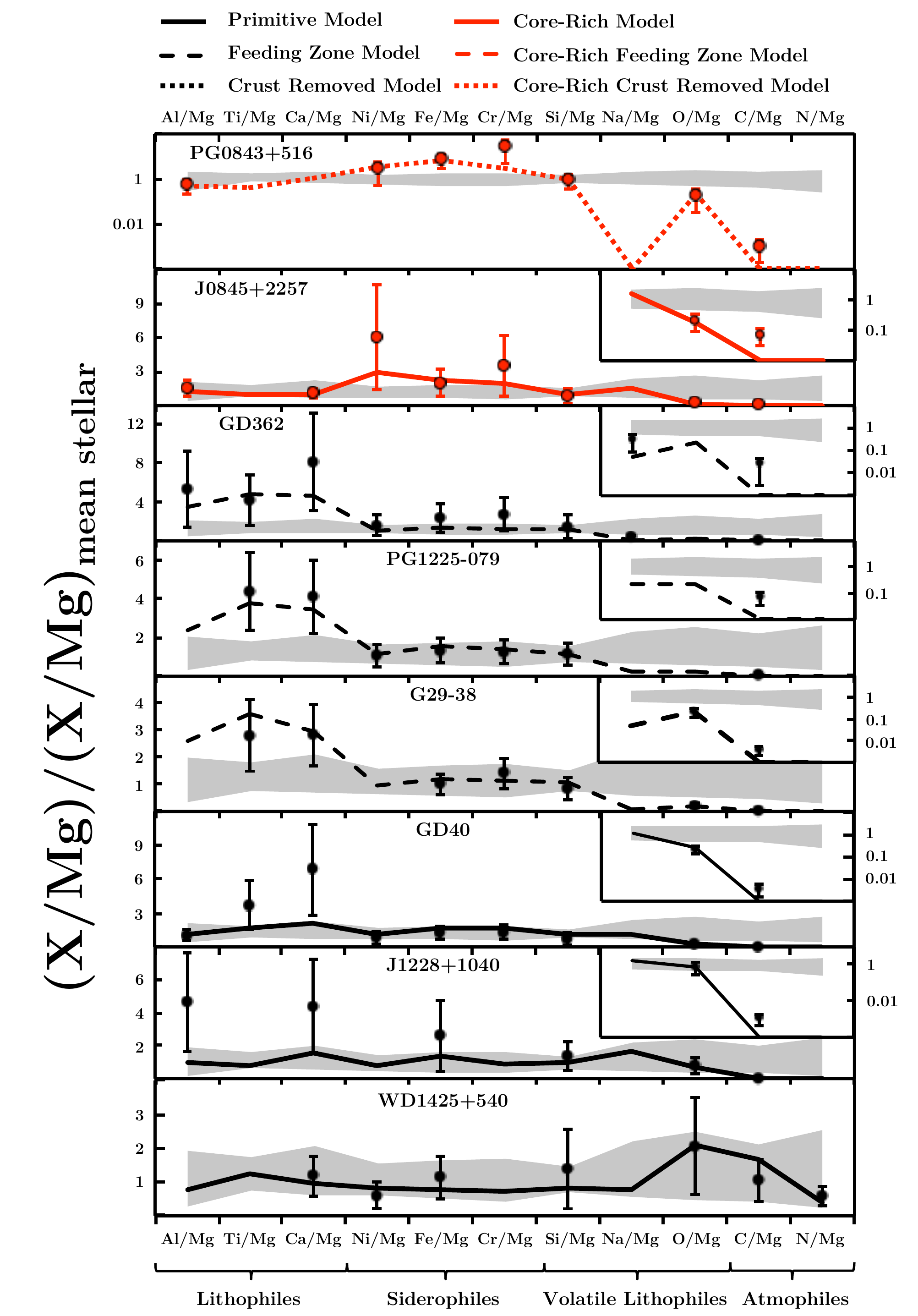}
    \caption{The steady state adjusted polluted white dwarf data and the models with the highest p-values over plotted onto the stellar range from Figure \protect\ref{fig:1}. The abundances shown for GD\,362, PG\,1225--079, and WD\,1425+540 have not been adjusted for steady state as we find our models produce a much better fit assuming accretion is in a pre-steady state phase. For J1228+1040 we plot the model with the second highest p-value, this is because the range in parameter space which provides the optimal model is unlikely to be produced in the reality.}
    \label{fig:pwd3}
\end{figure*}

\begin{figure*}
	% To include a figure from a file named example.*
	% Allowable file formats are eps or ps if compiling using latex
	% or pdf, png, jpg if compiling using pdflatex
	\includegraphics[height=0.95\textheight]{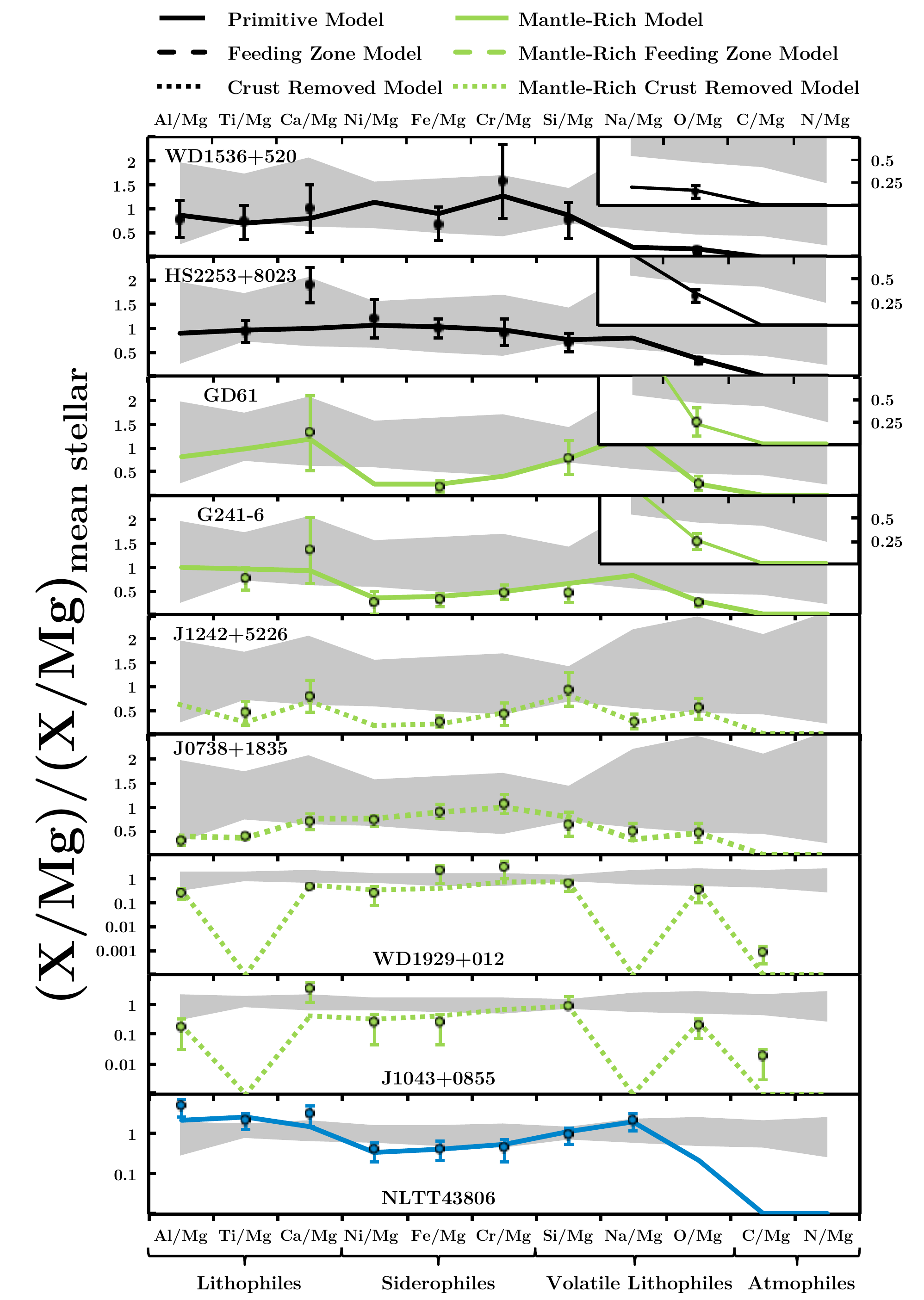}
    \caption{The steady state adjusted polluted white dwarf data and the models with the highest p-values over plotted onto the stellar range from Figure \protect\ref{fig:1}. The abundances shown for HS\,2253+8023, J0738+1835, G241-6, and J1242+5226 have not been adjusted for steady state as we find our models produce a much better fit assuming accretion is in a pre-steady state phase.}
    \label{fig:pwd4}
\end{figure*}

Figures \ref{fig:pwd3} and \ref{fig:pwd4} show 17 subplots in which the observed pollutant abundances, converted to assume steady state accretion, for each system are plotted with the stellar range presented in Figure \ref{fig:1}. The abundances shown for GD\,362, PG\,1225--079, HS\,2253+8023, J0738+1835, G241-6, WD\,1425+540 and J1242+5226 have not been converted to assume a steady state of accretion as our models match the abundances considerably better if we assume they are in a pre-steady state phase, and given they correspond to the coolest helium white dwarfs analysed in this work this is not unexpected.  The lines shown are the models with the highest p-values for each system when applied to one star in the sample. For J1228+1040 we plot the model with the second highest p-value, this is because we expect that the region of parameter space which provides the model with the largest p-value is unlikely to be produced in the reality (our reasoning will be discussed in Sections \ref{individual} and \ref{diss}).

\subsubsection{WD\,1425+540}
When fitting the pollutant abundances of WD\,1425+540 we include, and sum over, only the stars in the stellar sample which have an O/Mg ratio greater than that of the pollutant. We do this because the O/Mg ratio of the pollutant is extremely high, and given the pollutants other elemental abundances, our model suggests that this must be because the host star has an intrinsically large O/Mg ratio. The best fit model for the system is one which involves the accretion of stellar material, however, this conclusion is not robust because we do not include H in our model. If we included H in our model the quality of the fit to stellar material would drastically decrease. This is because if stellar material was accreted we would expect much more H in the system's atmosphere than is present. Otherwise, we find that the pollutant is best reproduced by a model where the accreting body is a ice rich primitive planetesimal. The pollutant is modelled best, p-value of 0.90, when the system is assumed to be in a pre-steady state, which given the considerable sinking timescales is not unlikely.  We find that the pollutant of this system contains ice species to a statistical significance of 3$\sigma$. Therefore our analysis supports the conclusions of \cite{Xu2017}.


\subsubsection{WD\,1536+520}
The atmospheric abundances of WD\,1536+520 can be fitted to a p-value of 0.93 by the accretion of a primitive planetesimal in the steady state phase. The expected formation location is well inside of the water ice line (we have not included the trace H abundance in our analysis as its origin is still unclear), however, more volatile elemental abundances are required to constrain a more accurate estimate for the pollutants formation location. In \cite{Farihi2016} it was suggested that the pollutant was primitive, however, the Cr value was cited as possible evidence for the pollutant being core-rich. We find that this explanation is not necessary as the Cr value lies within the range possible for nearby stars, and thus a primitive pollutant best matches the abundances.
\subsubsection{GD\,40}
A scenario in which the pollutant is a primitive planetesimal accreting in steady state fits the observed abundances well, p-value of 0.57. The expected formation location lies in the inner part of the system, well inside the ice lines, as suggested in \cite{Jura2012}. This formation location is consistent with the abundances of P and S, P is approximately solar while S is heavily depleted relative to solar, however due to a lack of stellar data for these elements they were not used when fitting the models. Measurements of the Na abundance in the system would allow our model to further constrain the formation location. These conclusions do not change when we analyse the data assuming the system is in a non-steady state phase.

\subsubsection{HS\,2253+8023}
The atmospheric abundances of HS\,2253+8023 are difficult to explain in the context of our model due to the anomalously high Ca value, with small quoted uncertainty. We can only fit the abundances to a p-value above 0.01 if we assume the system is in a pre-steady state phase and if we do not include the Ca abundance in the calculation. A scenario where the pollutant is a primitive planetesimal which formed outside the water ice line produces the best fit to the abundances, p-value of 0.55. We find that the pollutant requires water ice and that the pollutant must be in a pre-steady state phase of accretion to a statistical significance of 3$\sigma$. \cite{Klein2011} also noticed the difficulty in explaining the Ca abundance and similarly they concluded that the pollutant could probably be explained by nebula condensation and that the Ca abundance was likely anomalous. In the plots displayed in this work any fit to the abundances of the pollutant of HS\,2253+8023 will not include the Ca abundance due to the uncertainty on the accuracy of the measurement and its error estimate.

\subsubsection{G29-38}
A scenario in which the pollutant of G29-38 is a primitive planetesimal produces a fit to the observations with a p-value of 0.30. This can be drastically improved to a p-value of 0.70 by allowing the pollutant to have formed from a feeding zone. Allowing the pollutant to be a fragment of a differentiated body provides no improvement to the fit of the observations. Our best fit model suggests the majority of the pollutant formed at a temperature around 1400\,K, and therefore shows a strong volatile depletion trend. These results may suggest that the pollutant was modified post-formation. \cite{Xu2014a} suggested that the pollutant was non-primitive, due to a poor match to bulk Earth, however it was stated that the exact nature of the body was not clear. We find that the pollutant could in fact be primitive and that the poor match to bulk Earth is due to a difference in the temperatures experienced by the bodies.

\subsubsection{PG\,1225--079}
When the average $\chi^{2}$ value is calculated using all of the stars in the stellar sample our models cannot fit the abundances to a p-value greater than 0.317 for any of the possible accretion scenarios. However when only the stars with Ca abundances above average are selected a p-value of 0.45 is obtained when the pollutant is modelled as a primitive planetesimal which formed in the severe volatile depletion zone and is currently accreting onto the white dwarf in a pre-steady state phase. This suggests that the white dwarf progenitor was a Ca rich star in comparison to the average star in our sample, which is not unexpected given that there may be an observational bias in the sample of polluted white dwarfs chosen for this study due to the fact that the Ca lines are often used as an indication of pollution. Due to the refractory abundances the formation temperature of the body is expected to be of the order 1400\,K. \cite{Xu2013} suggested that the pollutant had no solar system analogue and therefore they drew no conclusions about the nature of the pollutant. Our analysis suggests that the pollutant was a body which was heated to extreme temperatures either during formation or post-formation, and thus, as \cite{Xu2013} suggested it has no solar system meteoritic analogue.

\subsubsection{GD\,362}
When the average $\chi^{2}$ value is calculated using all of the stars in the stellar sample our models cannot fit the abundances to a p-value greater than 0.317 for any of the possible accretion scenarios. However when only the stars with Ca abundances above average are selected a p-value of 0.36 is obtained when the pollutant is modelled as a primitive planetesimal which formed in the severe volatile depletion zone and is currently accreting onto the white dwarf in a pre-steady state phase. This suggests that the white dwarf progenitor was a Ca rich star in comparison to the average star in our sample, which is not unexpected given that there may be an observational bias in the sample of polluted white dwarfs chosen for this study due to the fact that the Ca lines are often used as an indication of pollution. \cite{Xu2013} suggested that the pollutant is a mesosiderite analogue, a type of stony-iron meteorite whose formation mechanism is still debated, in our model this would correspond to a fragment which is simultaneously core-rich and crust-rich, we find that such a model produces at best a p-value of 0.31 even when only stars with above average Ca are considered and, therefore, suggest that it is more likely that the pollutant is primitive and experienced temperatures of the order 1400\,K.

\subsubsection{GD\,61}
Assuming the system is in a steady state of accretion, our analysis fits the observations with a p-value of 0.11 for a scenario in which the pollutant of GD\,61 is a primitive planetesimal, however to achieve a non-negligible p-value this model requires the body to have formed at temperatures between of 1300\,K and 1270\,K to reproduce the Fe depletion. This p-value can be drastically improved to 0.93 when the pollutant is considered to be a mantle-rich fragment of a differentiated body which formed outside the water ice line. If measurements of the abundance of Ni were found in this system, and they were shown to be heavily depleted it would reduce the fit of a primitive planetesimal by ruling out formation temperatures between 1300\,K and 1270\,K. In the p-value maximised model the pollutant is expected to contain water ice, however, due to the large uncertainties on the O abundance the pollutant could still be fitted well without water ice (we have not included the trace H abundance in our analysis as its origin is still unclear). These conclusions reinforce those suggested in \cite{Farihi2011}. However, if we assume the system is not in a steady state of accretion these conclusions change dramatically. When the system is assumed to be in a declining phase we find that a model where the pollutant is a primitive planetesimal fits the data with a p-value of 0.90. If measurements of more elemental abundances or if more precise measurements of the existing abundances were found one could rule out the possibility of the system being in a declining phase. \footnote{All results presented in this work for GD\,61 were found using the data given in \cite{Farihi2011}. New optical and UV spectra were obtained for GD\,61 and the atmospheric abundances were updated in \cite{Farihi2013}. Since publication we have analysed the data presented in \cite{Farihi2013} and found no change in the best fit formation scenario. However the constraints on the core mass fraction, formation location, and the phase of accretion are tighter. We find that the abundances are not consistent with the system being in a declining phase of accretion, and to a statistical significance of at least 3$\sigma$ the pollutant contains water ice and is a mantle-rich fragment of a differentiated body. }

\subsubsection{SDSS\,J0845+2257}
A scenario in which the pollutant is a primitive planetesimal which is accreting onto the white dwarf in a steady state phase produces a fit to the observed abundances with a p-value of 0.80. The best fit model, p-value of 0.88, requires the pollutant to be a fragment with an excess of core-like material as concluded in \cite{Wilson2015}. The expected formation location is well inside the water ice line, however, the exact formation temperature is poorly constrained due to the lack of a Na, S, or P abundance measurement. These conclusions are not altered when we assume the system is in a declining phase and we can rule out to 3$\sigma$ scenarios where accretion stopped more than two sinking timescales ago.

\subsubsection{G241-6}
Assuming the system is in a steady state of accretion, a model where the pollutant is a primitive planetesimal provides a fit to the observations with a p-value of 0.20, this can be improved to 0.22 by allowing the pollutant to be a mantle-rich fragment of a differentiated body. If it is assumed that the system is in a pre-steady state phase the fit can be improved to one with a p-value of 0.75 when the pollutant is considered to be a mantle-rich fragment of a differentiated body, this suggests that the system is likely in a phase of pre-steady state accretion which given the considerable sinking timescales for this system is not unlikely. The best fit models suggest that the pollutant formed outside the water ice line and to a statistical significance of 1$\sigma$ the pollutant contains clathrate species, this agrees with the abundances of P and S which are approximately solar, however due to a lack of stellar data P and S were not used when fitting the models.~\cite{Jura2012} suggested that the pollutant was primitive, due to a reasonable match to bulk Earth. We find that it is much more likely that the pollutant is a mantle rich body. 

\subsubsection{SDSS\,J1228+1040}
The atmospheric abundances of SDSS\,J1228+1040 can be modelled to a p-value of 0.47 by the steady state accretion of a primitive planetesimal. A p-value of 0.74 is obtained when the pollutant is modelled as a simultaneously crust-rich and core-rich fragment. Due to the inability of collisional models to produce such fragments \citep{Carter2017}, we expect that the latter model is not a realistic history for the pollutant. \cite{Gansicke2012} suggested that the pollutant was most likely a primitive planetesimal which matches our results. The formation location of the pollutant is expected to be outside the water ice line to a statistical significance of 2$\sigma$, observations of more volatile species could help to reinforce this conclusion.

\subsubsection{NLTT\,43806}
Steady state accretion of a primitive planetesimal provides a fit to the observations with a p-value of 0.01. The fit can be improved to a p-value of 0.22 when the pollutant is allowed to be a mantle-rich fragment of a differentiated body, and if crustal differentiation is taken into account we can improve the fit to a p-value of 0.53. This dramatic increase suggests that the body may well be an exo-lithosphere as concluded by \cite{Zuckerman2011}, especially given its low core mass fraction which can readily be reproduced via collisional models \citep{Carter2017}. The key element in this conclusion is Na, as an enhanced Na abundance relative to the expected primitive level can only be modelled by an excess of crustal material due to its volatility relative to Mg. These conclusions are not altered when we assume the system is in a declining phase. We can rule out to 3$\sigma$ scenarios where accretion stopped more than five sinking timescales ago.

\subsubsection{SDSS\,J1043+0855}
The atmospheric abundances of SDSS\,J1043+0855 can be well reproduced by the steady state accretion of a primitive planetesimal, p-value of 0.38, this fit can be improved to a p-value of 0.57 when the pollutant is a mantle-rich fragment of a differentiated body which had a large crust to mantle mass ratio but collisional processing removed the crustal component. The expected formation location is in a region where clathrate species contribute to the composition of the body, however the large uncertainty in the O abundance means this conclusion is not robust. \cite{Dufour2016} suggested a similar most likely formation history. Remeasuring the Ca and Al abundances could constrain whether the pollutant is crustally depleted.

\subsubsection{SDSS\,J1242+5226}
Assuming the system is in a steady state of accretion, a model where the pollutant is a primitive planetesimal provides a fit to the observations with a p-value of 0.001, this can be improved to 0.13 by allowing the pollutant to be a mantle-rich fragment of a differentiated body which had a large crust to mantle mass ratio. If it is assumed that the system is in a pre-steady state phase, the fit can be improved to one with a p-value of 0.57 when the pollutant is considered to be a mantle-rich fragment of a differentiated body which had a large crust to mantle mass ratio, this suggests that the system is likely in a phase of pre-steady state accretion which given the considerable sinking timescales for this system is not unlikely. To a statistical significance of 3$\sigma$ the pollutant is required by our model to be a fragment of a differentiated body which formed outside the water ice line. These conclusions agree with those published in \cite{Raddi2015}, where it was suggested that the pollutant was an icy mantle-rich body. The pollutant of this system may indicate that some asteroids have icy upper mantles, this would help to explain how ice survives and pollutes white dwarfs. Although not analysed in this work, a large amount of trace hydrogen is present in the atmosphere of the white dwarf which could reinforce the requirement for the pollutant to be icy.

\subsubsection{WD\,1929+012}
The atmospheric abundances of WD\,1929+012 can be fitted by the steady state accretion of a primitive planetesimal to a p-value of 0.21, this can be improved to a p-value of 0.50 when the pollutant is a mantle-rich fragment of a differentiated body which had a large crust to mantle mass ratio but collisional processing removed the crustal component. The expected formation location is in a region were water ice contributes to the composition of the body, in fact the pollutant requires water ice to a statistical significance of 3$\sigma$. The S abundance, which is not modelled in our work due to a lack of stellar data, when adjusted for the mantle-rich nature of the pollutant suggests that the pollutant has super-solar S, reinforcing the conclusion of cool formation conditions. As the pollutant is expected to be a mantle-rich icy fragment it may suggest that asteroids possibly have icy upper layers, this conclusion would help to explain how ice survives and pollutes white dwarfs. \cite{Gansicke2012} and \cite{Xu2013} were unsure of the nature of the pollutant due to the clash between the siderophiles and the anomalously low refractory abundances. Our analysis indicates that given the uncertainties on Fe and Cr it is more probable that the pollutant is mantle-rich rather than core-rich and the refractory abundances are well modelled by differentiation involving a parent body with a substantial crust to mantle mass ratio.

\subsubsection{SDSS\,J0738+1835} \label{J0738}
Assuming the system is in a steady state of accretion, a model where the pollutant is a primitive planetesimal provides a fit to the observations with a p-value of 0.15, this can be improved to 0.35 by allowing the pollutant to be a mantle-rich fragment of a differentiated body which had a large crust to mantle mass ratio. If it is assumed that the system is in a pre-steady state phase the fit can be improved to one with a p-value of 0.48 when the pollutant is considered to be a mantle-rich fragment of a differentiated body which had a large crust to mantle mass ratio, this suggests that the system is likely in a phase of pre-steady state accretion which given the considerable sinking timescales for this system is not unlikely. The pollutant is required to form outside the water ice line to a statistical significance of 3$\sigma$. Similarly \cite{Dufour2012} suggested that the pollutant is icy and that it is potentially a fragment of a planetary body which has had its crust stripped.

\subsubsection{PG\,0843+516}
Accretion of a primitive planetesimal in the steady state phase provides a fit to the observations with a p-value of 0.02. This can be drastically improved to a p-value of 0.41 when the pollutant is allowed to be a core-rich fragment of a differentiated body, and if the mantle mass fraction of the parent body is taken as a free parameter we can improve the fit to a p-value of 0.42. Therefore, to a statistical significance of 2$\sigma$ the pollutant is a core-rich fragment of a differentiated body. The formation location is expected to be inside the water ice line. The S abundance, when corrected for the core-rich nature of the body, is sub-solar reinforcing the expected non-icy nature of the pollutant. These conclusions match those of \cite{Gansicke2012} which suggested that the pollutant is most likely a non-icy, core-rich fragment.

\subsection{Constraints on Collisions, Differentiation, and Fragmentation}

In this section we consider whether the abundances observed in the atmospheres of polluted white dwarfs require the pollutants to be the fragments produced in collisions between differentiated planetary bodies to be accurately modelled.

\begin{figure}
	% To include a figure from a file named example.*
	% Allowable file formats are eps or ps if compiling using latex
	% or pdf, png, jpg if compiling using pdflatex
	\includegraphics[width=\columnwidth]{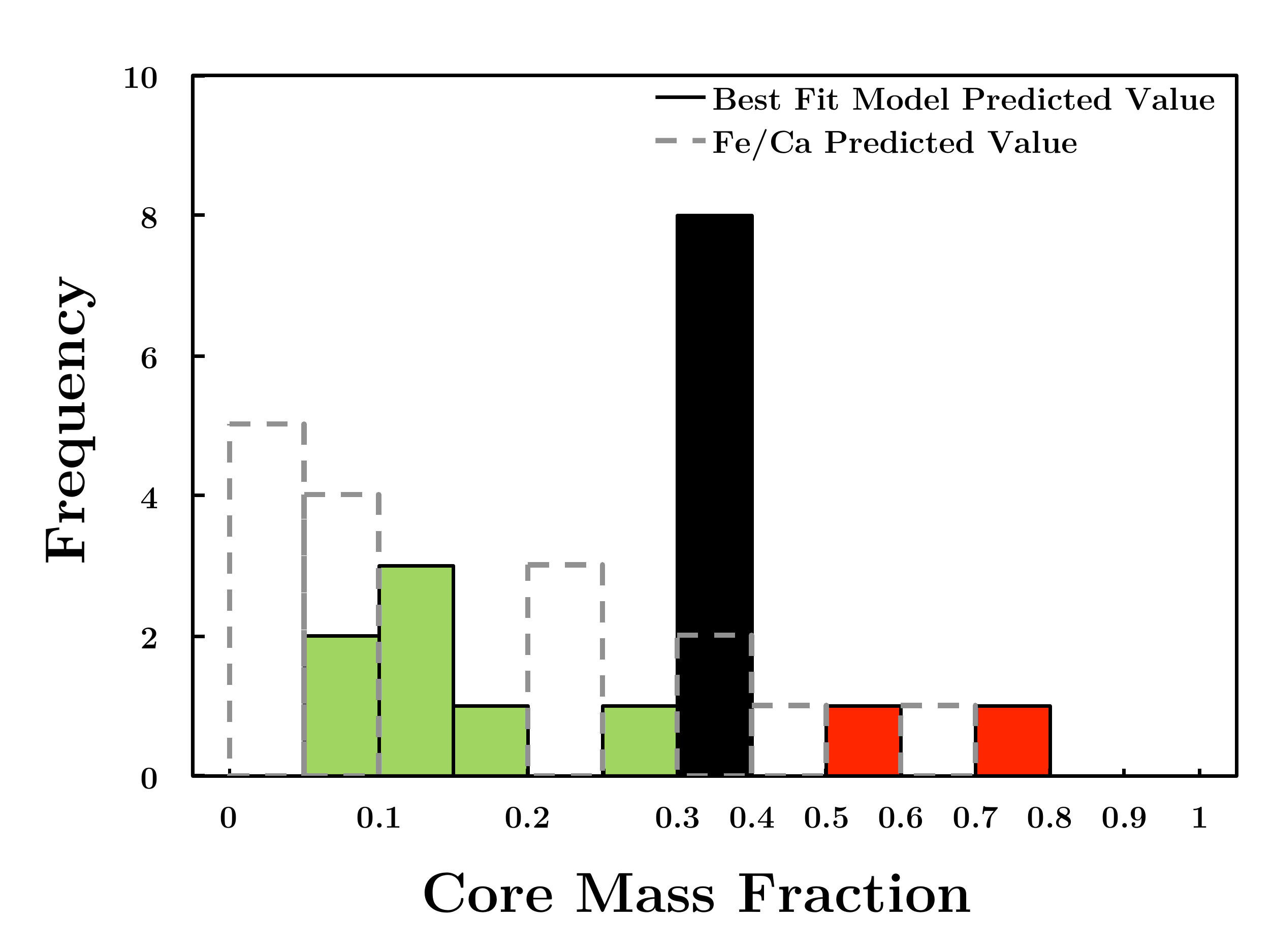}
    \caption{ A histogram of the best fitting core mass fractions for the polluted white dwarfs considered in this work (solid line. The red fill indicates core-rich and the green fill indicates mantle-rich). Simply taking the pollutant Fe/Ca ratio and finding the predicted core mass fraction from those values using bulk Earth data (dashed line) generates a drastically different spread due to the lack of consideration for the uncertainties across multiple elements and the possibility of host star chemical variance.}
    \label{fig:8}
\end{figure}

Figure \ref{fig:8} shows how the core mass fractions which produce the highest p-values are distributed in the 17 systems analysed (solid line). We compare this to the distribution found when one calculates the core mass fractions by simply finding the core mass fraction of earth-like material which best matches the steady state Fe to Ca ratio of the pollutant (dashed line). We indicate in red the systems whose best fit model has an excess of core-like material, in green the systems whose best fit model has an excess of mantle-like material, and we indicate in black the systems which do not require an excess in either.

\begin{figure*}
	% To include a figure from a file named example.*
	% Allowable file formats are eps or ps if compiling using latex
	% or pdf, png, jpg if compiling using pdflatex
	\includegraphics[width=\textwidth]{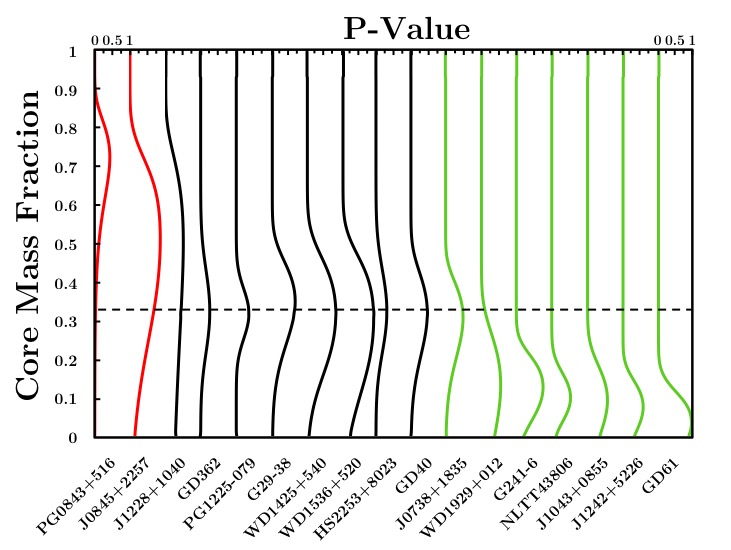}
    \caption{The p-value for the fit between the model compositions and the observed compositions for each white dwarf pollutant when the core mass fraction is varied from 0 to 1 while the formation location and feeding zone parameter remain fixed to the values which produce the best fit. The black dashed line displays the core mass fraction of bulk Earth. The systems in red are systems where the best fit model has an excess of core-like material, the systems in green are systems where the best fit model has an excess of mantle-like material, and the systems in black do not require an excess in either.}
    \label{fig:7}
\end{figure*}

In Figure \ref{fig:7} we display how the quality of the fit (p-value) between the observed atmospheric abundances and the modelled compositions of planetary fragments changes as we vary the core mass fraction of the fragments between 0 and 1, while the formation location and feeding zone parameter remain fixed to the values which produced the best fit for each system. The phase of accretion assumed is also fixed for each system to the one which produces the highest p-value. We note on the figure that Earth's core mass fraction is 0.32 \citep{McDonough2003} and therefore a mass fraction of this value indicates that the body need not be differentiated nor collisionally processed.

Figures \ref{fig:8}  and \ref{fig:7} highlight how planetary differentiation, collisions, and fragmentation are required to fit the abundances measured in the atmospheres of at least one, but possibly up to nine, of the 17 white dwarf systems studied in this work. This result reinforces the conclusions previously established in the literature which suggest that the atmospheric abundances of externally polluted white dwarfs can probe the geological and collisional activity in exoplanetary systems. Our model also predicts that a spread in the core mass fraction of these pollutants is required to reproduce the white dwarf atmospheric abundances. When this distribution is estimated by simply using the steady state Fe to Ca ratio of the pollutant and comparing this to the Fe to Ca ratio in various layers of the Earth a vastly different spread is generated. The main reasons for this is that the simpler model does not take into account the abundances of multiple elements simultaneously or their relative errors. It also does not take into account host star chemical variance or the possibility that the Fe to Ca ratio can be altered independently of the core mass fraction by the bodies formation conditions. The spread generated from our model roughly matches the spread expected if the pollutants of the white dwarfs are randomly sampled from the distribution of core mass fractions predicted by collisional models \citep{Carter2015, Carter2017}. This suggests that using the Fe to Ca ratio alone is ineffective in constraining the geological and collisional history of white dwarf pollutants.

\begin{figure*}
	% To include a figure from a file named example.*
	% Allowable file formats are eps or ps if compiling using latex
	% or pdf, png, jpg if compiling using pdflatex
	\includegraphics[width=\textwidth]{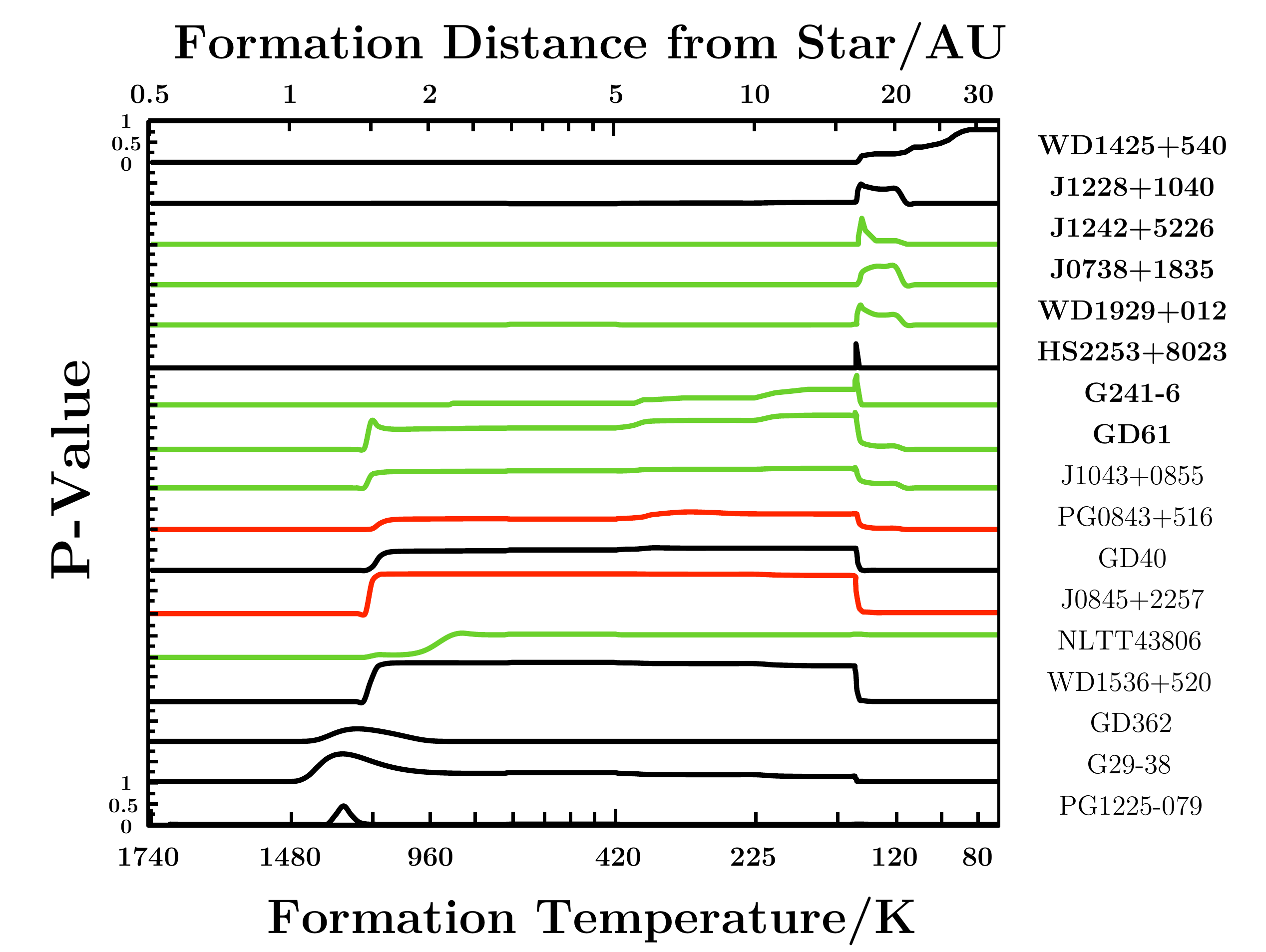}
    \caption{The p-value for the fit between the model compositions and the observed compositions for each white dwarf pollutant when the formation location is varied from between 0.05\,AU and 33\,AU while the core mass fraction and parent mantle mass fraction remain fixed at the value which produces the best fit. The systems in red are systems where the best fit model has an excess of core-like material, the systems in green are systems where the best fit model has an excess of mantle-like material, and the systems in black do not require an excess in either. The systems whose best fit models require the pollutants to contain water ice are indicated in bold. The formation location displayed on the upper axis is derived assuming formation around a solar mass host star at t=0\,Myrs \protect\citep{Chambers2009}. }
    \label{fig:9}
\end{figure*}

\subsection{Constraints on Heating During Formation or Subsequent Evolution}

In this section we consider whether the diversity in the elemental abundances observed in the atmospheres of polluted white dwarfs could be explained by the heating of planetary material during formation that depletes the planetesimals of volatile species, potentially moderately volatile species, and possibly causes enhancements in the refractory species.

Our analysis finds the range of formation temperatures consistent with the observed abundances. These are plotted in Figure \ref{fig:9} which shows how the quality of the fit (p-value) to the observed abundances of the best fit model for each system changes as the formation location is varied from 0.05\,AU to 33\,AU in a solar-like protoplanetary disc at time equals 0 \,Myrs (corresponding to a temperature range between 3000\,K and 75\,K). The fragment core mass fraction and the parent mantle mass fraction is kept fixed at the value predicted by the best fit model for each system along with the assumed phase of accretion. The colour code used is the same as in Figure \ref{fig:7} and the systems whose best fit models require the pollutants to contain water ice are indicated in bold.

\begin{figure*}
	% To include a figure from a file named example.*
	% Allowable file formats are eps or ps if compiling using latex
	% or pdf, png, jpg if compiling using pdflatex
	\includegraphics[width=\textwidth]{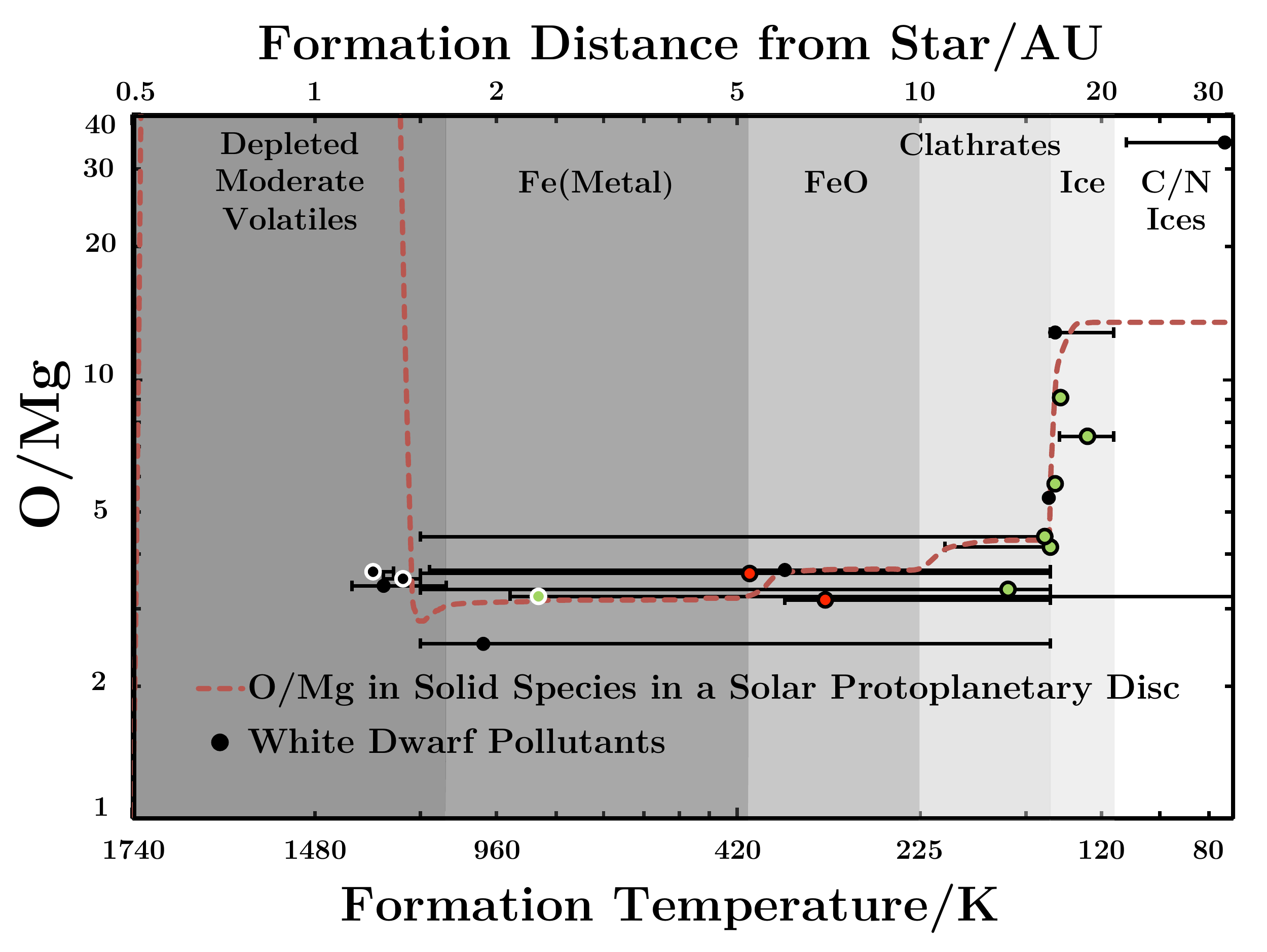}
    \caption{The spread in the predicted formation temperatures/locations of the pollutants versus the O/Mg ratio of the pollutants (points with a white outline have no oxygen abundance measurement therefore the value plotted is the one predicted by our models using the other elemental abundances). The error bars relate to the range of values which cannot be rejected to a statistical significance of 1$\sigma$ when considering all possible models and phases of accretion. The formation location on the upper x axis is derived assuming formation around a solar mass host star at t=0\,Myrs \protect\citep{Chambers2009}. }
    \label{fig:10}
\end{figure*}
\begin{figure*}
	% To include a figure from a file named example.*
	% Allowable file formats are eps or ps if compiling using latex
	% or pdf, png, jpg if compiling using pdflatex
	\includegraphics[width=\textwidth]{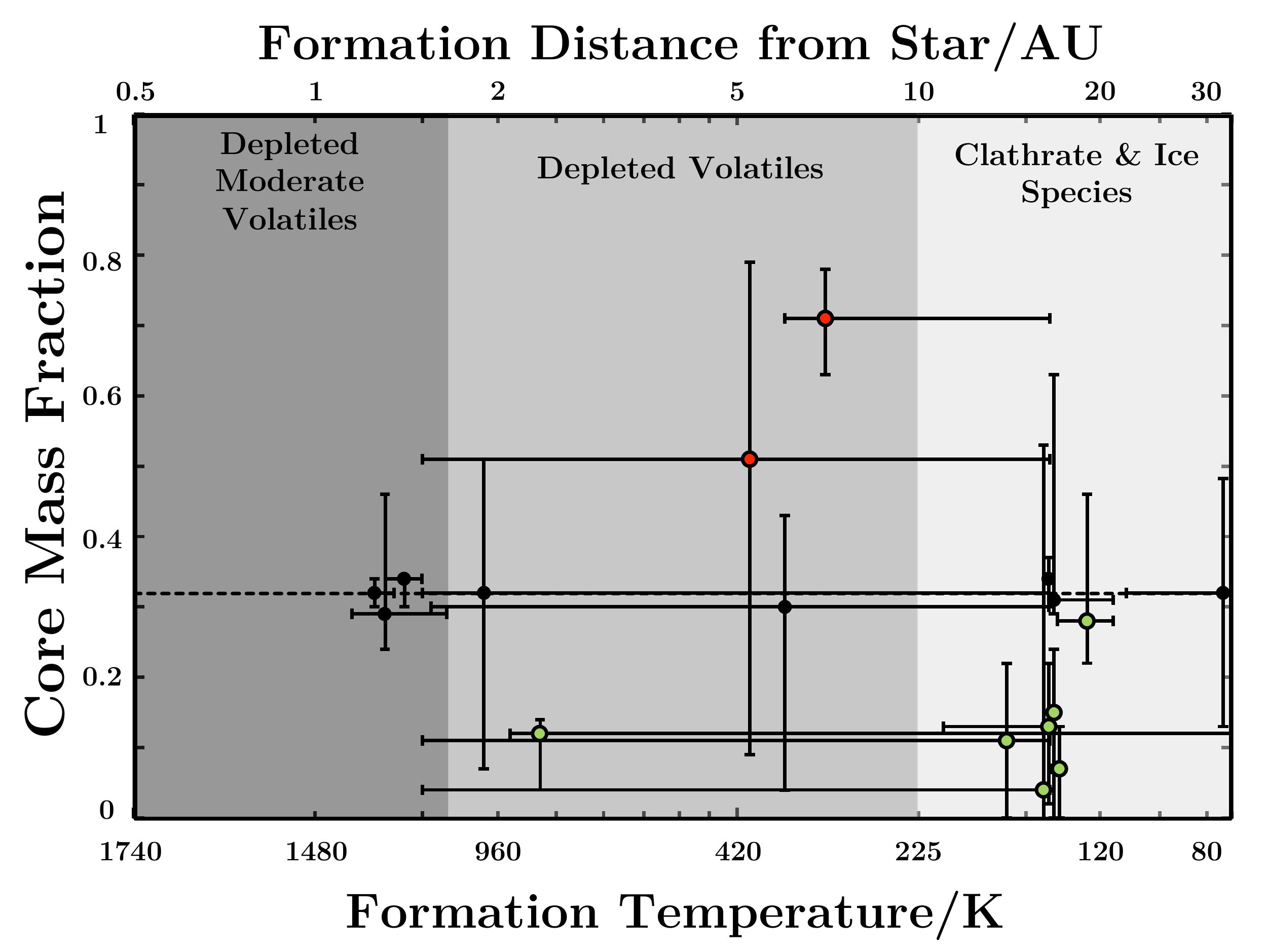}
    \caption{The position of the best fitting solution for the pollutant of each system in core mass fraction-formation temperature/location space. The positions of the black points have been moved away from their actual best fitting core mass fraction value, 0.32, in order to clearly display the formation temperature constraints of multiple systems on one plot. The error bars relate to the range of values which fit the observations such that the p-value of the fit cannot be rejected to a statistical significant of 1$\sigma$ when considering all possible models and phases of accretion. The formation location displayed on the upper axis is derived assuming formation around a solar mass host star at t=0\,Myrs \protect\citep{Chambers2009}. }
\label{fig:11}
\end{figure*}

Figure \ref{fig:10} shows the O/Mg steady state or pre-steady state abundance ratio (depending on which phase of accretion produces the best fit for the system) for each of the systems plotted against the formation distance (temperature) which produces the highest p-value from Figure \ref{fig:9}. The error bars correspond to the range of values which cannot be rejected with a statistical significance of 1$\sigma$ (p-value greater than 0.317) when considering all possible  phases of accretion and all possible models. The colour coding of the points is the same as in Figure \ref{fig:7}. The points with a white outline represent systems with no measured O abundance value, therefore the predicted O abundance from our best fit model is shown. The dashed red line represents the O/Mg number ratio of solid species which would be expected to condense out of a solar nebula at a time equals 0\,Myrs as a function of formation location. The shaded areas indicate the important chemical regions in the disc. There are three main regions. Firstly, the depleted moderate volatile region, where the temperatures are sufficient enough to partially vaporise many Ni, Mg, Si, Fe, and Cr species and therefore the solid species present can have strong enhancements in the refractories and depletions in the moderate volatiles. Secondly, the intermediate region (Fe(Metal), FeO), where the temperatures are such that the refractory and moderate volatile composition of the solids represent that of the host star, however the volatiles (Na,S,O,C,N) are still depleted. Thirdly, the volatile rich region (Clathrates, Ice, and C/N Ices), where clathrate species and water ice species start to condense out and help compose some of the solids formed, thus increasing the O to Mg ratio. At the coldest temperatures in the volatile rich region C ices and N ices condense out into solid form until the solid species have abundance ratios equivalent to that of the host star. We plot the horizontal axis as both temperature and distance from the host star, because the formation distance is heavily model dependent and is degenerate in formation time, therefore we expect the constraints on the formation temperature to be a much more robust prediction.

Figures \ref{fig:9} and \ref{fig:10} highlight that whilst our models are occasionally degenerate across a large range in formation temperature, they generally indicate which of the three regions the pollutant most likely formed in, thus, allowing constraints to be placed on the formation conditions of the white dwarf pollutants. The atmospheric abundances observed in five white dwarf systems require their pollutants to contain water ice to a statistical significance of 3$\sigma$ and the atmospheric abundances in three white dwarf systems require their pollutants to be severely depleted in volatiles to a statistical significance of 1$\sigma$. This suggests that there is a large range in the temperature conditions experienced by the pollutants, indicating that pollutants arrive in white dwarf atmospheres with a roughly equal efficiency from a wide range of radial locations. These results reinforce the conclusions reached previously in the literature which suggest that there is evidence for both volatile rich and volatile poor pollutants \citep{JuraYoung2014}, however, importantly our analysis allows quantitative constraints to be placed on the pollutants formation location.

Figure \ref{fig:11} shows the position of the optimal (highest p-value) solution for the pollutant of each system in core mass fraction-formation temperature space. The positions of the black points, which correspond to systems with primitive pollutants, have been moved away from their actual best fitting core mass fraction value, 0.32, in order to clearly display the formation temperature constraints of multiple systems on one plot. The error bars were found by allowing the core mass fraction and formation location to vary over the range displayed in the plot for each of the three accretion scenarios and finding the places where the p-value was greater than 0.317 and, thus, could not be rejected.

Figure \ref{fig:11} highlights that for the 17 systems analysed in this work there appears to be an overabundance of icy mantle-rich pollutants and a lack of icy core rich pollutants. A possible explanation for this result may be that water ice is sequestered in the upper mantles of planetary bodies as previously suggested in the literature \citep{JuraXu2010, Malamud2016}.

\subsection{Results Summary}

The key results presented in this paper are:

\begin{itemize}[leftmargin=*]
\setlength\itemsep{1em}
\item For 94 percent of the systems analysed in this work we can rule out pollutants with compositions similar to stellar material (model 1) with a confidence of at least 3$\sigma$. However, when we assume that the pollutants are similar to rocky planetary material (models 2 to 7) consistent fits which cannot be ruled out are produced for all 17 systems analysed. This reinforces the hypothesis that for the white dwarf systems analysed in this work it is much more probable that the pollutants are external rocky planetary bodies, rather than stellar-like material \citep{JuraYoung2014}.

\item The chemical abundances of the pollutants of 11 of the systems are accurately reproduced (p-values greater than 0.3173) by the accretion of volatile depleted material with compositions similar to nearby stars which require no post formation processing (models 2 and 3). However, six systems require differentiation, collisions, and fragmentation to have occurred (models 4, 5, 6, and 7) in order to fit their chemical abundances to a p-value greater than 0.3173. This reinforces the conclusions reached in the literature which suggest that polluted white dwarfs offer unique insights into the geology and collisional activity in exoplanetary systems \citep{JuraYoung2014}. 

\item Although the exact core mass fractions predicted here are dependent on the differentiation model used, the conclusions regarding whether the pollutant is mantle-rich, core-rich, or requires no collisional processing are robust. Collision models predict an approximate even spread in the core content of collision fragments centred around primitive \citep{Carter2015, Carter2017}. As we find a similar spread in the predicted core mass fractions of the pollutants studied here we suggest that differentiation and collisional processing is altering the abundances of rocky bodies in exoplanetary systems. The pollutant of NLTT\,43806 is the only one which requires crustal material (model 6) to reproduce its abundance pattern to a statistical significance of 1$\sigma$ and its abundance of core-like material is severely depleted, supporting the conclusion that the pollutant is equivalent to an exo-lithosphere. Higher than chondritic abundances of Ca can be mistaken for crustal material, however, we show in this work that hot formation temperatures and host star compositional variation can readily explain these abundances. Therefore, our models remove the need for pollutants that are simultaneously rich in core-like and crust-like material, an unlikely consequence of collisional evolution \cite{Carter2017}, to fit the observations.

\item Four of the six systems which require their pollutants to be fragments of differentiated bodies are best fitted by a model which allows asteroidal-like differentiation, modelled here to consider the production of planetary bodies with large crust to mantle mass ratios before collisions and fragmentation (model 7). This adds extra support to the conclusion that the pollutants are equivalent to smaller planetary bodies such as asteroids or fragments of minor planets \citep{JuraYoung2014}.

\item The Fe to Ca ratio is not a good proxy for the core mass fraction of a pollutant. The core mass fractions predicted by these abundances are often misleading due to the large uncertainties on the values, the lack of consideration of other observed abundances, and the fact that the abundances of Fe and Ca can be modified independently of changes in the core mass fraction. The ratio of Fe to Mg is a better proxy however the large uncertainties and lack of consideration of other elemental abundances still provide problems, therefore we suggest using a more complete model such as the one outlined in this work.

\item The white dwarf pollutants analysed in this work exhibit evidence for a wide range of formation temperatures, they can be classified into three categories: severely volatile depleted, volatile depleted, and volatile rich. The even spread of pollutants across these three categories indicates that pollutants arrive in white dwarf atmospheres with a roughly equal efficiency from a wide range of radial locations supporting the predictions of \cite{bonsor11}.

\item A strength of our model is that it allows quantitative constraints to be placed on whether the pollutant material contains water ice a much discussed topic in the literature. We find that five of the 17 white dwarf pollutants analysed in this work require water ice to a statistical significance of 3$\sigma$ reinforcing the idea that the pollutants are often volatile rich and these volatiles can survive to pollute the white dwarf.

\item Three of the 17 pollutants analysed are only well modelled if they formed in the severely volatile depleted and refractory enhanced region. This may suggest that the heating of planetary bodies on the giant branches is changing the pollutant compositions. This is because one would not expect bodies which formed so close to the host star to survive until the white dwarf phase due to the expansion of the host star. We instead suggest that the bodies could have formed further from the host star but as the luminosity increased on the giant branches the bodies heated up and the evaporation of some species caused a post formation volatile trend to develop. However, this hypothesis requires further study.

\item The lack of core-rich pollutants and the abundance of mantle-rich pollutants in the volatile rich region could support the conclusion that water ice is protected from vaporisation as it is sequestered in the upper mantles of planetary bodies \citep{JuraXu2010, Malamud2016}. These conclusions are tentative due to the small sample size however could be reinforced by future observations and analysis of white dwarf pollutants.

\item We find that constraints can be placed on the phase of accretion that a polluted white dwarf system is in by fitting our models to its observed elemental abundances. seven of the nine systems which we fitted our models to assuming they were in a pre-steady state phase produced a more consistent fit than when we assumed a steady state phase and for one of these systems we can rule out a steady state of accretion to a statistical significance of 3$\sigma$. For all 11 systems which we fitted our models to assuming a declining phase of accretion we can rule out to 3$\sigma$ a declining phase which has lasted longer than five sinking timescales. These constraints could be placed on future observations to offer insights into the lifetimes of white dwarf accretion discs and the mechanisms which govern white dwarf accretion.
\end{itemize}

\FloatBarrier 
\section{Discussion}\label{diss}
\subsection{Discussion of Caveats}
The aim of this work was to improve our understanding of the origin of white dwarf pollutants by modelling their expected abundances in a variety of formation scenarios for many elements simultaneously. In the initial set up described in this work many basic assumptions have been made and, therefore, our models are subject to many caveats. In this section we will discuss the notable caveats and how they affect the conclusions one can draw about the origin of white dwarf pollutants, noting that due to the large uncertainties on the observationally derived abundances, more complex models are not necessary.

In this work we have assumed that the pollution present in the white dwarfs atmospheres is the result of the accretion of one polluting body. We note that the pollution could arise from multiple bodies \citep{Wyattstochastic}; however we do not anticipate this would affect our conclusions, as the abundances are expected to be dominated by the largest body. As sinking timescales are short \citep{Koester09} in comparison to scattering timescales, it is most likely that multiple scattered bodies originated from similar locations in the system. If multiple bodies were accreted and the bodies did have different geological histories one would expect the signatures of differentiation to be washed out, and the pollutants to appear more primitive in nature, hence our conclusions regarding the requirement for some pollutants to be differentiated fragments are valid and in fact hint at abundances dominated by single bodies.

In our modelling we assume that the pollutants can only accrete onto the white dwarf in one of three phases; a pre-steady state phase, a steady state phase, and a declining phase. It is expected that these phases capture the way in which the atmospheric abundances could be related to the pollutant abundances \citep{Koester09}. However, if accretion modes were predicted which related the atmospheric abundances to the pollutant abundances in a different way to the accretion modes modelled here our model could easily be extended to investigate their plausibility.

We note that the errors calculated for the polluted white dwarf abundances may not take into account systematics and potential degeneracies. Possible dependencies between the observed abundances and their uncertainties of multiple elements could influence our conclusions. If the error bars were underestimated, one would expect an increase in the degeneracies between models and less systems would require non-primitive pollutants to accurately model their abundances. The tightness of our constraints on the formation location of bodies would also decrease. The uncertainties on the measurements are not expected to be overestimates, however if they were this would decrease the quality of the fit for many of our models.

Using as many elements as possible when constraining the formation location of the pollutant is useful as contradictions allow volatile dependent trends or Earth-like differentiation trends to be ruled out. Our method requires at least one lithophile, one siderophile, one volatile, and Mg to have observed abundances. Ideally each system analysed would have two lithophiles, two siderophiles, and two volatiles as well as Mg and Si as this would allow possible contradictions to the trends expected from severe volatile depletion and Earth-like differentiation. Therefore, in this work we have only analysed polluted white dwarfs which have at least 5 elemental abundance measurements. However, we note that the degeneracies present in the best fit models for many of the analysed pollutants could be reduced by further observations finding the abundances of more elements.

We note that it is possible that the stellar chemical compositions used could be dissimilar to the actual abundances of the white dwarf progenitors and the planet forming material around them. The main effect of this would be to predict differentiation and volatile depletion trends when they are not necessary. However as the stellar compositions appear to be spread around a mean star which is chemically similar to the Sun we suggest that finding a more realistic stellar chemical catalogue for the progenitors would not make a dramatic difference. The inclusion of metal poor stars could have a larger effect on the conclusions reached, however the ability of metal poor stars to produce planetary systems is currently poorly understood.

A major assumption in our work is that the pollutants are planetesimals whose abundances are dictated when they condense out of a protoplanetary disc in equilibrium. This model can recreate the bulk composition of all rocky bodies in the solar system to first order, and evidence from the Earth suggests a nebula origin to the volatile depletion trends observed and one which is due to condensation rather than volatilisation \citep{Palme2003}. The only major inconsistency is in the C value for the chondrites and this is mainly due to the incompleteness of the HSC chemistry database and a lack of understanding of the cosmochemistry of solid C species. This does not detract from the strong reproduction of the other elemental abundances. For these reasons we do not use the C values of the pollutants when fitting the observations to our models. Planetesimal formation is clearly significantly more complex than our model suggests. An incomplete understanding of how dust grains grow make it hard to predict the composition of planetesimals, however, the nucleation around pre-solar dust grains is often considered a major component in the formation of planetesimals. This process has not been modelled in our work. Another process which may be important and which we have not modelled is planetary migration and dust migration \citep{Desch2017b,Desch2017a}. Although this almost certainly occurs in all exoplanetary systems, we do expect that any trends present in the planetary bodies composition due to the formation conditions will be preserved and will offer insights into the rough formation location, at least to first order.

Our model can robustly predict the relative distances at which the white dwarf pollutants most likely formed. However, we do not consider the absolute values to be robust. This is because of the simplistic viscous irradiated protoplanetary disc model used, the requirement of a single formation time, the absence of migration, and the assumption of a one solar mass host star. Also because the volatile depletion signatures could have developed post-formation due to giant branch heating the absolute values of formation location predicted here could be incorrect, however, the relative distance would again remain robust. 

The final major caveats of our work involve the differentiation model used. As a first approximation we considered differentiation into a crust, mantle, and core with Earth-like enhancements in each layer relative to the bulk composition of the body. The bodies which differentiated and fragmented to form the white dwarf pollutants most likely did not differentiate in an Earth-like manner. They are most probably much smaller bodies and, therefore, the pressures and temperatures at which core formation occurred would have been drastically different. It is also not known whether all crusts and cores have an Earth-like composition or whether the geology of the solar system is universal. However, we can only assume, and model, the geology we observe in the solar system, and therefore we expect the general trend in siderophiles and lithophiles to be the same regardless of the size of the parent body. As we model the majority of pollutants analysed thus far accurately we can conclude that there is currently no evidence from polluted white dwarf atmospheric compositions for extra solar geology that is dissimilar to the geology seen in the solar system. We note that even though the exact core mass fraction predicted by our model will most likely not be exact, the overall trend (core-rich or mantle-rich) will be a robust result. 

It may be possible that the composition of the rocky planetary bodies in exo-systems are determined by more than the three processes outlined in this work (initial composition of the planetesimal forming disc, heating during formation or subsequent evolution, and differentiation, collisions, and fragmentation). However, as our model is currently consistent with the observations analysed thus far, we expect that any other mechanism does not alter the composition to first order.

This work could be easily repeated for another stellar catalogue, differentiation model, or condensation series if it was found that the abundances expected were not sufficiently reproduced by any of the chosen models.

\subsection{Discussion of Results} \label{dissres}

Our analysis suggests that the majority of white dwarf pollutants are depleted in at least one volatile species, and many are heavily depleted, supporting the hypothesis that the pollutants are external rocky planetary bodies rather than stellar material. This is a robust conclusion as we find that for all but one system a model involving the accretion of stellar material produces a fit with a p-value which we can rule out to at least 3$\sigma$. If atmospheric H abundances in DBZ white dwarfs were taken to be upper limits for pollutant H abundances, and the pollutant H abundances were included in our modelling, we could then rule out a model involving the accretion of stellar material for all 17 systems. This adds further support to the literature preferred model of external white dwarf pollution via accretion of rocky planetary material \citep{JuraYoung2014}.

We find that the pollutants of 11 of the 17 systems analysed can be explained by the accretion of material that matches the composition of nearby stars in combination with a condensation temperature dependent trend, without any need for any further processing. Our model includes the key addition of considering potential variability in the composition of the initial planet-forming discs. This inclusion reduces the number of white dwarfs which require differentiation and fragmentation to accurately fit their observed abundances. Primitive pollutants are not unexpected given that not all rocky bodies will differentiate and that not all collisions between differentiated bodies will produce fragments with a notable excess of core-like or mantle-like material.

The white dwarf observations provide evidence that the differentiation and collisional fragmentation of exoplanetary bodies is common. Nine of the 17 pollutants are best fitted using models which involve differentiation, collisions, and fragmentation, however only six of these pollutants require differentiation, collisions, and fragmentation to achieve a fit which cannot be ruled out to a statistical significance of 1$\sigma$. The requirement for exoplanetary differentiation has been discussed previously in the literature. Our work finds that the abundance patterns present in some of the pollutants cannot be well modelled without the inclusion of differentiation, collisions, and fragmentation, however, it is also important to note that we have shown that it is not necessary in most polluted white dwarf systems.

The expected core mass fractions of the white dwarf pollutants analysed in this work are, given the small sample size, roughly evenly distributed between mantle-rich and core-rich. This conclusion can be well explained by the collisional processes expected to occur in a planetesimal belt. \cite{Carter2015} modelled collisions between differentiated bodies during planet formation, and although core-like and mantle-like fragments experienced different "fates" in collisions, the final planetesimal distribution had a broad and approximately even spread in core mass fractions. We also find that planetesimals which are rich in both crustal material and core material are not necessary to explain the observations. In the literature pollutants with enhancements in both siderophiles and refractory lithophiles have often been cited as being simultaneously core-rich and crust-rich, and thus, mantle depleted. This state is difficult to reproduce in collisional models \cite{Carter2017}. Our results find that these systems can all be explained either using strong volatile depletion trends or by variance in the material that comprises the planetesimal forming disc, which we expect are more realistic histories.

Interestingly we have shown that the requirement for differentiation and the core mass fractions expected from simply utilising the Fe to Ca ratio relative to bulk Earth do not provide a good proxy for the results we found. This is because many pollutants have Ca enhancements or depletions, due to the formation temperatures experienced or the initial composition of the planet-forming disc, which are independent of the core mass fraction. If only a few elemental abundances are available, the Fe to Mg ratio provides a better proxy to the expected core mass fraction, as the Mg abundance is not enhanced in the crust, nor is it enhanced by the temperature of formation, and the uncertainties on the Mg abundance are generally less than those on the abundances of Ca. However, the lack of analysis involving the large uncertainties and the abundances of multiple elements means that this proxy can still often lead to different results. The results of our model, which includes many elemental abundances, highlight the need to observe and analyse as many elements present in the atmospheres of white dwarfs as possible if one would like to constrain their formation history.

The only pollutant which requires crustal material to reproduce its atmospheric abundance pattern to a statistical significance of 1$\sigma$ is NLTT\,43806.  Its abundance of core-like material is severely depleted, supporting the conclusion that the pollutant is equivalent to an exo-lithosphere \citep{Zuckerman2011}. One out of 17 pollutants being enhanced in crustal material is not an unreasonable occurrence rate, as collisional models show crustal fragments are rare \citep{Carter2017}. One other system, SDSS J1228+1040, has an optimised p-value when a large amount of crustal material is present. However, unlike for NLTT\,43806, we can also produce fits with p-values greater than 0.317 using other models. The best fit produced by model 6 suggests that the pollutant is simultaneously core-rich and crust-rich, due to the inability of collisional models to produce simultaneously crust-rich and core-rich bodies we suggest that model 2 is a more realistic solution to the formation history of this pollutant.

Four out of the six systems which require their pollutants to be fragments of differentiated bodies are best fitted by a model which involves the production of planetary bodies with large crust to mantle mass ratios before collisions and fragmentation. The required crust to mantle mass ratios are much larger than that of a terrestrial planet but are similar to those expected for a large asteroid or minor planet \citep{Ceres2005, Vesta2012}. This supports the conclusion that the pollutants are small rocky planetary bodies as predicted by the frequency of pollution, and the low convective zone masses \citep{Jura2006, JuraYoung2014}. Unless a mechanism for refractory lithophile depletion other than crustal stripping during collisions can be found, this offers a robust conclusion on the nature of these bodies.

The range in formation temperatures from ~1400K to ~80K, and their apparent even spread, suggest that pollutants can be scattered into the tidal radius from many different radial locations. This is possibly as expected, because scattering simulations predict that although the scattering efficiency of planetesimals decreases with increasing distance from the white dwarf, the planetesimal belts expected to occupy the outer regions of the system are expected to be less collisionally evolved and therefore retain more mass. Therefore, pollutants may be expected to have an equal probability of accreting onto the white dwarf, regardless of formation location \citep{bonsor11}. Even though the formation distances predicted in this work are heavily model dependent, the necessity for a spread is formation temperatures is a robust conclusion, as regardless of the model chosen, the volatile depletion trends of the elements remain roughly the same. 

The scattering of bodies from outside the ice lines hints at the requirement for planetary systems to be rich in planets. This is because the required scattering mechanisms usually involve "full" planetary systems \citep{bonsor_tiss, WyattMarino2017}. The possible presence of water ice in pollutants has been extensively studied in the literature \citep{Farihi2011, Gansicke2012, Dufour2012, Raddi2015, Dufour2016, Xu2017}. Our results match those predicted using a simple O excess argument however the strength of our method is that we can estimate formation distances rather than simply whether the body formed inside or outside of the water ice line as well as finding the statistical significance to which we require water ice. We can also take into account multiple elements simultaneously and this is important as it allows conclusions to be derived regardless of the large uncertainties on the abundances.

Five of the eight systems predicted to have formed outside the water ice line are also predicted to be mantle-rich. This, and the lack of core-rich bodies so far predicted to have formed in this region, may suggest that water ice does indeed become sequestered in the upper layers of the mantle \citep{JuraXu2010, Malamud2016}. This hypothesis is frequently cited when attempting to explain how ice could reach the atmosphere of a white dwarf without being sublimated before the pollution process.

Our model suggests that three systems are best reproduced by pollutant bodies which have undergone severe volatile depletion. It was not expected that pollutants of white dwarfs would show signatures of formation close to the host star. This is because if a body was located within a few AU of the star it would be engulfed on the giant branch and therefore would be unable to pollute the white dwarf \citep{schrodersmith08, Veras_review}. Due to the uncertainty on the formation time and the nature of the host star it is possible that these signatures were produced during formation and that the bodies formed sufficiently far from the star to avoid destruction, however we theorise that this signature is most likely to be produced when a body is heated on the giant branch and its outer layers are partially vaporised. This gives the upper layers a steep condensation temperature dependence. This hypothesis requires further testing but on first inspection seems to explain the unexpected frequency of these signatures. Abundances of the volatile species and measurements of the abundances in more polluted white dwarf systems will help to analyse the validity of this conclusion.

The atmospheric O/Mg ratio offers useful insights into the formation location of the pollutant body, however, due to the nature in which the O/Mg ratio varies with temperature and the large uncertainties on the measurements many systems still have poorly constrained pollutant formation locations. Inclusion of S, P, Mn, and other volatile species could constrain the formation location of the pollutants to a much narrower temperature range, these elements were not included in the current model due to a lack of observed stellar abundances. If Na abundances were found for the 13 systems which currently have no measurement we could drastically improve the uncertainties on the formation locations predicted by our model. Our estimates could be used in the future to constrain dynamical models which predict the frequency of pollutants expected from certain scattering locations.

Using our models and the elemental abundances present in the atmospheres of the white dwarf we can place constraints on the phase of accretion that a system is in. Of the 11 white dwarfs with the longest sinking timescales seven produce more consistent fits to our models when we assume they are in a pre-steady state phase of accretion, and for one of these systems we can rule out a steady state of accretion to a statistical significance of 3$\sigma$. We can also rule out being in a declining phase which has lasted for longer than three sinking timescales for eight of the 11 systems and a declining phase which has lasted for longer than five sinking timescales for all 11 systems, as none of our models can produces p-values greater than 0.003 for these systems in these cases. However, for GD\,61 we find that the pollution could be fitted to a p-value of 0.90 if the system is now in a declining phase and had previously accreted a primitive planetesimal. These results are not unexpected, one would imagine that due to the long sinking timescales DBZ white dwarfs are likely to be found in any of the three phases of accretion. Our models could be used in the future to place constraints on the accretion phase of white dwarfs and thus offer insights into the disc lifetimes in these systems along with the mechanisms which govern white dwarf accretion.

\FloatBarrier

\section{Conclusions}
In this work we present a method for determining the most probable formation history of the rocky planetary material that pollutes the atmospheres of some white dwarfs. Our method attempts to match the abundance patterns observed in the externally polluted white dwarfs' atmospheres to the bulk chemical abundances expected from the accretion of planetesimals which could have formed in protoplanetary discs with a range of initial compositions, at various locations, and with various geological and collisional histories. 

The abundance patterns present in the atmospheres of externally polluted white dwarfs are found to be inconsistent with the accretion of stellar material, as previously suggested \citep{JuraYoung2014}. Our models show that all 17 of the polluted white dwarf systems analysed in this work can be well fitted (p-values greater than 0.317) by a scenario in which they have accreted rocky planetary material.

The best fit model for nine of the 17 systems analysed required them to have accreted fragments of differentiated bodies. For one of these systems this is the only explanation to a statistical significance of 3$\sigma$. Crust-rich fragments are not generally required to accurately fit most abundance patterns; in fact only in the system NLTT\,43806 does it produce a significantly better fit. Our model finds an approximately even spread in the core mass fraction of the pollutant material accreted by the white dwarfs, ranging from a depletion in core-like material to an excess in core-like material. A lack of crustal material and an even spread in core mass fraction is an expected outcome of the collisional processing of planetesimals \citep{Carter2015, Carter2017}. When the core mass fractions are predicted by the Fe to Ca ratio they do not match what is expected from collisional models and are often misleading. This is because the abundances of Fe and Ca have large uncertainties and can be altered independently of the core mass fraction.

The compositions of five of the nine pollutants which are optimally modelled by scenarios in the white dwarf has accreted a fragment of a differentiated body require the parent body to have had a low mantle mass fraction in comparison with bulk Earth, suggesting that the bodies which are accreting onto white dwarfs are often asteroidal in size. This has previously been concluded in the literature due to the frequency of pollution, the low convective zone masses, and the C abundances present in the polluted white dwarfs \citep{Jura2006, JuraYoung2014}.

Our analysis suggests that white dwarf pollutants originate from a wide range of formation locations, suggesting that the scattering of planetary bodies into a white dwarf's tidal radius is equally efficient at all semi-major axes. The optimal solutions for eight of the pollutants include water ice in their compositions, and five of these systems require the presence of water ice in their pollutants to a statistical significance of 3$\sigma$. Three of the pollutants are best matched by severely volatile depleted planetesimals; this may suggest that giant branch heating is vaporising the outer layers of bodies before the pollution process takes place. 

The lack of core-rich bodies predicted to have formed in the volatile rich zone, and the abundance of mantle-rich planetesimals in this region, may suggest that water ice is sequestered in the upper mantle of planetary bodies, and, therefore, can avoid sublimation until it is accreted onto the white dwarf as suggested by \cite{JuraXu2010} and \cite{Malamud2016}.

The chemical abundances measured in the atmospheres of polluted white dwarfs can allow constraints to be placed on the phase of accretion that the system is in. It is often assumed that pollutants are accreting onto the white dwarf in a steady state phase. However, we find evidence seven systems are accreting in a pre-steady state phase, while one system is possibly in a declining phase.

\section*{Acknowledgements}

We would like to thank Rik van Lieshout, Ted von Hippel, Mark Wyatt, Mihkel Kama, Laura Rogers, Oliver Shorttle, Ed Gillen, and Matthew Auger for useful and insightful discussions. We are grateful to the Science \& Technology Facilities Council, and the Royal Society - Dorothy Hodgkin Fellowship for funding the authors of this paper.

%%%%%%%%%%%%%%%%%%%%%%%%%%%%%%%%%%%%%%%%%%%%%%%%%%

%%%%%%%%%%%%%%%%%%%% REFERENCES %%%%%%%%%%%%%%%%%%

% The best way to enter references is to use BibTeX:

\bibliographystyle{mnras}
\bibliography{/Users/AislinnMcdonagh/Downloads/JHref}

%%%%%%%%%%%%%%%%%%%%%%%%%%%%%%%%%%%%%%%%%%%%%%%%%%

%%%%%%%%%%%%%%%%% APPENDICES %%%%%%%%%%%%%%%%%%%%%
\FloatBarrier
\appendix

\section{FGK Stellar Abundance Data} \label{AppA}

\FloatBarrier
\begin{figure*}
\centering
\includegraphics[width=0.95\textwidth]{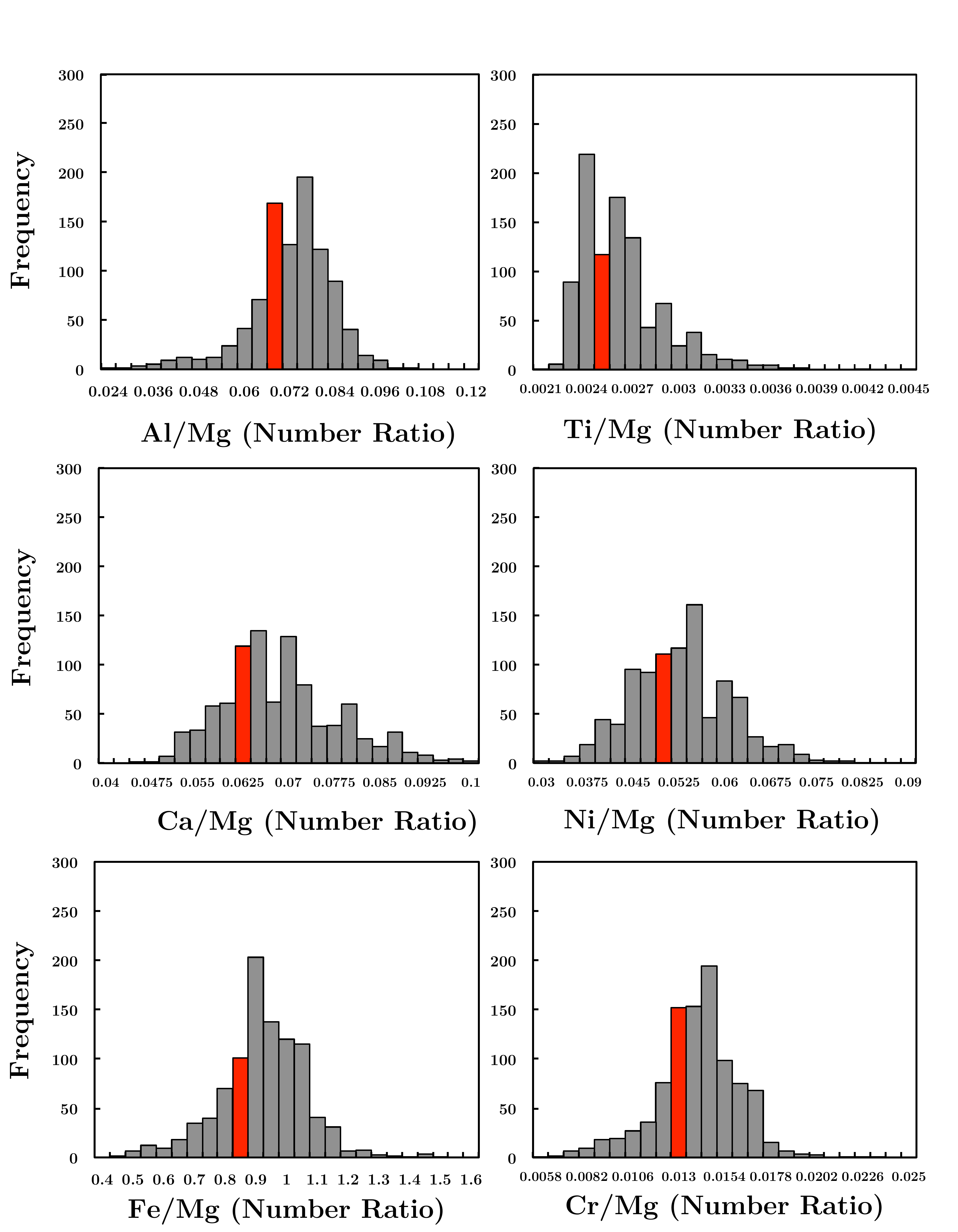}
    \label{fig:B1}
\caption{A group of histograms displaying the data obtained by \protect\cite{FischerBrewer2016}. The red coloured bars indicate the abundances present in the solar photosphere. The compositions of nearby stars are on average similar to the composition of the sun, however, individual elemental abundance ratios can vary from as low as half solar to as high as twice solar.}
\end{figure*}
\FloatBarrier

\FloatBarrier
\begin{figure*}
\centering
\includegraphics[width=0.95\textwidth]{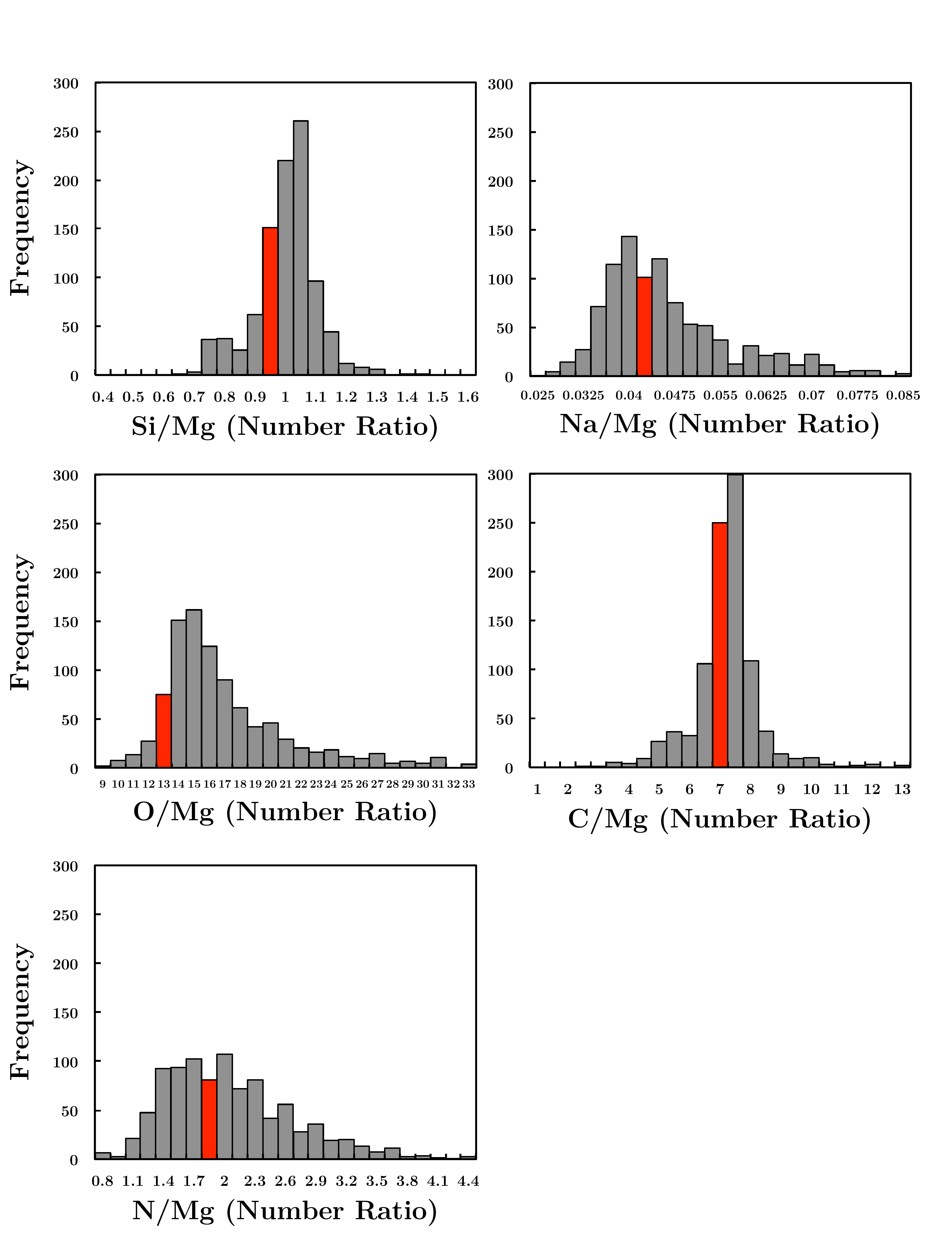}
    \label{fig:B2}
\caption{A group of histograms displaying the data obtained by \protect\cite{FischerBrewer2016}. The red coloured bars indicate the abundances present in the solar photosphere. The compositions of nearby stars are on average similar to the composition of the sun, however, individual elemental abundance ratios can vary from as low as half solar to as high as twice solar.}
\end{figure*}
\FloatBarrier

\section{Viscous Irradiated Protoplanetary Disc Model} \label{disc}

The pressure-temperature space in which to perform the equilibrium chemistry calculations was determined by using the theoretical model derived in \cite{Chambers2009}, which models the pressure-temperature space in an evolving, viscous, irradiated disc. This model has been previously used in the literature for the modelling of planetesimal formation in protoplanetary discs \citep{Moriarty2014}. The Chambers model is a disc model with an alpha parameterisation which divides the disc into three sections; an inner viscous evaporating region, an intermediate viscous region, and an outer irradiated region. For all calculations in this work we used $M_{0} = 0.1M_{\odot}$ , $s_{0} = 33\,AU$ , $R_{*} = 3R_{\odot}$ , $T_{*} = 4200\,K$ , $\kappa_{0} = 0.3\,m^{2}kg^{-1}$ , $\alpha = 0.01$ , $\gamma = 1.7$ , $\mu = 2.4$ , and $M_{*} = M_{\odot}$ . Using these values corresponds to the second example in \cite{Stepinski1998}, which is consistent with a planetesimal forming disc around a solar mass star. A possible improvement to our model, which would allow the absolute distances predicted to become more robust, would be to modify these input parameters to the values expected for the white dwarf progenitors.
\par
\noindent \textbf{The Inner Viscous Evaporating Region}
\par
\noindent The inner viscous evaporating region has a surface density given by 
\begin{equation}
\Sigma(r,t) = \Sigma_{evap} \left(\frac{r}{s_{0}}\right)^{-\frac{24}{19}}  \left( 1 + \frac{t}{\tau_{vis}}\right)^{-\frac{17}{16}}
\end{equation}
where 
\begin{equation}
\Sigma_{evap} = \Sigma_{vis} \left(\frac{T_{vis}}{T_{e}}\right)^{\frac{14}{19}}
\end{equation}
and $ T_{e} = 1380\,K$.
The opacity in the inner viscous evaporating region follows the power law described in \cite{Stepinski1998}.
\par
\noindent The temperature in the viscous evaporating inner region is given by
\begin{equation}
T(r,t) = T_{vis}^{\frac{5}{19}} T_{e}^{\frac{14}{19}}  \left(\frac{r}{s_{0}}\right)^{-\frac{9}{38}}  \left( 1 + \frac{t}{\tau_{vis}}\right)^{-\frac{1}{8}}
\end{equation}
and the transition radius to the intermediate viscous region is
\begin{equation}
r_{e} (t) = s_{0} \left(\frac{\Sigma_{evap}}{\Sigma_{vis}}\right)^{\frac{95}{63}} \left( 1 + \frac{t}{\tau_{vis}}\right)^{-\frac{19}{36}}
\end{equation}
\textbf{The Intermediate Viscous Region}
\par
\noindent The surface density in the intermediate viscous region is
\begin{equation}
\Sigma(r,t) = \Sigma_{vis} \left(\frac{r}{s_{0}}\right)^{-\frac{3}{5}}  \left( 1 + \frac{t}{\tau_{vis}}\right)^{-\frac{57}{80}}
\end{equation}
where
\begin{equation}
\Sigma_{vis} = \frac{7M_{0}}{10 \pi s_{0}^{2}}
\end{equation}
and the temperature in the intermediate viscous region is
\begin{equation}
T(r,t) = T_{vis} \left(\frac{r}{s_{0}}\right)^{-\frac{9}{10}}  \left( 1 + \frac{t}{\tau_{vis}}\right)^{-\frac{19}{40}}
\end{equation}
where 
\begin{equation}
T_{vis} = \left(\frac{27 \kappa_{0}}{64 \sigma}\right)^{\frac{1}{3}}  \left(\frac{ \alpha \gamma k }{\mu m_{H}}\right)^{\frac{1}{3}}   \left(\frac{7 M_{0}}{10 \pi  s_{0}^{2}}\right)^{\frac{2}{3}}  \left(\frac{G M_{*} }{s_{0}^{3}}\right)^{\frac{1}{6}} 
\end{equation}
and
\begin{equation}
\tau_{vis} = \frac{1}{16 \pi} \frac{\mu m_H \Omega_{0} M_{0}}{\alpha \gamma k \Sigma_{vis} T_{vis}}
\end{equation}
and the transition radius between the intermediate viscous region and the outer irradiated region is
\begin{equation}
r_{t} (t) = s_{0} \left(\frac{\Sigma_{rad}}{\Sigma_{vis}}\right)^{\frac{70}{33}} \left( 1 + \frac{t}{\tau_{vis}}\right)^{-\frac{133}{132}}
\end{equation}
\textbf{The Outer Irradiated Region}
\par
\begin{figure}
	\includegraphics[width=\columnwidth]{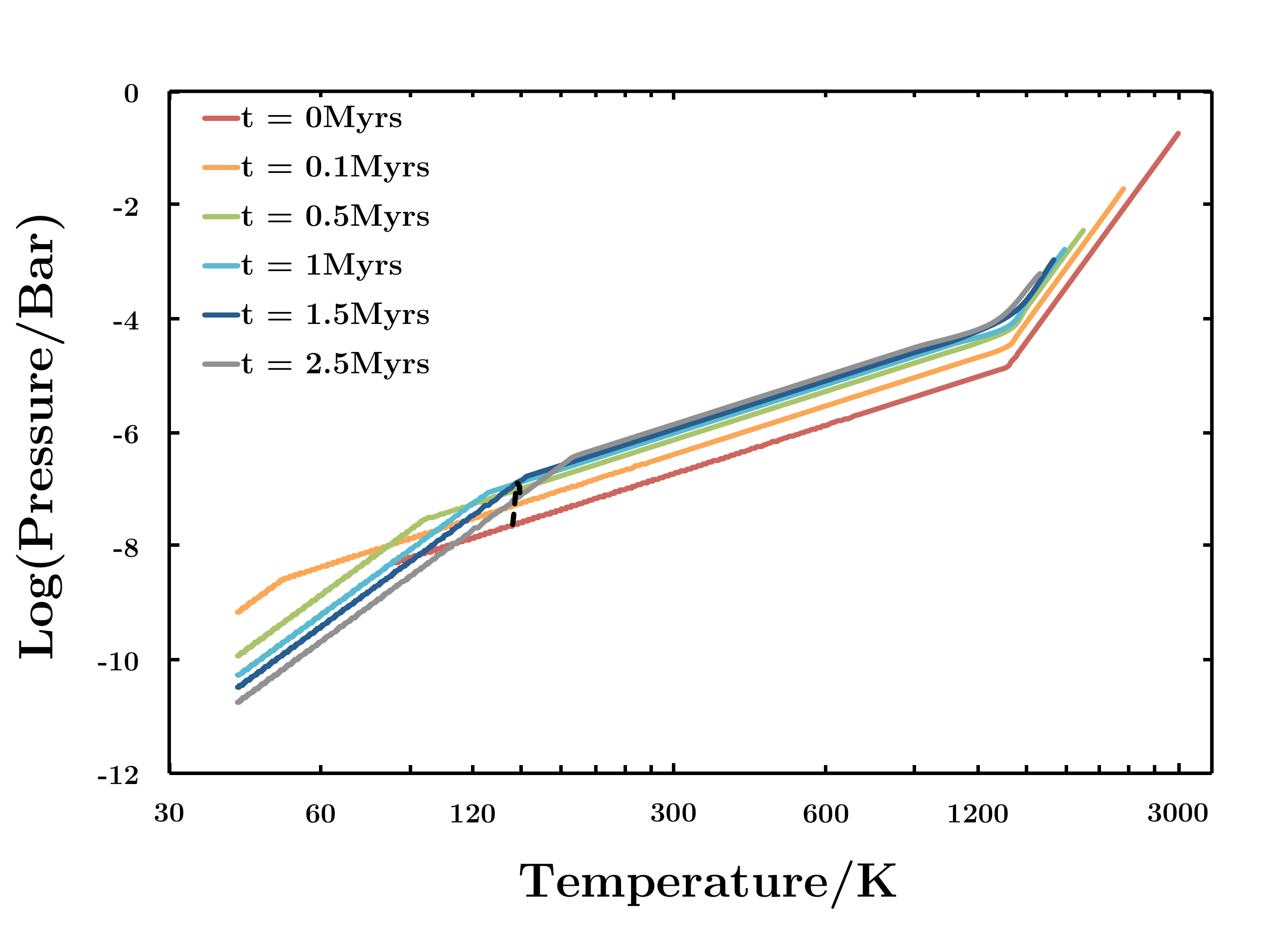}
\caption{The Pressure-Temperature space mapped out by the \protect\cite{Chambers2009} protoplanetary disc model as a function of radius and time. The black dashed line indicates the position of the water ice line as a function of time.}
    \label{fig:D3}
\end{figure}

\noindent The surface density in the outer irradiated region is
\begin{equation}
\Sigma(r,t) = \Sigma_{rad} \left(\frac{r}{s_{0}}\right)^{-\frac{15}{14}}  \left( 1 + \frac{t}{\tau_{vis}}\right)^{-\frac{19}{16}}
\end{equation}
where 
\begin{equation}
\Sigma_{rad} = \Sigma_{vis} \frac{T_{vis}}{T_{rad}}
\end{equation}
and
\begin{equation}
T_{rad} = \left(\frac{4}{7}\right)^{\frac{1}{4}}  \left(\frac{ T_{*} R_{*} k }{G M_{*} \mu m_{H}}\right)^{\frac{1}{7}}  \left(\frac{R_{*}}{s_{0}}\right)^{\frac{3}{7}}  T_{*}
\end{equation}
and the temperature in the outer irradiated region is
\begin{equation}
T(r,t) = T_{rad} \left(\frac{r}{s_{0}}\right)^{-\frac{3}{7}}  
\end{equation}
To convert the surface density profile into a pressure profile we have assumed the disc is an ideal gas with a Gaussian density profile.  The surface density is converted into a pressure as follows:
as
\begin{equation}
P = \frac{k \rho T}{\mu m_{H}}
\end{equation}
and
\begin{equation}
\int \rho \,dz = \Sigma 
\end{equation}
and we assume that
\begin{equation}
\rho = \rho_{0} e^{-\frac{z^{2}}{2H^{2}}}
\end{equation}
we therefore find that
\begin{equation}
\Sigma = \rho_{0} \sqrt{2H^{2} \pi }
\end{equation}
hence the pressure at the midplane is
\begin{equation}
P = \frac{k \Sigma T}{\mu m_{H} H \sqrt{2 \pi} }
\end{equation}
Using the standard formulae
\begin{equation}
c^{2}_{s} = \frac{kT}{\mu m_{H}}
\end{equation}
 \begin{equation}
H = \frac{c_{s}}{\Omega}
\end{equation}
\begin{equation}
\Omega = \sqrt{\frac{G M_{*}}{r^{3}}}
\end{equation}
we find that the relationship between the pressure profile and the surface density profile is
\begin{equation}
P = \sqrt{\frac{G M_{*}  k\Sigma^{2}T}{2 \pi \mu m_{H} r^{3}}}
\end{equation}
The Pressure-Temperature space mapped out by this evolving disc is displayed in Figure \ref{fig:D3}.
\FloatBarrier

\begin{figure}
\centering
	\includegraphics[width=\columnwidth]{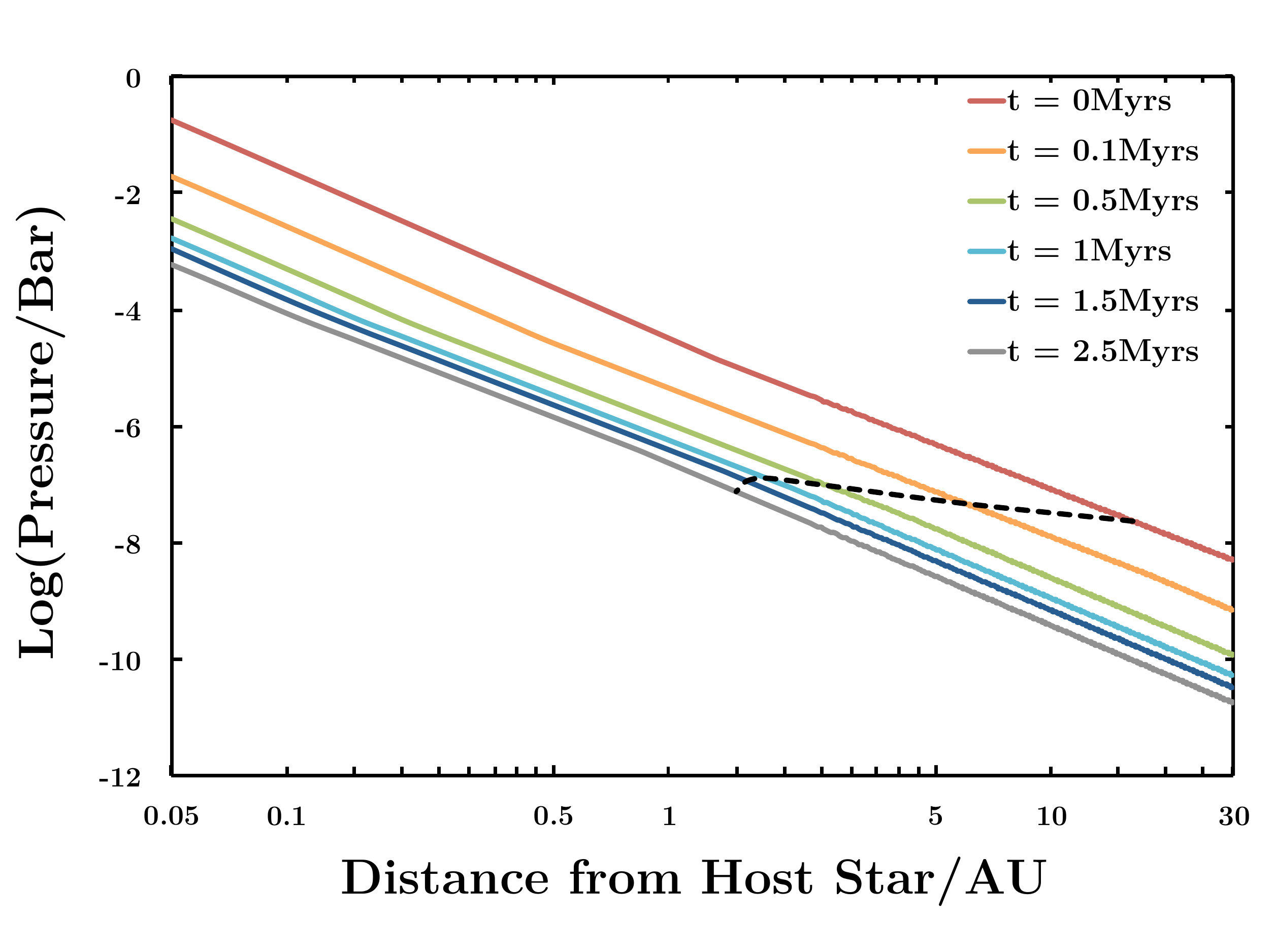}
\end{figure}
\begin{figure}
\centering
	\includegraphics[width=\columnwidth]{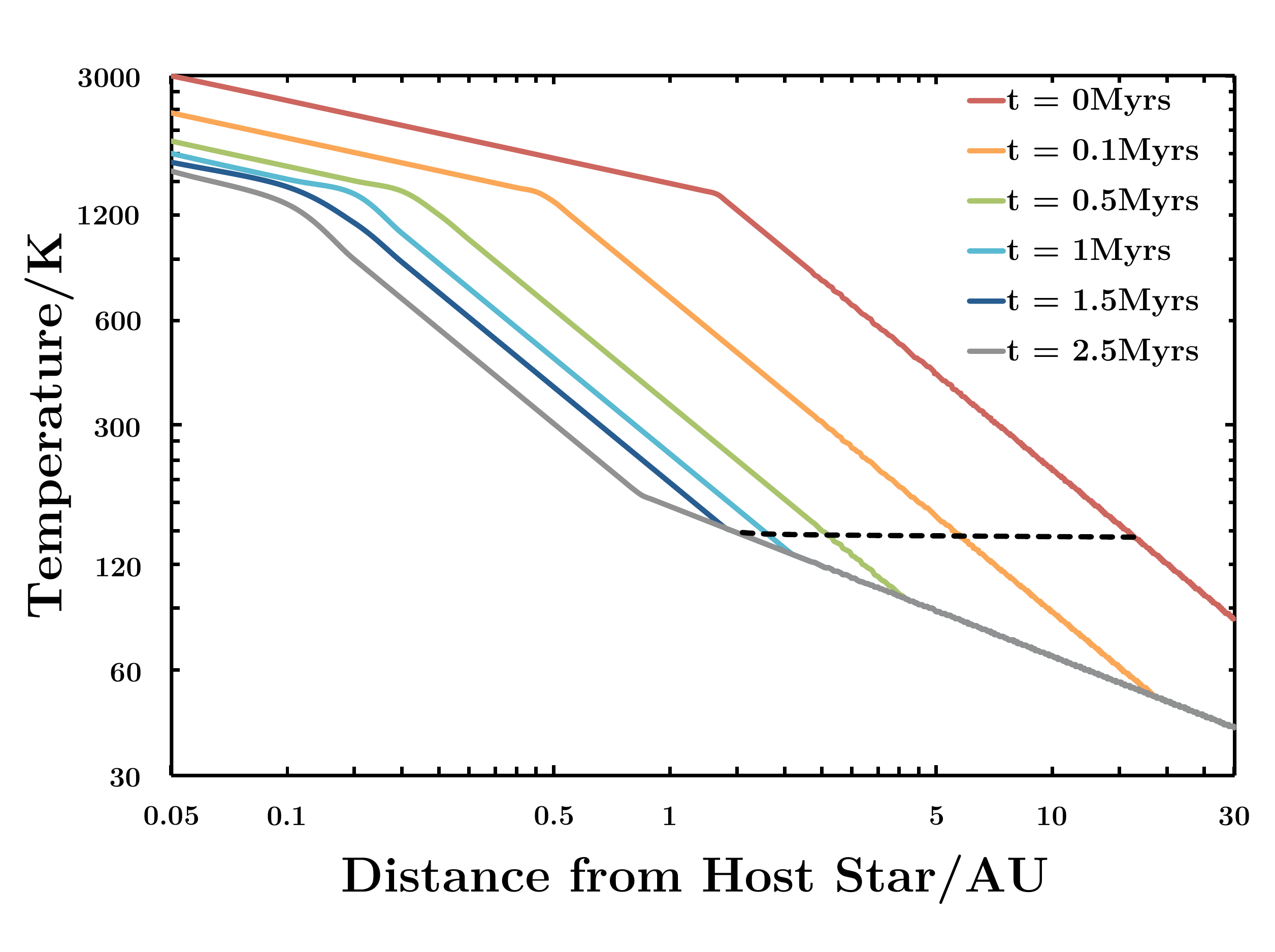} 
\end{figure}

\section[Equilibrium Chemistry Planetesimal Condensation Model]{Equilibrium Chemistry Planetesimal Condensation Model} \label{gibbs}

To recreate the expected abundance ratios present in extrasolar primitive planetesimals we employ a Gibbs free energy minimisation model at the pressures and temperatures expected to be present in a protoplanetary disc (as derived in Appendix \ref{disc}). We assume that the composition of the solid bodies that condensed out of the disc, at various times and at various locations according to the equilibrium chemistry model, have abundances similar to extrasolar primitive planetesimals. 

Our method assumes that the white dwarf pollutants were solid bodies, and is equivalent to a scenario in which solid planetesimals condense out of the nebula gas during the protoplanetary disc phase, then, after the star has evolved past the main sequence, are scattered into the white dwarf's tidal radius and pollute the white dwarf's atmosphere. Other processes involved in the formation of planetesimals during grain growth, like migration, are ignored as it is not expected that they will drastically change the bulk chemistry of the planetesimal. The caveats of this assumption are discussed in Section \ref{diss}. Our assumptions are validated by the fact that our method can reproduce the bulk composition of the terrestrial planets in the solar system and the abundances of solar system meteorites and comets.

The equilibrium chemistry model used to find the composition of the material that condenses out of a protoplanetary disc was HSC chemistry version 8. It was set up in the same way as \cite{Moriarty2014} and \cite{Bond2010a}, which both used the software to model planetesimal compositions. The gaseous elements inputted, the list of gaseous species included in the model, the list of solid species included in the model, and the initial gaseous abundances are displayed in Table \ref{tab:3}, Table \ref{tab:4}, Table \ref{tab:5}, and Table \ref{tab:6} respectively. We then altered the gaseous input abundances from the values presented in Table \ref{tab:6} to the values of the stars referenced in Section \ref{HSCV}. This allowed us to estimate the expected abundance ratios present in the primitive extra solar planetesimals one might expect to find in the polluted white dwarf systems. 

The species included are the same as those considered in \cite{Bond2010a}. All solids are in pure form and no solid solutions were included. The elements chosen for this study are expected to be the most abundant elements in the galaxy. Our list also contains the 14 most abundant elements in the rocky debris in the solar system. We assume that these elements will be the most important when forming extrasolar rocky bodies. The list of compounds included was selected by \cite{Bond2010a} after the included species were found to be the most commonly occurring and important over the pressure-temperature space expected in protoplanetary discs. The list is limited by HSC chemistry's database. The only major species missing, that would possibly alter the results, are the complex carbon macromolecules which are found in many asteroids and meteorites \citep{pizza2006} and many of the most common ice species: \ce{NH3}, \ce{N2}, \ce{CO}, \ce{CO2}, \ce{CH3OH}, \ce{CH4}, and \ce{H2S}. However, as outlined in \cite{Marboeuf2014} water ice condenses out of the nebula at much higher temperatures than any of the aforementioned ices (even if clathrate species of those ices are included). This suggests our model should predict the elemental abundances of planetesimals accurately, unless they formed in a region further from the star than the water ice line when the predicted abundances of S, N, and C may not be accurate. The formation mechanism of carbonaceous matter in asteroids, especially complex macromolecules, is not yet understood. Organic material identified in meteorites suggest they were formed in radiation shielded environments and in the presence of liquid water \citep{GlavinDworkin2009}. Taking this into account we must be careful when predicting abundances of solid state C as it may be present in solid species when our model suggests it would not be. We, therefore, do not use the abundance of C when fitting our modelled chemical abundances to the data. However, in our method we use the \cite{Marboeuf2014} model to predict the locations/temperatures when one would expect all C and all N to be in the solid phase and these locations/temperatures can be used to find a limit on formation conditions.  

\begin{table}
	\centering
	\caption{The gaseous elements which were included in the equilibrium chemistry code, HSC chemistry v. 8.}
	\label{tab:3}
\begin{tabular}{ c c c c c c c c}
\hline
\multicolumn{8}{c}{Gaseous Elements Included}\\
\hline
H & He & C & N & O & Na & Mg & Al \\
Si & P & S & Ca & Ti & Cr & Fe & Ni \\
\hline
\end{tabular}
\end{table}

\begin{table}
	\centering
	\caption{The list of possible gaseous species which could form in the equilibrium chemistry code, HSC chemistry v. 8.}
	\label{tab:4}
\begin{tabular}{ c c c c c c c c  }
\hline
\multicolumn{8}{c}{Gaseous Species Included}\\
\hline
Al     &   CrO &        MgOH &        PN  &       AlH  &      CrOH   &      MgS  &      PO                 \\
NS    &    SO       &    \ce{ CH4}      &    FeS     &     Na      &    SO2      &      CN    &    HC        \\             
Ca    &   HPO      &     NiH     &      SiP    &     CaH     &     HS     &        Cr   &     MgH       \\
 P     &   \ce{TiO2}    &       CrN    &     MgO    &    CaS    &      Mg    &         O  &        TiN          \\            
 CrS &       C  &      FeOH &     \ce{ H2O}    &   Ni     &    SiO &     TiO &     CrH    \\
\ce{N2}   &   \ce{Al2O}     &    AlOH    &     FeH    &   \ce{ NH3}      &   \ce{ S2}        &    \ce{ Na2}     &   Si           \\             
\ce{CO2} &  HCN   &        NaO     &     SiH   &      NiO     &    \ce{SiP2}      &    CaO      & \ce{H2S}         \\        
NiS     &  Ti        &       PH    &        TiS     &    AlS   &       FeO       &    NO    &    SN        \\
 PS   &      Fe    &       S &     \ce{H2}   &    NaH     &   SiC &  SiS  &    CaOH       \\
 HCO &    NaOH  &    SiN &     AlO &     S &   \ce{O2}   &     N &    MgN   \\
CO &   NiOH  &     CP &  He & & & & \\
\hline
\end{tabular}
\end{table}

\begin{table}
	\centering
	\caption{The list of possible solid species which could form in the equilibrium chemistry code, HSC chemistry v. 8.}
	\label{tab:5}
\begin{tabular}{ c c c c  }
\hline
\multicolumn{4}{c}{Solid Species Included}\\
\hline
\ce{Al2O3} &     \ce{ FeSiO3} &         \ce{CaAl2Si2O8} &           C         \\
SiC       &          \ce{Ti2O3}       & \ce{Fe3C}  &          \ce{Cr2FeO4}         \\
\ce{Ca3(PO4)2}  &     TiN          &      \ce{Ca2Al2SiO7}     &      Ni          \\
P       &               \ce{Fe3O4}          &         CaS         &            Si      \\
\ce{MgSiO3}      &       Cr         &              \ce{H2O}        &      \ce{CaMgSi2O6}     \\
  \ce{Fe3P}        &          \ce{CaTiO3}          &              Fe &                  AlN       \\
 \ce{MgAl2O4} &           \ce{Mg3Si2O5(OH)4}   &         MgS  &         \ce{CaAl12O19}  \\
        TiC      &               FeS   &      \ce{Mg2SiO4} &      \ce{Fe2SiO4} \\
    \ce{NaAlSi3O8} & & & \\
\hline
\end{tabular}
\end{table}

\begin{table}
	\centering
	\caption{The inputted gaseous elemental abundances, the values are in kmol and are representative of the solar nebula.}
	\label{tab:6}
\begin{tabular}{ c c   }
\hline
Element & Input \\
\hline
Al & $2.34 \times 10^{6}$ \\
C &  $2.45 \times 10^{8}$  \\
Ca &  $2.04 \times 10^{6}$ \\
Cr &  $4.37 \times 10^{5}$ \\
Fe &  $2.82 \times 10^{7}$ \\
H &  $1.00 \times 10^{12}$ \\
He &  $8.51 \times 10^{10}$ \\
Mg &  $3.39 \times 10^{7}$ \\
N &  $6.03 \times 10^{7}$ \\
Na &  $1.48 \times 10^{6}$ \\
Ni &  $1.70 \times 10^{6}$ \\
O &  $4.57 \times 10^{8}$ \\
P &  $2.29 \times 10^{5}$ \\
S &  $1.38 \times 10^{7}$ \\
Si &  $3.24 \times 10^{7}$ \\
Ti &  $7.94 \times 10^{4}$ \\
\hline
\end{tabular}
\end{table}

\begin{figure*}
\centering
	\includegraphics[width=\textwidth]{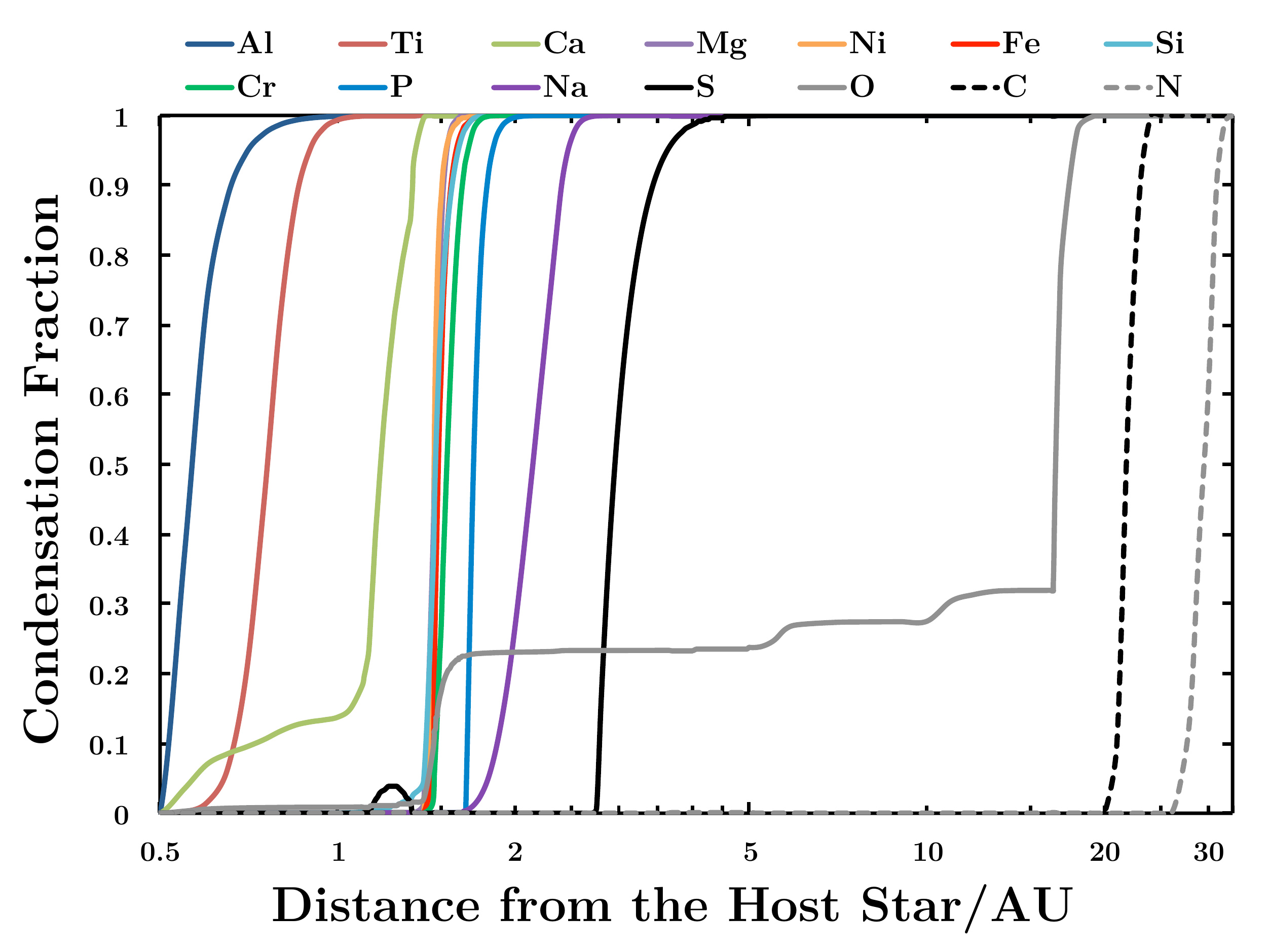}
    \label{fig:C12}
\caption{The abundance of each element which is present in solid species relative to each elements overall abundance at various locations in the protoplanetary disc at t=0\,Myrs. The solid lines were calculated using HSC chemistry v.8 with the inputs given in Appendix \protect\ref{gibbs}, whereas the dashed lines, whose main solid components are ice species, which are not modelled in HSC chemistry, were estimated using the ice lines for the relevant species given in \protect\cite{Marboeuf2014}.}
\end{figure*}

Figure \ref{fig:C12} shows how the percentage of each element in solid state relative to gaseous state changes with increasing radial separation from the host star at time equals zero in the protoplanetary disc for a solar chemistry and was produced by inputting a solar nebula composition (Table \ref{tab:6}) into the equilibrium chemistry model.

Figure \ref{fig:C12} illustrates how our model can reproduce the expected condensation series found in much more advanced simulations \citep{Lodders2003,Lodders2010}.  We find that in accordance with \cite{Lodders2003} the elements condense out of the disc into solid species at around the temperatures expected and in the correct order. The elements defined as refractory (Al, Ti, and Ca) are the first to fully condense out of the nebula. The moderate-volatiles (Mg, Ni, Si, Fe, and Cr) are the next species to fully condense out. The volatiles (P, Na, S, and O) and finally the atmophiles (whose only abundant solid species in our model are ices) condense out last (C, N, and the Noble Gases). We again note that the trend for C has many caveats and is only used as a guide for a limit on formation conditions. The physical reason the elements condense out in this order is due to the compounds that most readily form at the pressures and temperatures seen in a protoplanetary disc, and the readiness of those compounds to be in the gaseous or solid phase under those conditions. The condensation series presented here, for solar elemental abundances at time equals 0\,Myrs, holds over the stellar compositional range analysed in this work and over all formation times up to 2.5\,Myrs.

\FloatBarrier
\FloatBarrier

%%%%%%%%%%%%%%%%%%%%%%%%%%%%%%%%%%%%%%%%%%%%%%%%%%

% Don't change these lines
\bsp	% typesetting comment
\label{lastpage}
\end{document}